\documentclass[longauth, traditabstract]{aa} 

\usepackage{graphicx}
\usepackage{txfonts}
\usepackage{fixltx2e}
\usepackage{amssymb}
\usepackage{natbib}
\usepackage{rotating}
\usepackage[breaklinks=true]{hyperref}
\usepackage{lscape}
\usepackage{array}
\newcolumntype{H}{>{\setbox0=\hbox\bgroup}c<{\egroup}@{}}
\bibpunct{(}{)}{;}{a}{}{,}

\begin{document}
\title{ Linking low- to high-mass YSOs with \textit{Herschel}-HIFI observations
  of water\thanks{{\it Herschel} is an ESA space observatory with science instruments provided by European-led Principal Investigator consortia and with important participation from NASA.} }

\titlerunning{ Excited water emission in YSOs: from low- to high-mass}

\author{
  I.~San~Jos\'{e}-Garc\'{i}a\inst{\ref{inst1}}\thanks{\email{sanjose@strw.leidenuniv.nl}}
  \and J.~C.~Mottram\inst{\ref{inst1}}  
  \and E.~F.~van Dishoeck\inst{\ref{inst1},\ref{inst3}}
   \and L.~E.~Kristensen\inst{\ref{inst111}} 
   \and F.~F.~S.~van~der~Tak\inst{\ref{inst10},\ref{inst11}}
   \and J.~Braine\inst{\ref{inst6},\ref{inst46}} 
   \and F.~Herpin\inst{\ref{inst6},\ref{inst46}}
  \and D.~Johnstone\inst{\ref{inst7},\ref{inst8}}
  \and T.~A.~van~Kempen\inst{\ref{inst1}} 
   \and F.~Wyrowski\inst{\ref{inst30}}
}

\institute{
  Leiden Observatory, Leiden University, PO Box 9513, 2300 RA Leiden, 
  The Netherlands. \label{inst1} 
  \and
   Max Planck Institut f{\"u}r Extraterrestrische Physik, Giessenbachstrasse 2, 
  85478 Garching, Germany. \label{inst3}
  \and
   Harvard-Smithsonian Center for Astrophysics, 60 Garden Street, Cambridge, MA 02138, USA\label{inst111}
  \and
  SRON Netherlands Institute for Space Research, PO Box 800, 9700 AV 
  Groningen, The Netherlands\label{inst10}
  \and
  Kapteyn Astronomical Institute, University of Groningen, PO Box 800, 
  9700 AV Groningen, The Netherlands\label{inst11}
  \and
  Universit\'{e} de Bordeaux, Observatoire Aquitain des Sciences de l'Univers, 
  2 rue de l'Observatoire, BP 89, F-33270 Floirac Cedex, France \label{inst6}
  \and
  CNRS, LAB, UMR 5804, Laboratoire d'Astrophysique de Bordeaux, 2 rue de l'Observatoire, 
  BP 89, F-33270 Floirac Cedex, France\label{inst46} 
  \and
  National Research Council Canada, Herzberg Institute of Astrophysics, 
  5071 West Saanich Road, Victoria, BC V9E 2E7, Canada\label{inst7}
  \and
 Department of Physics and Astronomy, University of Victoria, Victoria, 
  BC V8P 1A1, Canada\label{inst8}
   \and
  Max-Planck-Institut f\"{u}r Radioastronomie, Auf dem H\"{u}gel 69, 53121 Bonn, 
  Germany\label{inst30}
}

\date{October 5, 2015}

\def \c2{cm$^{-2}$}
\def \cc{cm$^{-3}$}
\def \ccs{cm$^{3}$ s$^{-1}$}
\def \kms{km\ s$^{-1}$}
\def \Kkms{K kms$^{-1}$}
\def \nh2{n_{H_2}}
\def \nh1{n_{HI}}
\def \ng{n$_g$}

\def \Lbol{$L_{\rm{bol}}$}
\def \Lsun{$L_{\odot}$}
\def \Menv{$M_{\rm{env}}$}
\def \FWZI{{\it FWZI}}
\def \FWHM{{\it FWHM}}
\def \FWHMb{{\it FWHM}$_{\rm b}$}

\def \LCO{$L_{\rm{CO}}$}
\def \Lwater{$L_{\mathrm{H}_{2}\mathrm{O}}$}
\def \Teva{$T_{\rm{eva}}$}
\def \sixteen15{\mbox{$J$=16--15}}
\def \ten9{\mbox{$J$=10--9}}
\def \nine8{\mbox{$J$=9--8}}
\def \five4{\mbox{$ J=$ 5--4}}
\def \three2{\mbox{$J$=3--2}}
\def \four3{\mbox{$ J=$ 4--3}}
\def \six5{\mbox{$ J=$ 6--5}}
\def \seven{\mbox{2$_{11}$-2$_{02}$}}
\def \eight{\mbox{2$_{02}$-1$_{11}$}}
\def \tennine{\mbox{3$_{12}$-3$_{03}$}}

\def \nh3{NH$_3$}
\def \n2h{N$_2$H$^+$}
\def \twco{$^{12}$CO}
\def \thco{$^{13}$CO}
\def \cei{C$^{18}$O}
\def \csev{C$^{17}$O}
\def \htwo{H$_2$}
\def \water{H$_2$O}
\def \hheo{H$_2^{18}$O}
\def \otwo{O$_2$}
\def \ox18{$^{16}$O$^{18}$O}
\def \hco{HCO$^+$}


%

%
\def\FigAveragedWater{
  \begin{figure*}[!th]
    \centering
    \includegraphics[scale=0.65, angle=0]{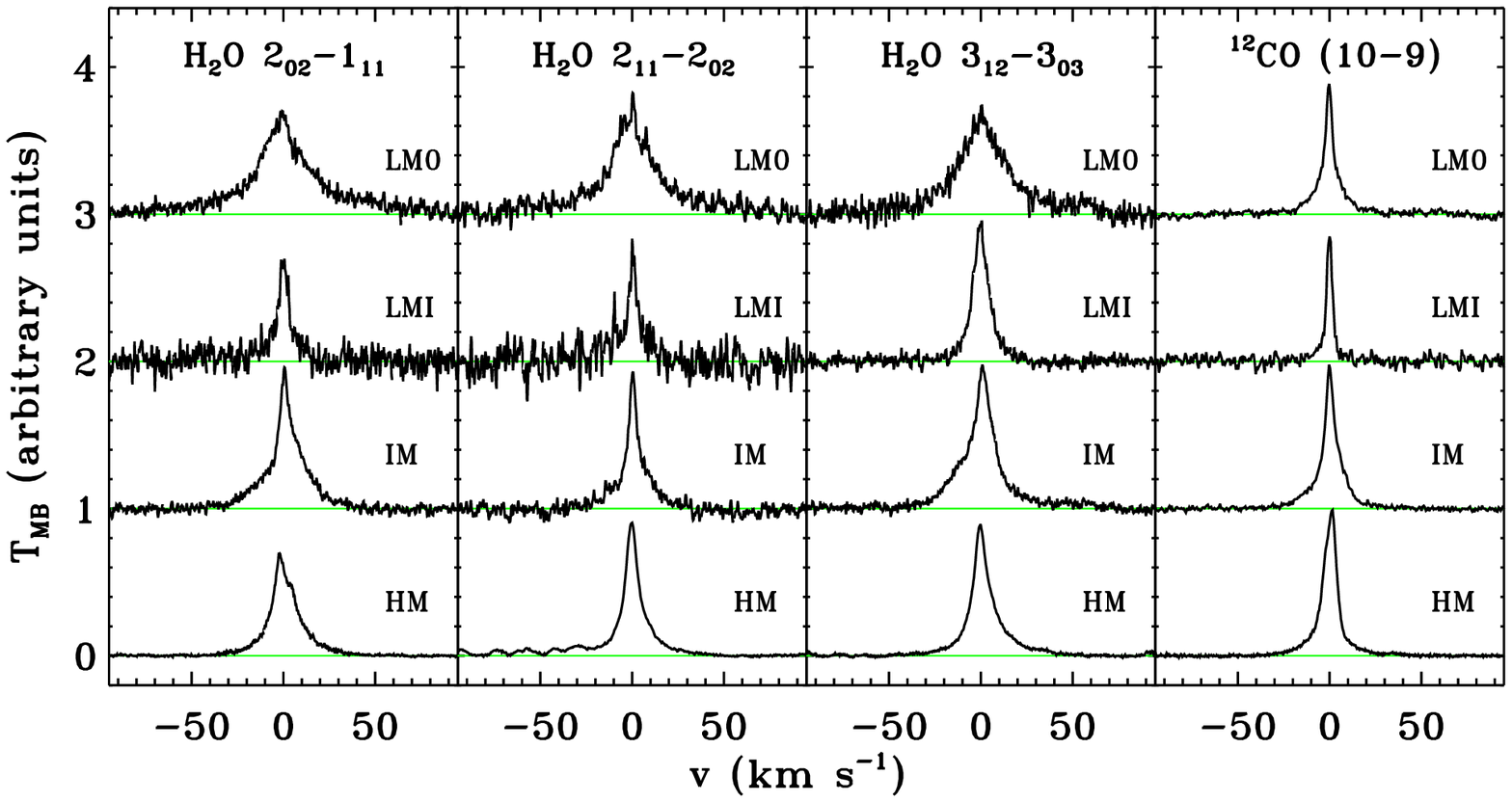}
     \includegraphics[scale=0.65, angle=0]{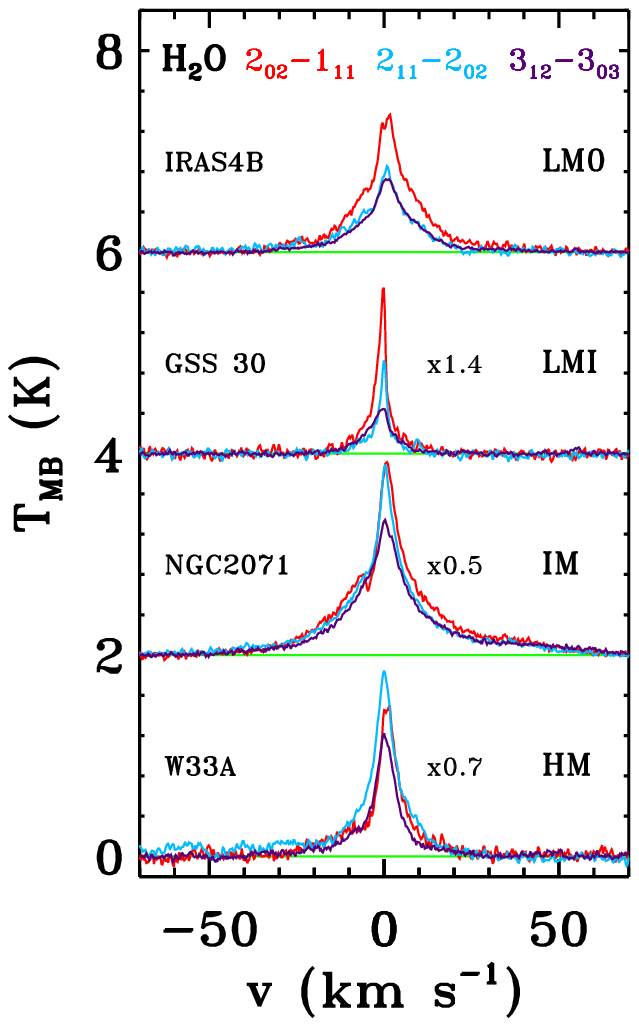}
    	\caption{ {\it Left} figure: Averaged and normalised spectrum calculated for the low-mass Class~0 (LM0) protostars, the low-mass Class~I (LMI), intermediate-mass YSOs (IM) and high-mass objects (HM) for the \water~\eight\ 988 GHz ({\it left} panel), \seven\ 752 GHz ({\it middle-left} panel), \tennine\ 1097 GHz ({\it middle-right} panel) transitions and the \twco\ \ten9 ({\it right} panel) spectra \citep[see][]{13SanJoseGarcia}. 
	The spectra of each sub-group of YSOs have been shifted vertically for visualisation purposes. 
	The low intensity features on the blue wing of the \water~\seven\ high-mass profile are due to methanol emission. 
	{\it Right} figure: \water\ \eight, \seven\ and \tennine\ spectra plotted in red, blue and purple respectively for NGC\,1333\,IRAS4B (LM0), GSS\,30 (LMI), NGC2071 (IM) and W33A (HM). 
	The horizontal light green lines in both figures represent the baseline level. }
    \label{fig4:AveragedWater}
  \end{figure*}
}

%
\def\FigFWZICOwatervsLbol{
  \begin{figure}[!t]
    \centering
     \includegraphics[scale=0.6, angle=0]{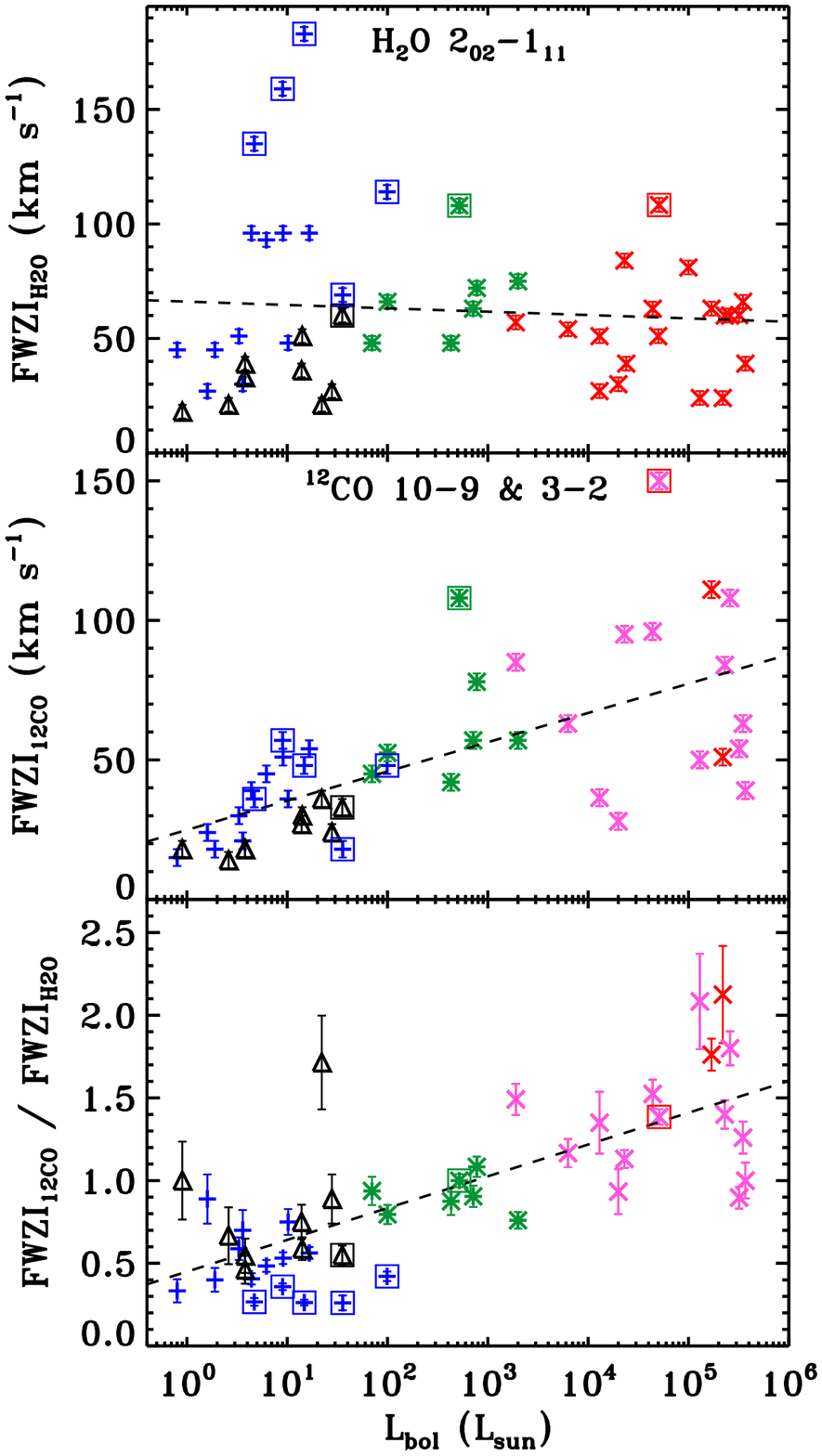}
    \caption{ ({\it Top}) \FWZI\ of the \water\ \eight\ (988 GHz) emission line versus the bolometric luminosity. 
	({\it Middle}) Same as {\it top} panel but for the \twco\ \ten9 and \three2 observations.
	({\it Bottom}) Ratio of the \twco\ and \water\ \eight\ \FWZI\ values as a function of \Lbol. 
	The blue plus symbols correspond to the low-mass Class~0 protostars, 
      the black triangles the low-mass Class~I, the green asterisks the intermediate-mass objects, the pink
      crosses the high-mass YSOs for which the $^{12}$CO~$J=3$--2 spectra are used, and the 
      red cross symbols the high-mass object for which $^{12}$CO~$J=10$--9 data are available \citep[see][]{13SanJoseGarcia}.
      The low- and intermediate-mass sources with detected EHV components are surrounded by a box, as well as the high-mass YSO with triangular water line profiles. }
    \label{fig4:FWZI_12CO_H2O_vsLbol}
  \end{figure}
}

%

%
\def\FigLwatervsLbol{
  \begin{figure}[!t]
    \centering
     	\includegraphics[scale=0.6, angle=0]{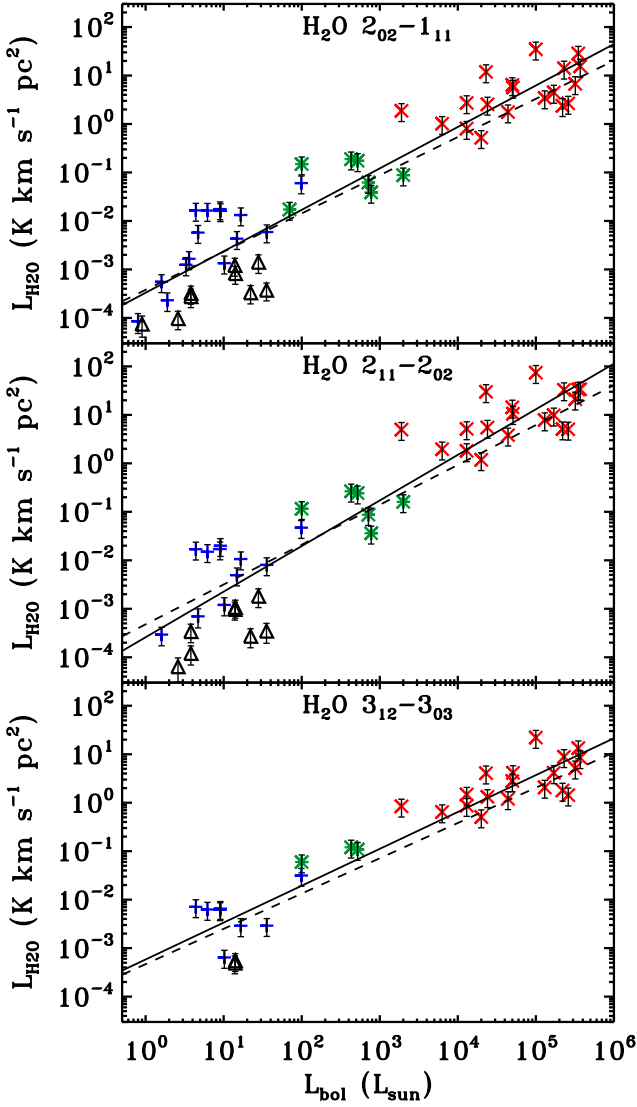}
    \caption{ Line luminosity of the H$_2$O 2$_{02}$-1$_{11}$ (988 GHz) line emission ($top$), the H$_2$O 2$_{11}$-2$_{02}$ (752 GHz) data ($middle$), and  H$_2$O 3$_{12}$-3$_{03}$ (1097 GHz) spectra ($bottom$) versus the bolometric luminosity of the source.
	The blue plusses correspond to the low-mass Class~0 protostars, the black triangles the low-mass Class~I, the green asterisks the intermediate-mass objects and the red cross symbols the high-mass YSOs.
        The solid line indicates the linear correlation of the logarithm of the total line luminosity, log(\Lwater), and log(\Lbol).
        The dashed line shows the log-log correlation of the luminosity measured for the broader Gaussian velocity component only (cavity shock emission; $L_{\mathrm{broad\,H}_{2}\mathrm{O}}$) and log(\Lbol).}
    \label{fig4:LH2O_vs_Lbol}
  \end{figure}
}

%

%
\def\FigRatiosLuminoLbol{
  \begin{figure*}[h]
    \centering
    \includegraphics[scale=0.6, angle=0]{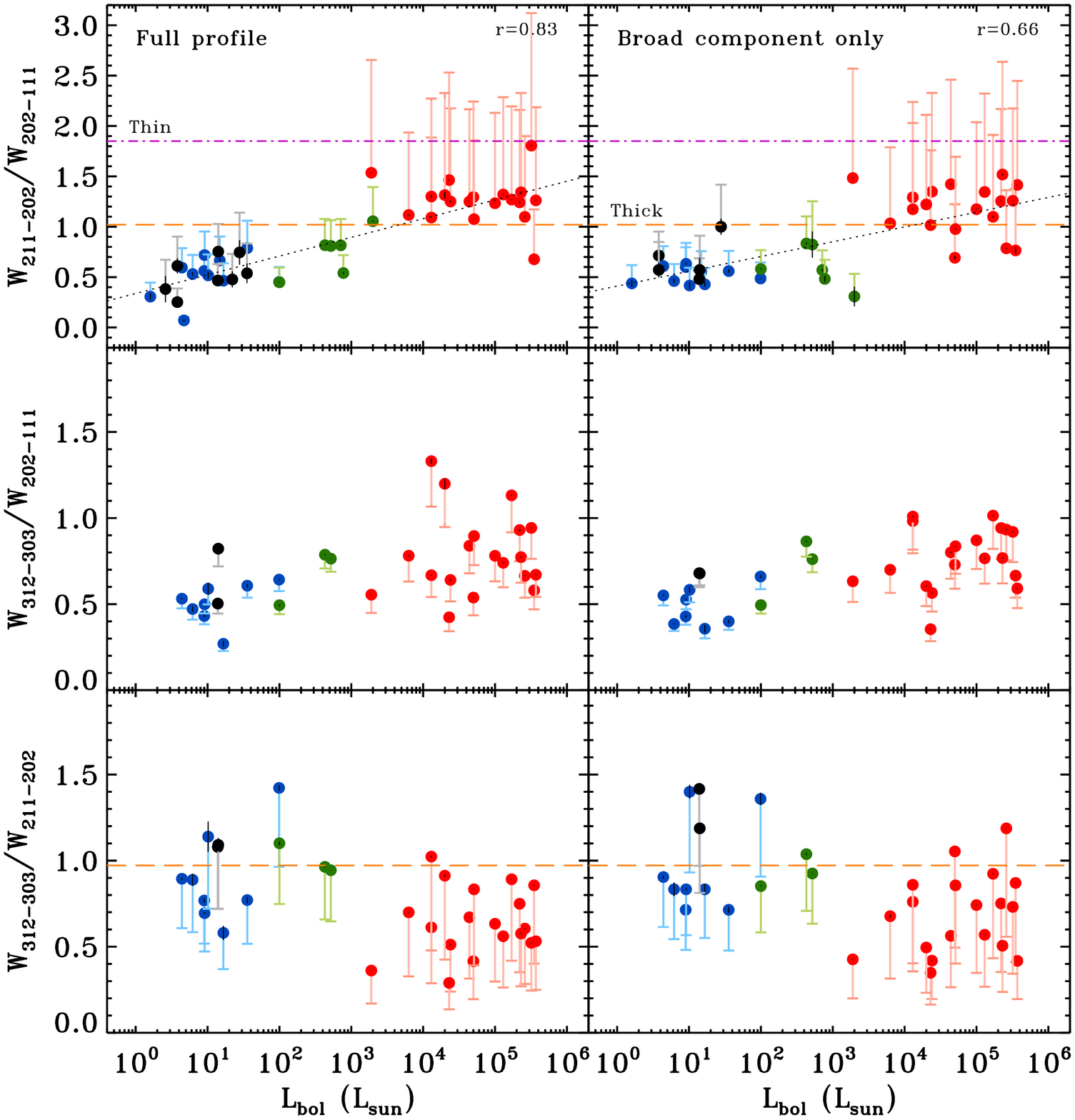}
    \caption{  (\textit{Top panels}) Ratio of the integrated intensities of the \seven\ (752 GHz) and \eight\ (988 GHz) water lines versus the bolometric luminosity of the source.   
    	(\textit{Middle panels}) Ratio of the integrated intensities of the \water\ \tennine\ (1097 GHz) and \eight\ (988 GHz) spectra versus \Lbol. 
	(\textit{Bottom panels}) Ratio of the integrated intensities of the \tennine\ (1097 GHz) and \seven\ (752 GHz) water emission lines as a function of \Lbol.
	The {\it left}-column presents the ratios calculated considering the entire line profile and the {\it right}-column shows the ratio of the integrated intensity calculated for the broad velocity component. 
	The value of the ratios with and without correcting by the different beam size factors are indicated by dashed lighter lines and darker dot symbols respectively. 
      Blue lines and symbols correspond to the low-mass Class~0 protostars, black to the low-mass Class~I, green to intermediate-mass objects and red lines and symbols to high-mass YSOs.
      The purple dashed-dotted horizontal lines and the orange dashed horizontal lines indicate the optically thin and optically thick limits, respectively, calculated assuming LTE and an excitation temperature of 300~K. 
      The linear correlation between the dot symbols (ratios not beam corrected) and the logarithm of the luminosity is shown by the dotted black lines in the {\it top} panels. }
    \label{fig4:ratioLH2OvsLbol}
  \end{figure*}
}

%
\def\FigIntensityRatios{
  \begin{figure}[!t]
    \centering
    \includegraphics[scale=0.6, angle=0]{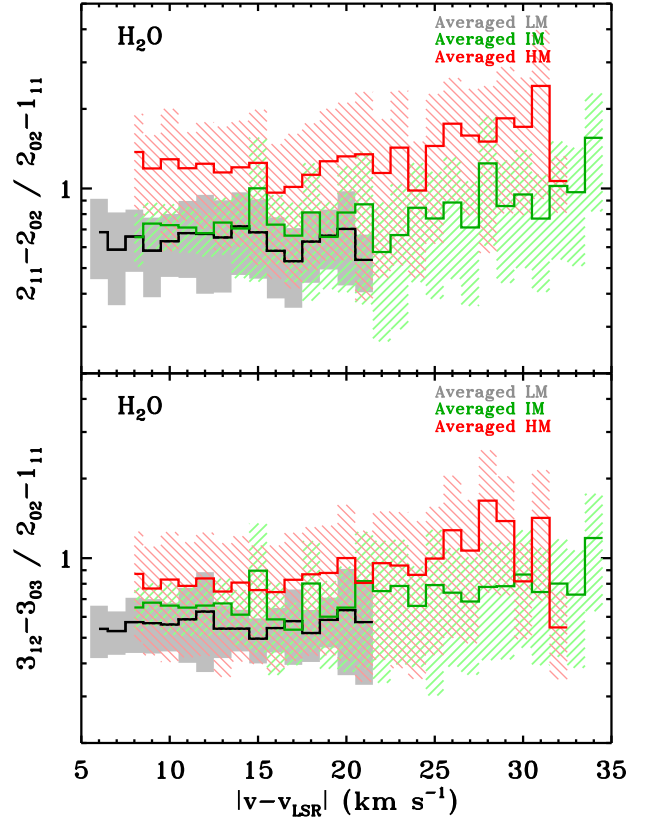}
       	\caption{ Averaged \water\ \seven\ / \eight\ ratio ($top$ panel) and \water\ \tennine/\eight line intensity ratio ($bottom$ panel) as a function of offset velocity for the low-mass protostars (LM; grey area from \citealt{14Mottram}), intermediate-mass objects (IM; green line and asterisks), and the high-mass YSOs (HM; red line and crosses).
	The dashed green and red regions indicate the calculated standard deviation of the line ratio for each velocity channel.}
    \label{fig4:Intensity_ratios}
  \end{figure}
}

%
\def\FigIntensityRatiosTwo{
  \begin{figure}[!t]
    \centering  
    \includegraphics[scale=0.6, angle=0]{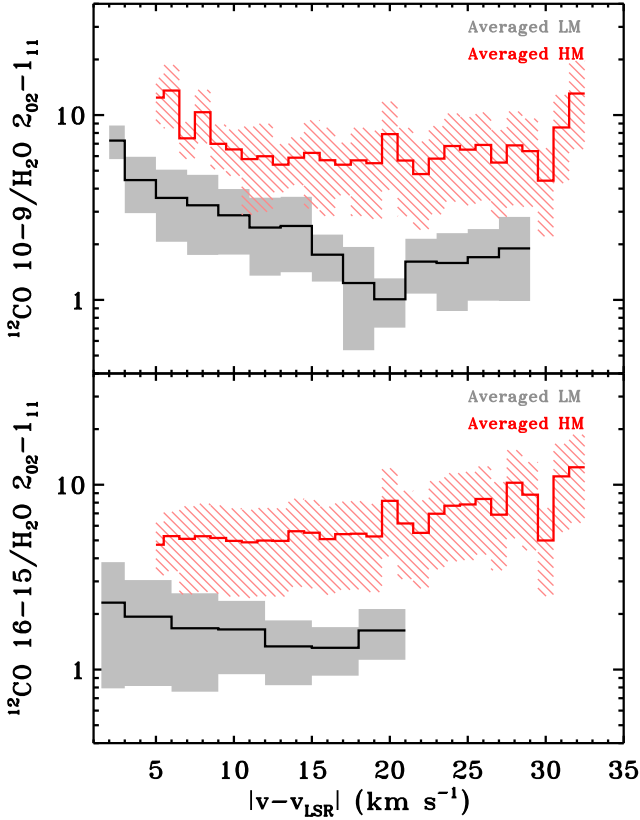} 
    	\caption{ Same as Fig.~\ref{fig4:Intensity_ratios} but for the \twco\ 10--9 ($top$ panel) and 16--15 ($bottom$ panel) transitions. }
    \label{fig4:Intensity_ratios2}
  \end{figure}
}

%
\def\FigExcitation{
  \begin{figure*}[t]
    \centering
   \includegraphics[scale=0.4, angle=0]{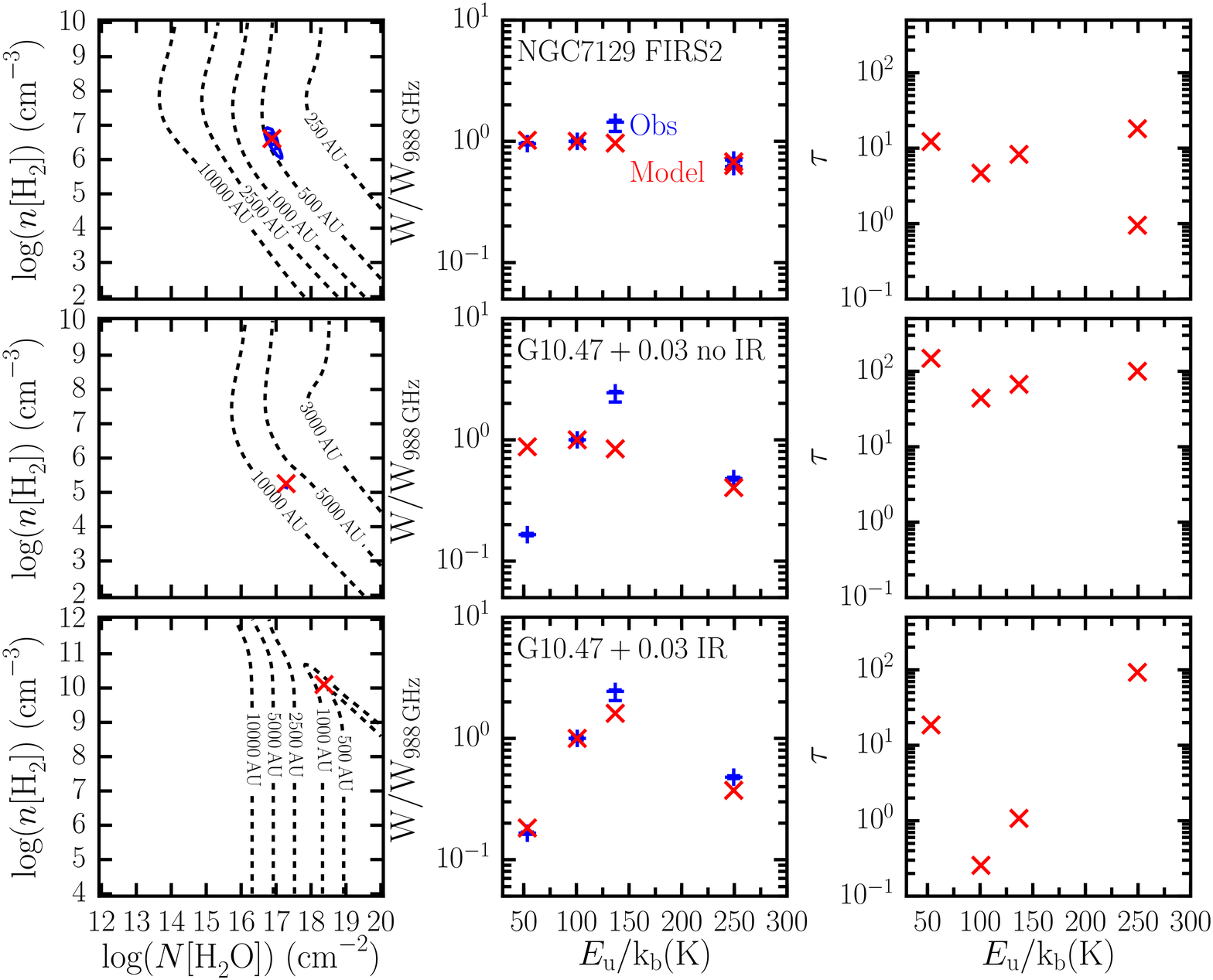}
    \caption{  {\sc radex} results for the average line ratios for the cavity shock components assuming a kinetic temperature of $T$ = 300~K. 
    The {\it top} panels corresponds to the intermediate-mass object NGC7129\,FIRS2 in which no infrared radiation field has been included. 
    The {\it middle} and {\it bottom} panels show the results for the high-mass YSOs G10.47+0.03 without and with radiation field respectively. 
    For each row, the {\it left}-hand panels show the best-fit (red cross), the 1, 3 and 5$\sigma$ confidence limits (blue contours) for a grid in $n_{\mathrm{H}_{2}}$ and $N_{\mathrm{H}_{2}\mathrm{O}}$ and the corresponding size of the emitting region in AU (black dashed lines). 
    The {\it middle} panels show a spectral line energy distribution comparing the observed and best-fit model with blue and red symbols respectively. 
    Finally, the {\it right}-hand panels present the optical depth, $\tau$, of the best-fit model for each \water\ line.}
    \label{fig4:Excitation}
  \end{figure*}
}

%
\def\FigcartoonOutflowCavityWall{
  \begin{figure}[!t]
    \centering
    \includegraphics[scale=0.76, angle=0]{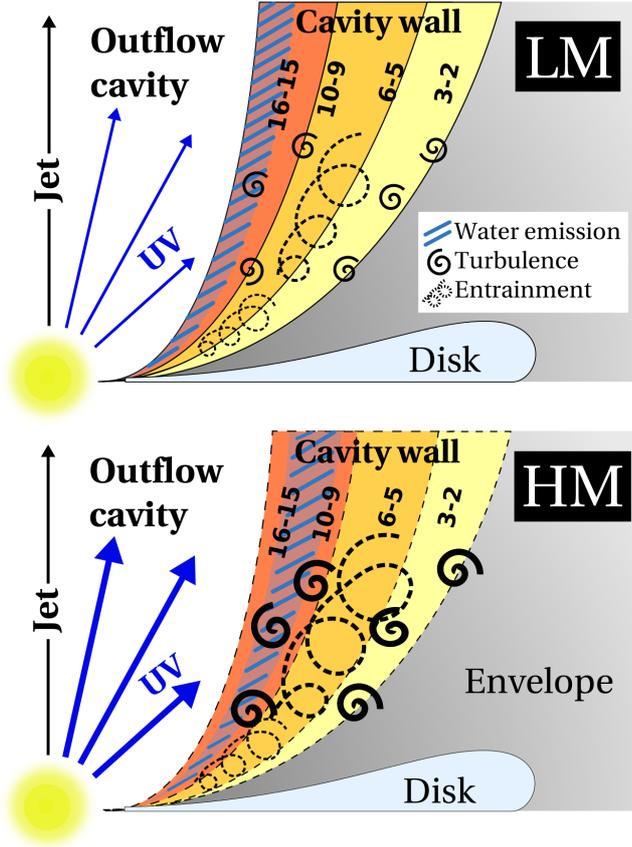}
    \caption{ Cartoon illustrating a scenario with a simplified physical structure of the different layers composing the outflow cavity wall for a representative low-mass protostar (LM; {\it top} panel) and for a high-mass YSO (HM; {\it bottom} panel). 
     The emitting area of the low-$J$ \twco\ transitions is shaded in yellow, that of the mid-$J$ transitions is orange and of the high-$J$ lines in red. Turbulent motions are represented with spiral symbols, the entrained material with swirls and the excited water emission is indicated with blue lines over the red region. }
    \label{fig4:cartoonOutflowCavityWall}
  \end{figure}
}

%
\def\FigLwatervsLbolExtrag{
  \begin{figure}[!t]
    \centering
    \includegraphics[scale=0.6, angle=0]{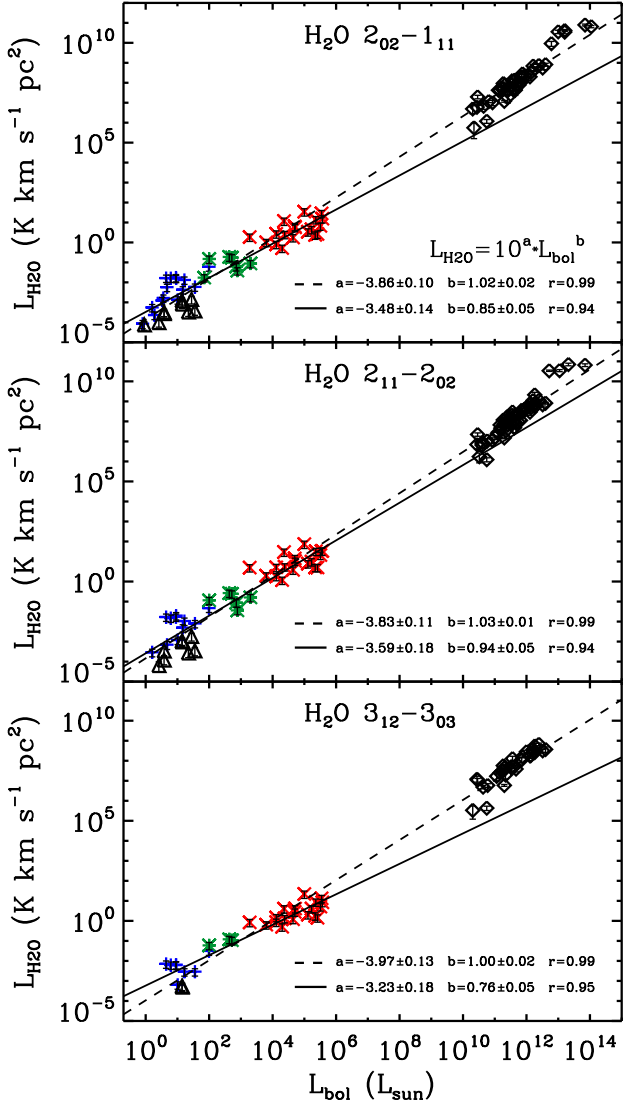}
    \caption{ Same as Fig.\ref{fig4:LH2O_vs_Lbol} but including the line luminosity observed for several extragalactic sources (black diamonds) taken from \citet{13Yang}. 
      The solid black line corresponds to the fit of the WISH YSOs and the dashed black line indicates the correlation including the values of extragalactic sources. }
    \label{fig4:LH2O_vs_Lbol_extra}
  \end{figure}
}

%
\def \FigLwaterNorvsEuExtrag{
  \begin{figure}[!t]
    \centering
    \includegraphics[scale=0.6, angle=0]{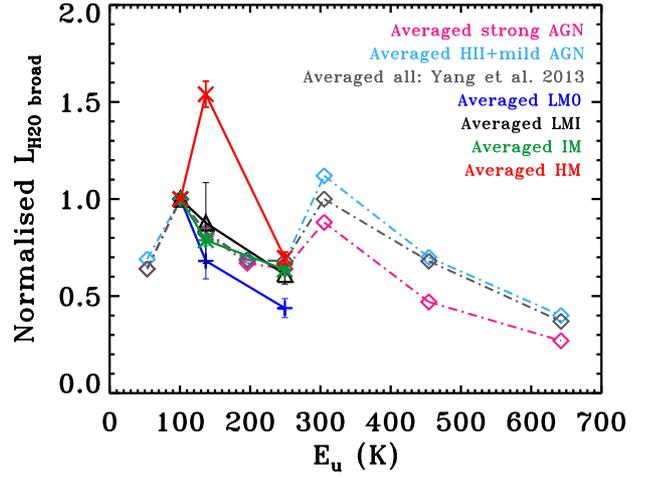}
      \caption{ Line luminosity of the water transitions normalised to the luminosity of the \water~\eight\ (988 GHz) line as a function of the upper energy level ($E_{\rm u}$) of each transition considering only the contribution from the cavity shock component. 
      The solid blue, black, green and red lines corresponds to the averaged value of the normalised intensity for the low-mass Class~0, Class~I, intermediate-mass and high-mass YSOs respectively. 
      The grey dashed line represents the average values of the sample presented in \citet{13Yang}, the pink and light blue dashed lines indicate the values of the strong AGN- and H{\sc ii}+mild-AGN- dominated galaxies, respectively.
      The \seven\ (752 GHz) line has an $E_\mathrm{u}/k_{\mathrm{B}}$ = 137~K and the \tennine\ (1097 GHz) an $E_\mathrm{u}/k_{\mathrm{B}}$ = 249~K.}
    \label{fig4:ratioLH2O_vs_Eu_extrag}
  \end{figure}
}


%
\def\FigNineGHz{
    \begin{figure*}[t]
      \centering
      \bigskip
      \includegraphics[scale=0.45, angle=0]{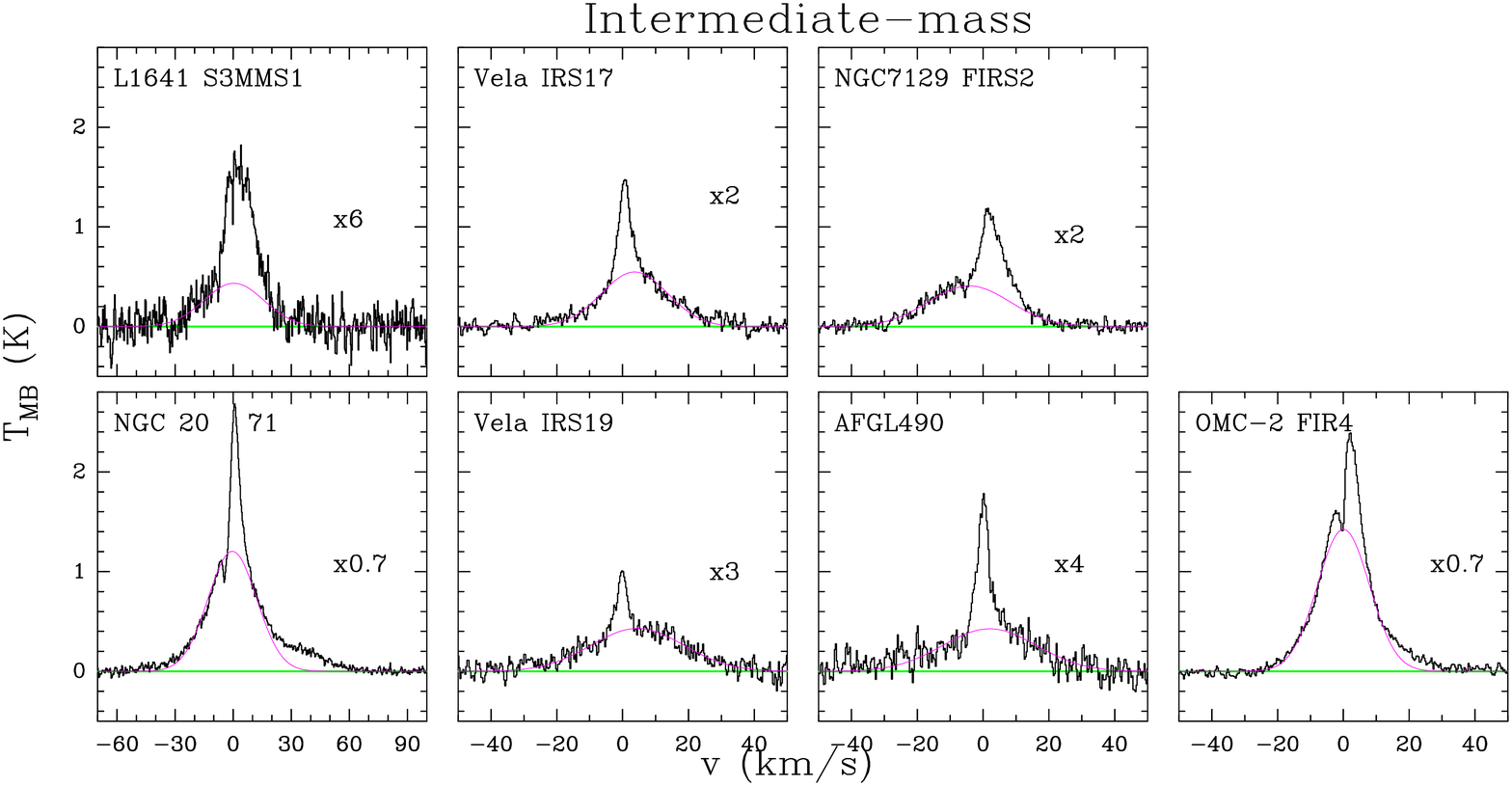}
      \bigskip
      \includegraphics[scale=0.45, angle=0]{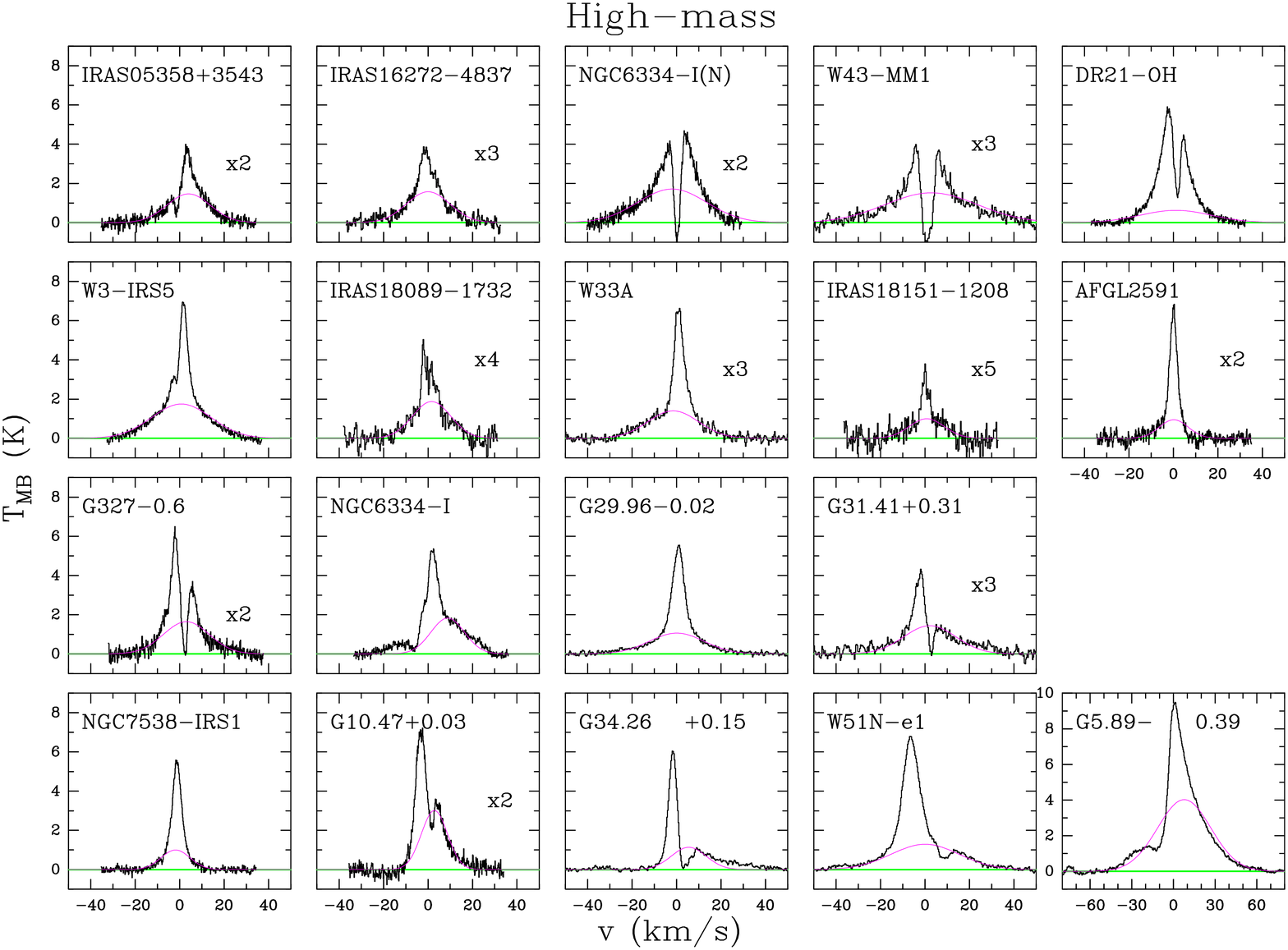}
      \caption{ H$_2$O 2$_{02}$-1$_{11}$ spectra for the intermediate-mass YSOs ({\it top}) and high-mass sources ({\it bottom}). 
        The green line represents the baseline level and the pink Gaussian the broad velocity component. 
        All spectra have been shifted to zero velocity. 
        The numbers indicate where the spectra have been scaled for greater visibility.}
      \label{fig4:988GHzspectra}
    \end{figure*}
}

%
\def\FigSevenGHz{
    \begin{figure*}[t]
      \centering
      \bigskip
      \includegraphics[scale=0.45,angle=0]{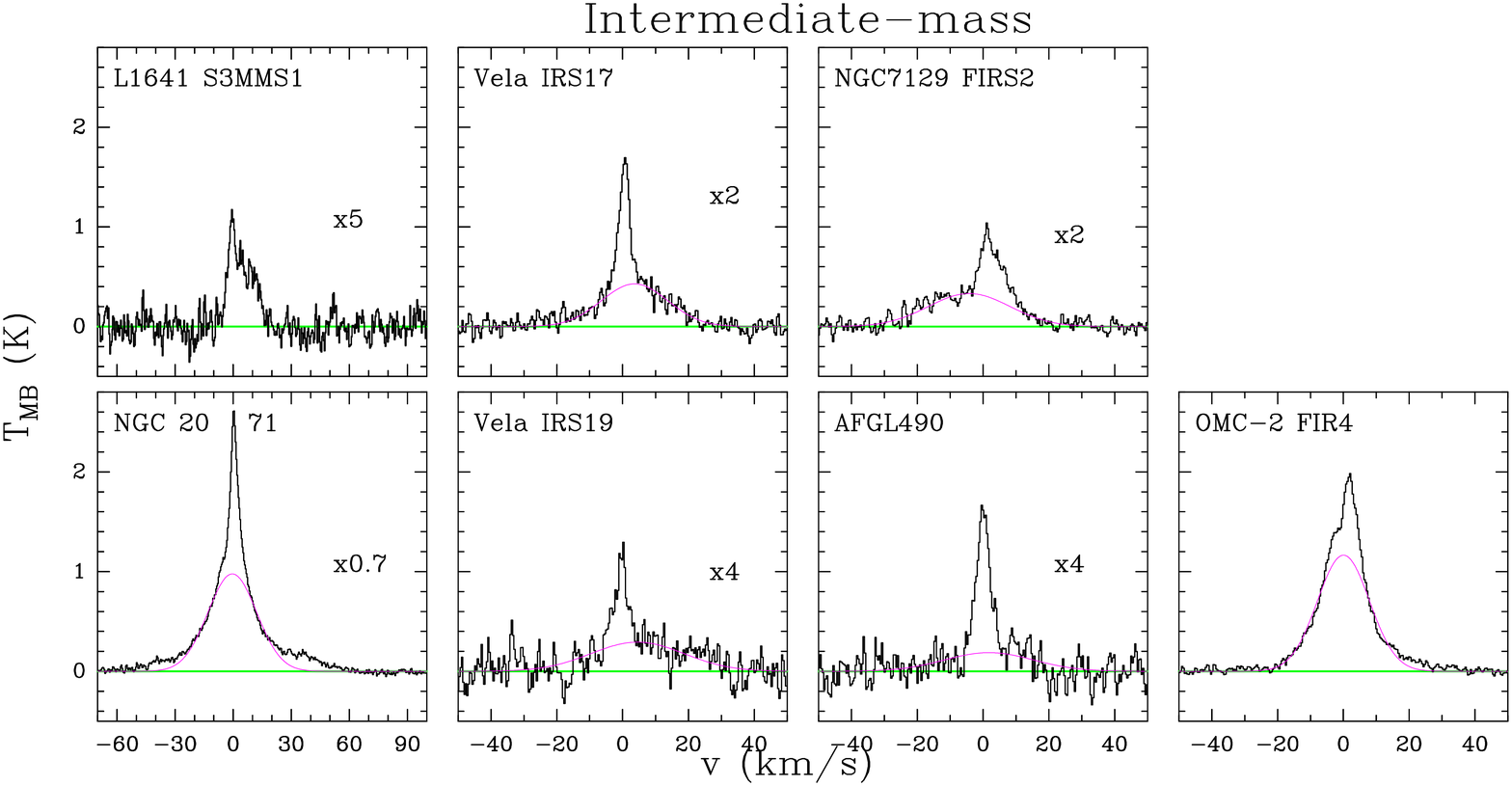}
      \bigskip
      \includegraphics[scale=0.45, angle=0]{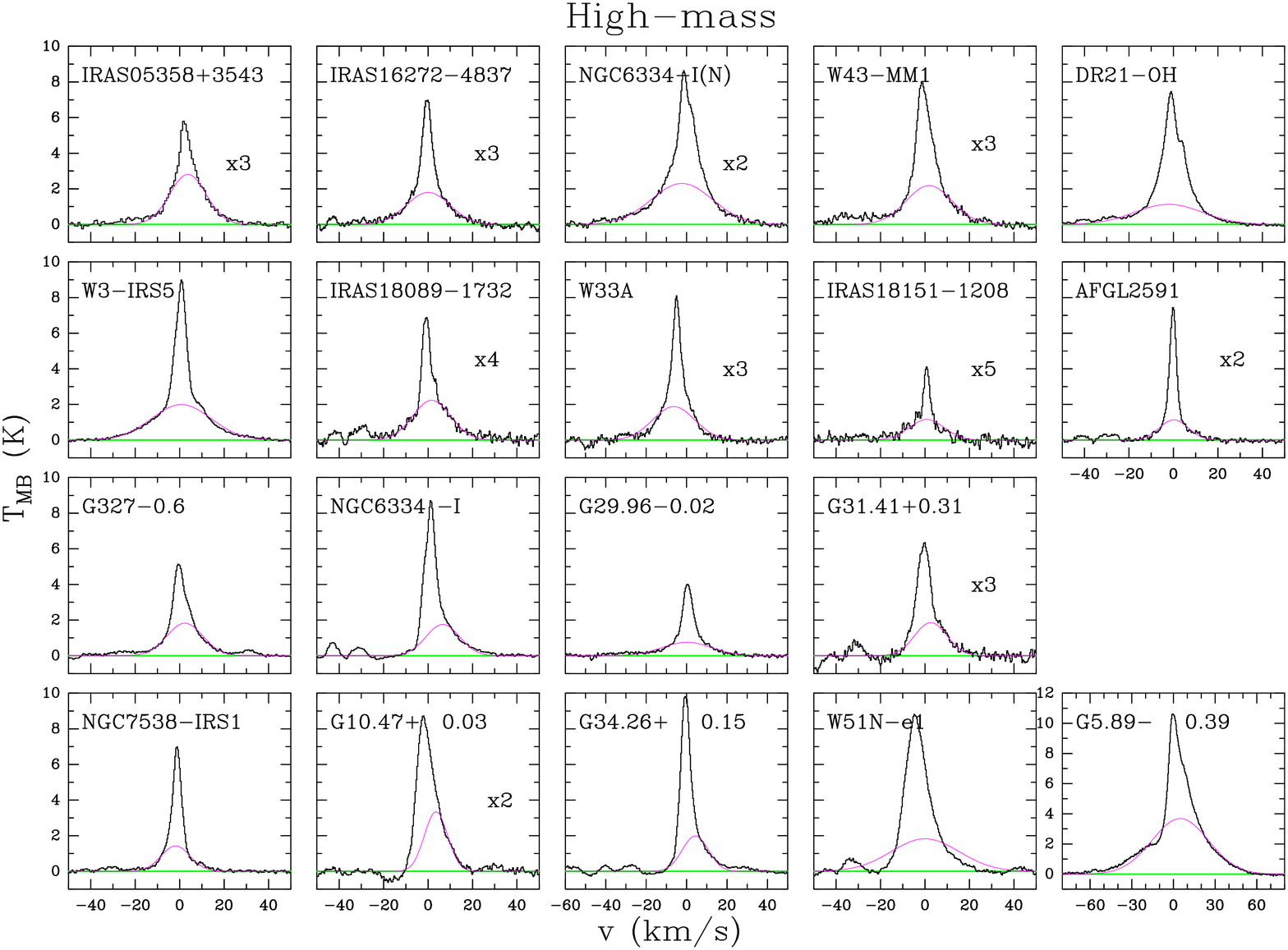}
      \caption{ H$_2$O 2$_{11}$-2$_{02}$ spectra for the intermediate-mass YSOs ({\it top}) and high-mass sources ({\it bottom}). 
        The green line represents the baseline level and the pink Gaussian the broad velocity component. 
        All spectra have been shifted to zero velocity. 
        The numbers indicate where the spectra have been scaled for greater visibility.}
      \label{fig4:752GHzspectra}
    \end{figure*}
}

%
\def\FigTenGHz{
    \begin{figure*}[t]
      \centering
      \bigskip
      \includegraphics[scale=0.45, angle=0]{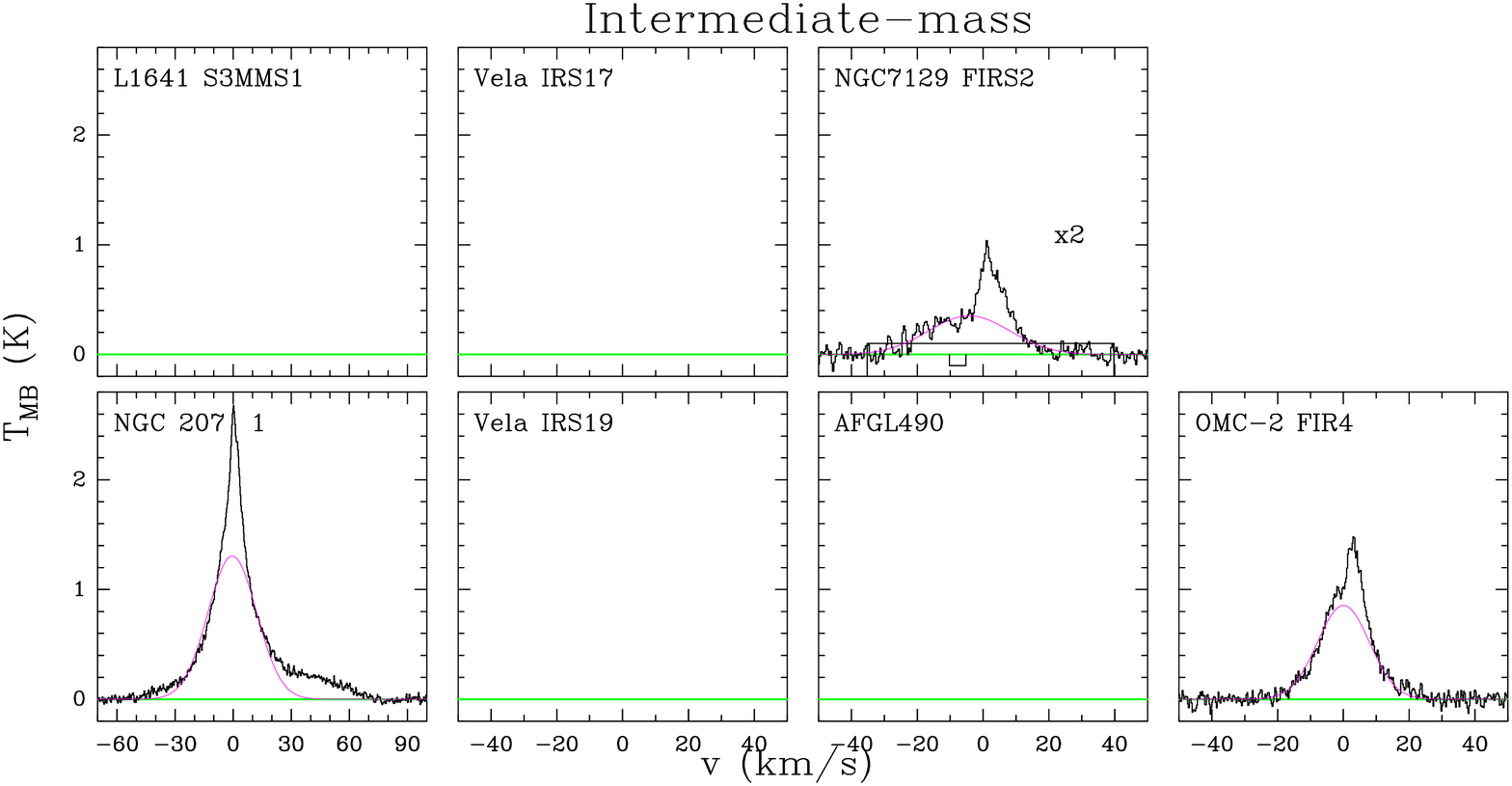}
      \bigskip
      \includegraphics[scale=0.45, angle=0]{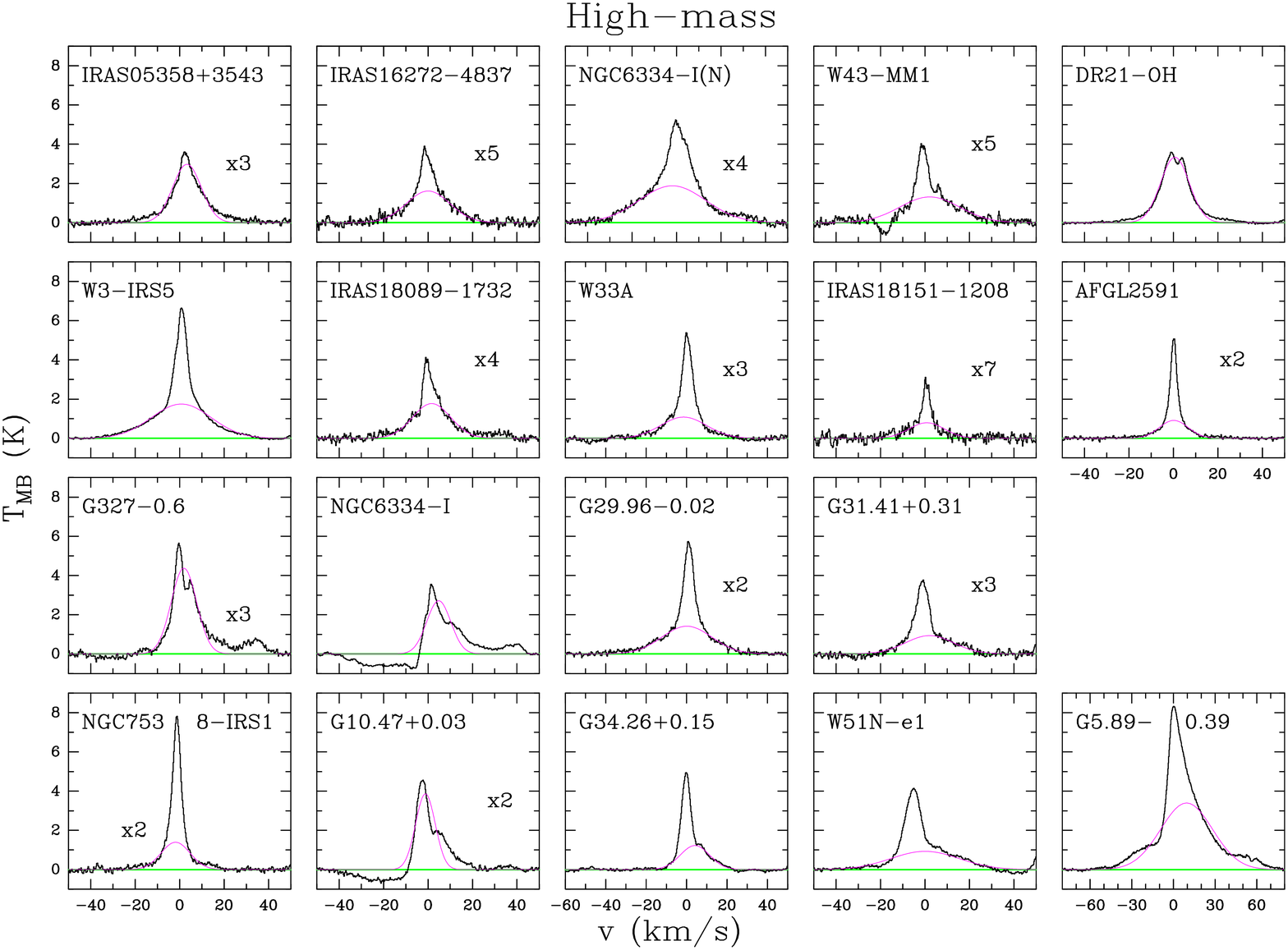}
      \caption{ H$_2$O 3$_{12}$-3$_{03}$ spectra for the intermediate-mass YSOs ({\it top}) and high-mass objects ({\it bottom}). 
        The green line represents the baseline level and the pink Gaussian the broad velocity component. 
        All spectra have been shifted to zero velocity. 
        The numbers indicate where the spectra have been scaled for greater visibility.}
      \label{fig4:1097GHzspectra}
    \end{figure*}
}


%
\def\FigFWHMvsMenvforCOwater{
  \begin{figure*}[!t]
    \centering
    \includegraphics[scale=0.6, angle=0]{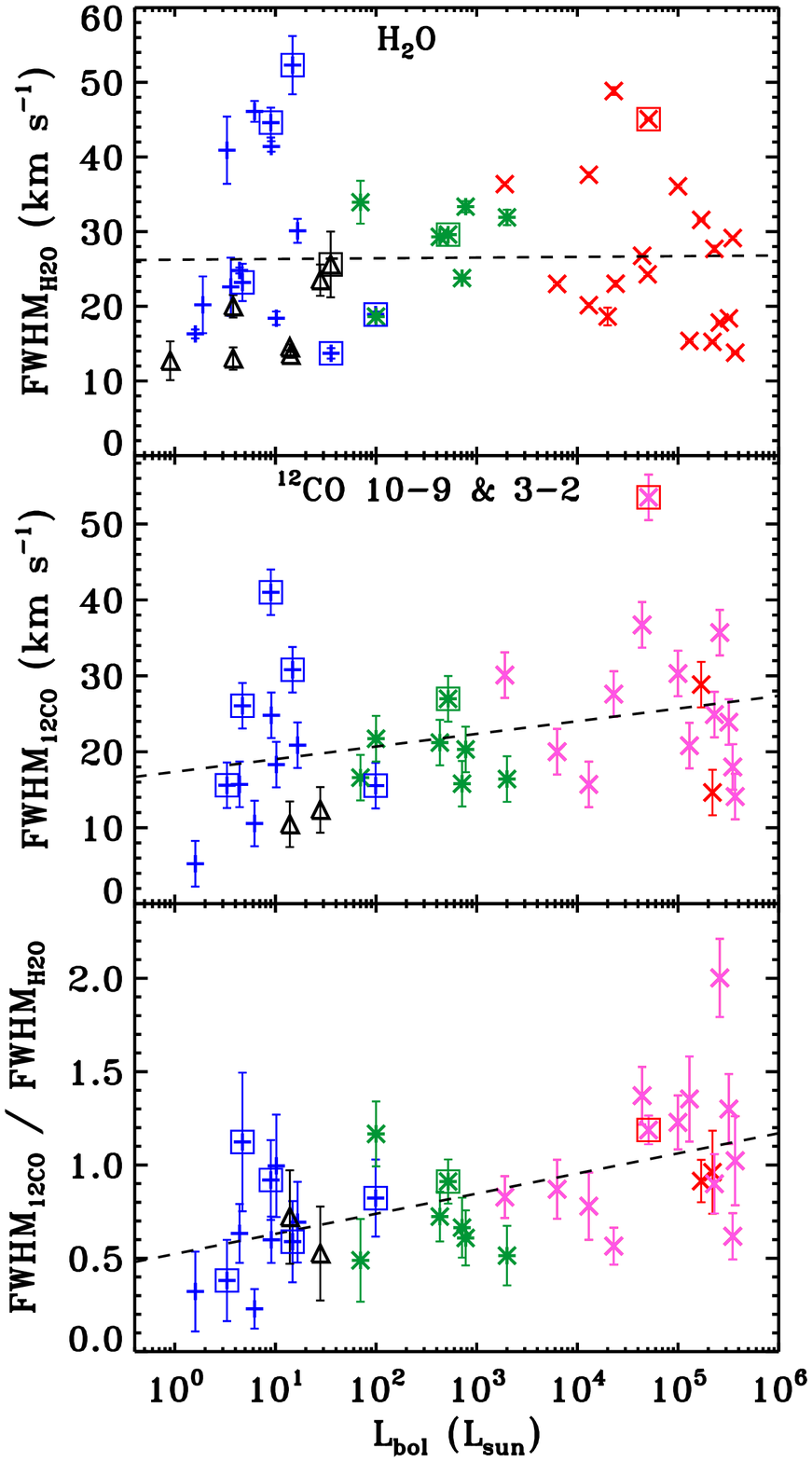}
    \includegraphics[scale=0.6, angle=0]{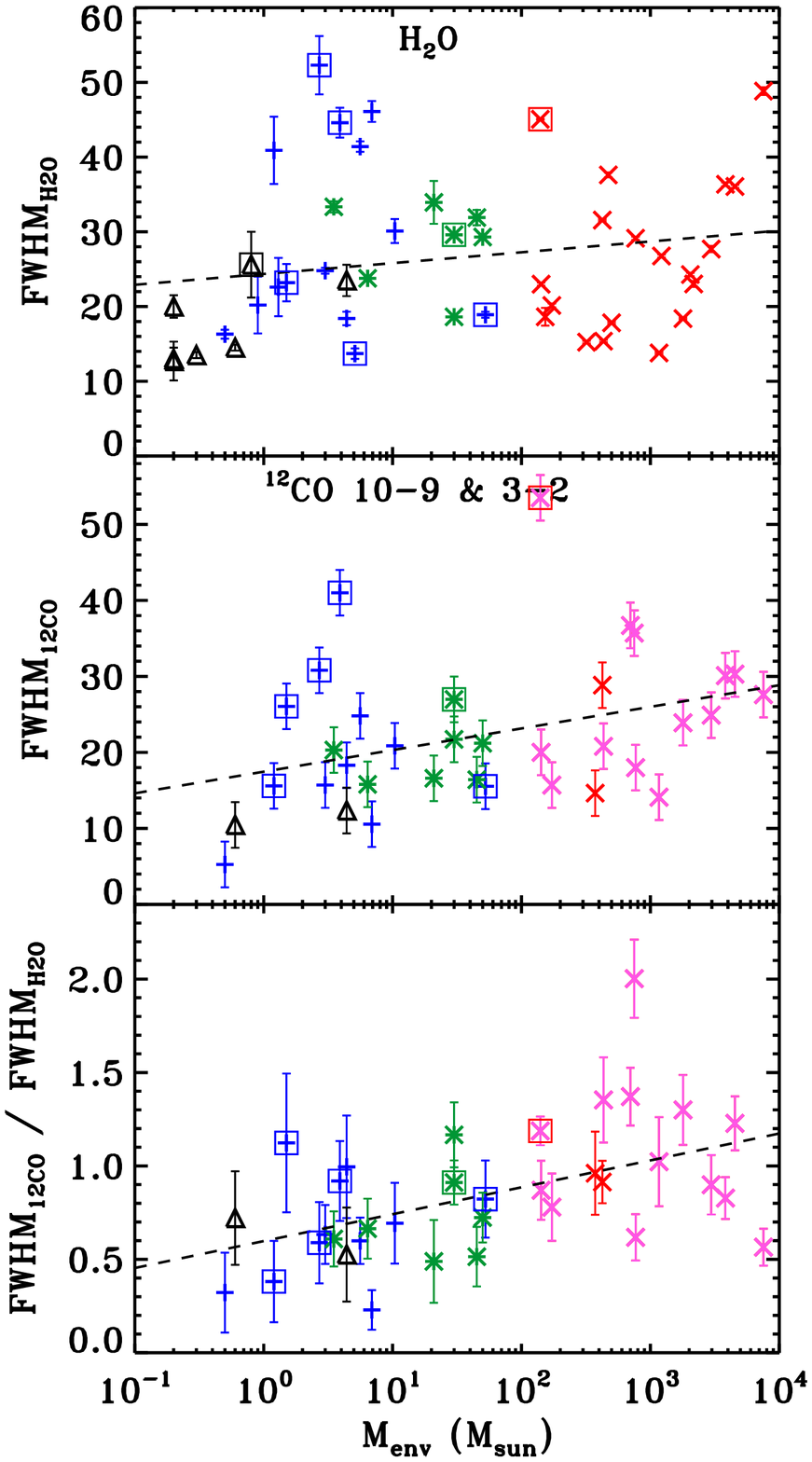}
    \caption{ ({\it Left-}column figure) Derived \FWHMb\ of the Gaussian profile fitted to the cavity shock component of the \water\ lines ({\it top}-panel)
     	as a function of bolometric luminosity.
	Constrained \FWHMb\ for the \twco\ \ten9 and \three2 observations ({\it middle}-panel) versus \Lbol. 
	Ratio calculated from the \twco\ \FWHMb\ divided by the \FWHMb\ of the \water\ lines ({\it bottom}-panel) as a function of \Lbol. 
	({\it Right}-column figure) Same as {\it left-} figure but plotted versus the envelope mass of the source, \Menv.
	The blue plus symbols correspond to the low-mass Class~0 protostars, 
      the black triangles the low-mass Class~I, the green asterisks the intermediate-mass objects, the pink
      crosses the high-mass YSOs for which the $^{12}$CO~$J=3$--2 spectra are taken, and the 
      red cross symbols the high-mass object for which $^{12}$CO~$J=10$--9 data are available \citep[see][]{13SanJoseGarcia}. 
      The low- and intermediate-mass objects with detected EHV components are surrounded by a box.
      Also the high-mass YSO with triangular water line profiles is surrounded by a box. 
      The value of the \FWHMb\ of the different velocity components is calculated as explained in Sect.~\ref{ch4_Decomposition_method} and these parameters are the same for the three water transitions.}
    \label{fig4:FWHM_12CO_H2O_vsLbol&Menv}
  \end{figure*}
}

%
\def\FigFWZIvsLbolMenv{
  \begin{figure*}[!t]
    \centering
    \includegraphics[scale=0.6, angle=0]{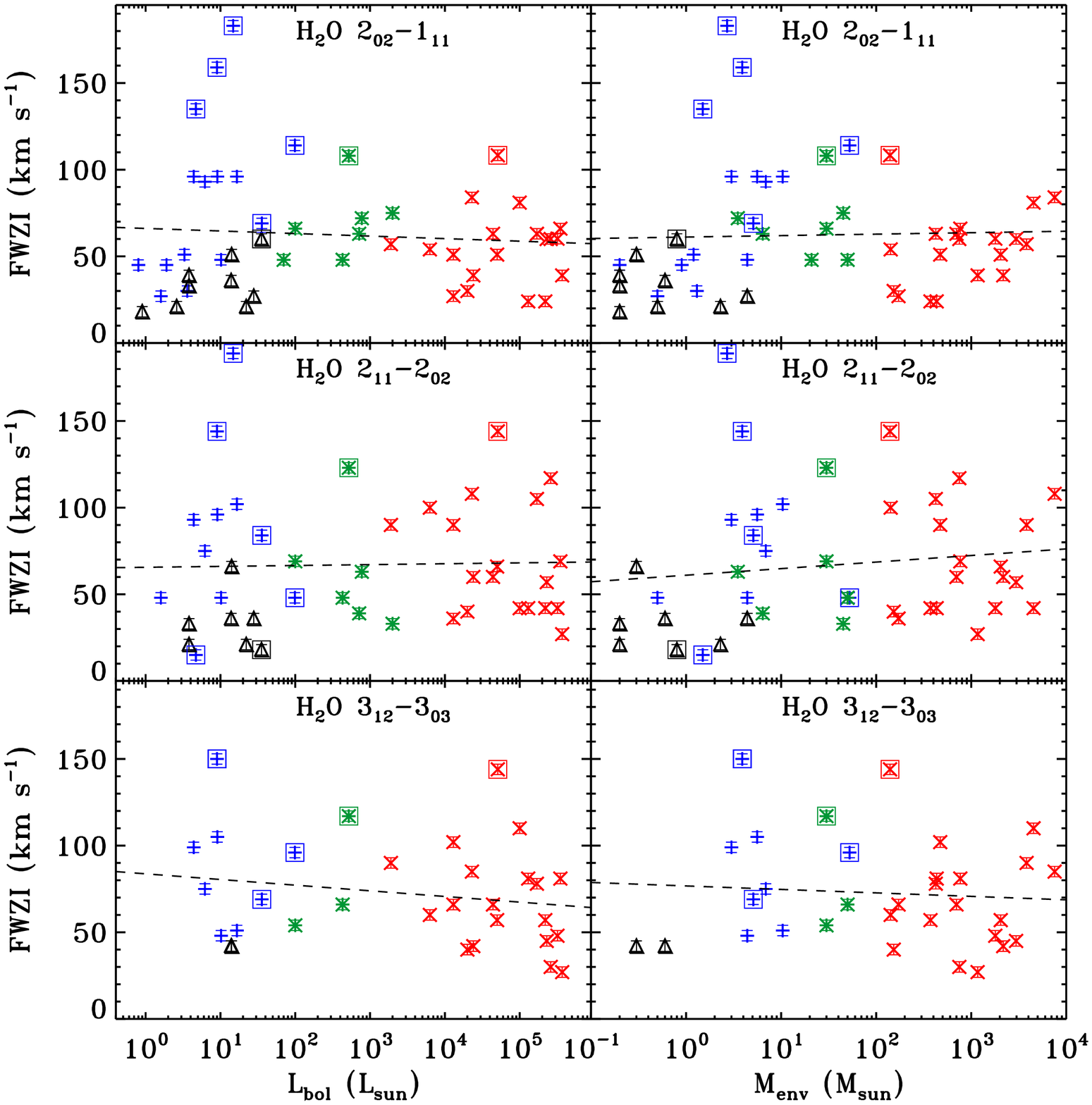}
    \caption{ ({\it Left}-column) \FWZI\ of the \water~\eight\ 988 GHz ({\it top} panel), \seven\ 752 GHz ({\it middle} panel) and \tennine\ 1097 GHz ({\it bottom} panel) transitions as a function of the bolometric luminosity of each source. 
    ({\it Right}-column) Same as {\it left}-column but versus the envelope mass of each YSO. 
    The blue plus symbols correspond to the low-mass Class~0 protostars, the black triangles the low-mass Class~I, the green asterisks to the intermediate-mass objects and the red crosses to the high-mass YSOs. 
      The low- and intermediate-mass objects with detected EHV components are surrounded by a box, as well as the high-mass YSO with triangular line profiles. 
      \FWZI\ is calculated by binning the spectra to 3~km\,s$^{-1}$.}
    \label{fig4:FWZIvsLbolMenv}
  \end{figure*}
}

%
\def\FigFWZIcovsLbolMenv{
  \begin{figure*}[!ht]
    \centering
	\includegraphics[scale=0.6, angle=0]{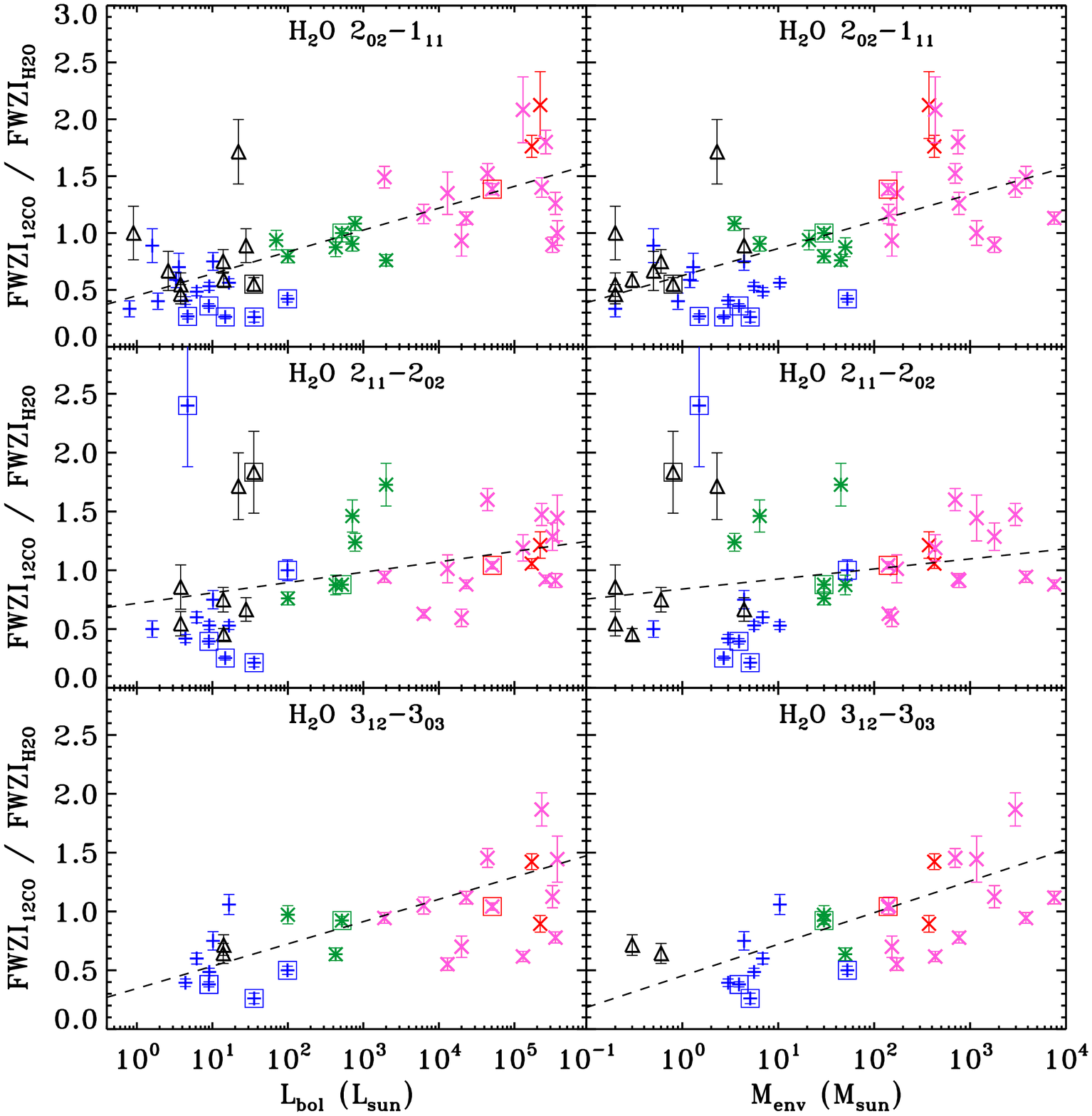}
    \caption{ ({\it Left}-column) Ratio of the \FWZI\ of the \water~\eight\ 988 GHz ({\it top} panel), \seven\ 752 GHz ({\it middle} panel) and \tennine\ 1097 GHz ({\it bottom} panel) transitions and the \FWZI\ of the \twco\ observations as a function of the bolometric luminosity.
       ({\it Right}-column) Same as {\it left}-column but versus the envelope mass of each YSO. 
       The low- and intermediate-mass sources with detected EHV components are surrounded by a box, and also the high-mass YSO with triangular line profiles. 
             Both values of \FWZI\ were calculated by binning the spectra to 3~km\,s$^{-1}$.
      The symbol and colour code is the same as in Fig.~\ref{fig4:FWZI_12CO_H2O_vsLbol}.}
    \label{fig4:FWZI_12COvsLbolvsMenv}
  \end{figure*}
}

%
\def\FigLwatervsMenv{
  \begin{figure}[!t]
    \centering
     \includegraphics[scale=0.6, angle=0]{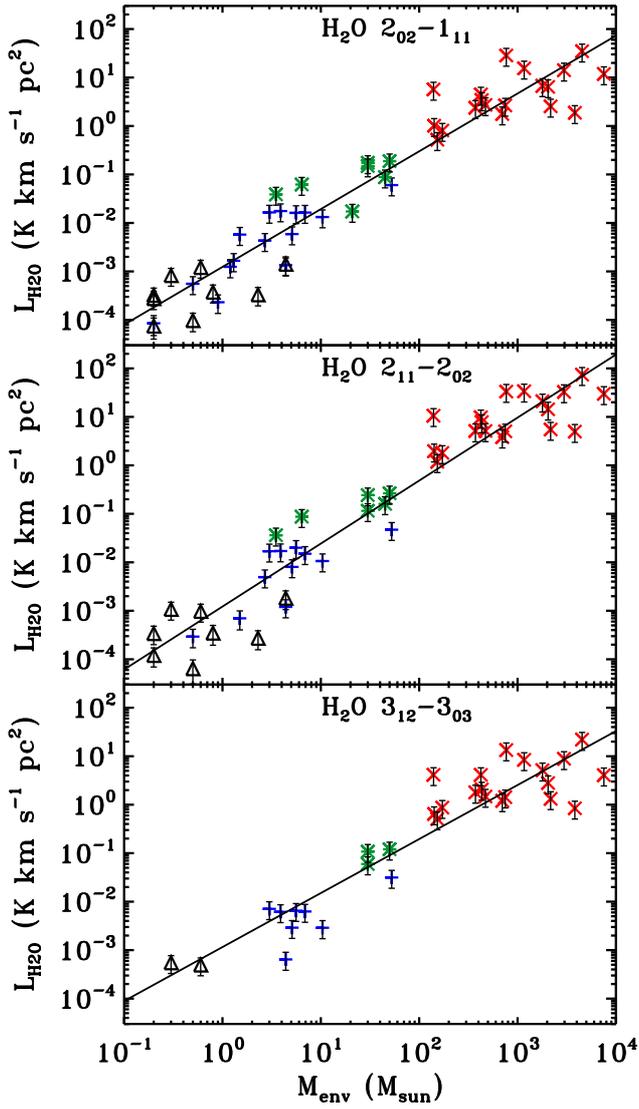}
    \caption{ Same as Fig.~\ref{fig4:LH2O_vs_Lbol} but plotted as a function of the envelope mass of the source.}
    \label{fig4:LH2O_vs_Menv}
  \end{figure}
}

%
\def\FigLbroadnorvsLbol{
  \begin{figure}[!ht]
    \centering
	\includegraphics[scale=0.6, angle=0]{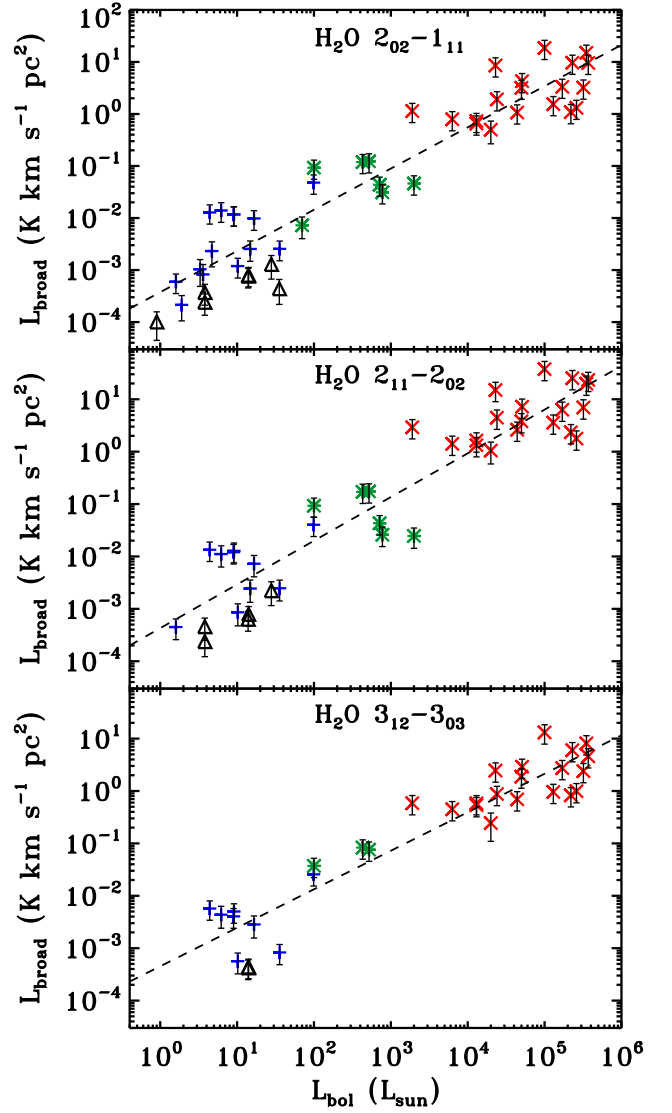}
    \caption{ Line luminosity of the broad velocity component (emission from shocked gas along the outflow cavity) versus the bolometric luminosity of the source.
    The symbol and colour code is the same as in Fig.~\ref{fig4:LH2O_vs_Lbol}.
    The dashed black line shows the log-log correlation of the luminosity measured for the cavity shock emission and \Lbol.}
    \label{fig4:LH2Onor_vsLbol}
  \end{figure}
}




\def\TableOverviewWATER{
  \begin{table*}[!h]
      \begin{center}
    \caption{\label{tbl4:H2O_lines} Overview of the main properties of the observed water lines with HIFI.}
      \begin{tabular}{l r r r c c c c c}
        \hline\hline
        \noalign{\smallskip}        
        Mol. & Trans.& $E_\mathrm{u}/k_{\mathrm{B}}$ & Frequency & Tel./Inst.-band&  $\eta_{\mathrm{MB}}$$^a$ & $\theta$$^a$ & Spec. Resol. & Correction\\
      \noalign{\smallskip}
        & & (K) & (GHz) &  & & ($\arcsec$) &  (km\,s$^{-1}$) & factor$^b$\\
        \hline       
        \noalign{\smallskip}         
        p-H$_2$O  	&  2$_{02}$-1$_{11}$	& 100.8 	&  987.927 	& HIFI-4a 	& 0.74 	&21.5 	& 0.15 	& 0.865\\
        p-H$_2$O 	&  2$_{11}$-2$_{02}$ 	& 136.9 	&  752.033 	& HIFI-2b 	&0.75 	& 28.2 	& 0.20 	&  0.853\\
        o-H$_2$O 	&  3$_{12}$-3$_{03}$ 	& 249.4 	&1097.365 	& HIFI-4b 	&0.74 	& 19.3 	& 0.14 	& 0.865\\[1pt]
       \hline
      \end{tabular}
    \end{center}
    {\bf Notes.} $^{(a)}$ Parameters calculated using equations 1 and 3 of \citet{12Roelfsema}, respectively. 
    $^{(b)}$ Beam efficiency correction factors of each excited water lines according to the updated values of $\eta_{\mathrm{MB}}$ for the different HIFI-bands$^1$. \\
  \end{table*}
}

\def\TableContributionCavity{
  \begin{table*}[!h]
      \begin{center}
    \caption{\label{tbl4:ContributionCavity} Average fraction of the integrated intensity that the narrow (envelope) and broad (cavity shock or entrained outflowing material) components contribute to the total integrated intensity of the \water\ \eight\ (988 GHz), \seven\ (752 GHz), \tennine\ (1097 GHz) and \twco\ \ten9 lines. } 
          \begin{tabular}{l c @{ } c @{ } | @{ } c @{ } c @{ } | @{ } c @{ } c @{ } || @{ } c @{ } c}
        \hline\hline
      \rule{0pt}{2.4ex} & \multicolumn{2}{@{ }c@{ }|@{ }}{ \water\ \eight} & \multicolumn{2}{@{ }c@{ }|@{ }}{ \water\ \seven} & \multicolumn{2}{@{ } c@{ }||@{ }}{ \water\ \tennine} &  \multicolumn{2}{@{ } c}{ \twco\ \ten9}\\
        \rule{0pt}{2.4ex} & $W_{\rm envelope}$/$W_{\rm tot}$$^a$\,\, &  $W_{\rm cavity\,shock}$/$W_{\rm tot}$& $W_{\rm envelope}$/$W_{\rm tot}$$^a$\,\, &  $W_{\rm cavity\,shock}$/$W_{\rm tot}$& $W_{\rm envelope}$/$W_{\rm tot}$$^a$\,\, &  $W_{\rm cavity\,shock}$/$W_{\rm tot}$ &  $W_{\rm envelope}$/$W_{\rm tot}$$^a$\,\, &  $W_{\rm entrainment}$/$W_{\rm tot}$\\
        \hline             
        \rule{0pt}{2.8ex}LM0  	&	0.0 $\pm$ 0.1  &	0.7 $\pm$ 0.1		&	0.0 $\pm$ 0.1  	&	0.7 $\pm$ 0.1		&	0.0 $\pm$ 0.1  	&	0.8 $\pm$ 0.1	&	0.5 $\pm$ 0.1  	&	0.5 $\pm$ 0.1\\
        LMI   				&	0.2 $\pm$ 0.1  &	0.8 $\pm$ 0.1		&	0.3 $\pm$ 0.2	&	0.7 $\pm$ 0.2		&	0.0 $\pm$ 0.1  	&	1.0 $\pm$ 0.1  	&	0.9 $\pm$ 0.1  	&	0.1 $\pm$ 0.1\\
        IM 					&	0.3 $\pm$ 0.1  &	0.6 $\pm$ 0.2		&	0.4 $\pm$ 0.2  	&	0.6 $\pm$ 0.2		&	0.3 $\pm$ 0.1  	&	0.6 $\pm$ 0.1  &	0.5 $\pm$ 0.1  	&	0.5 $\pm$ 0.1\\
        HM 				&	0.4 $\pm$ 0.2  &	0.6 $\pm$ 0.2		&	0.4 $\pm$ 0.2  	&	0.6 $\pm$ 0.2		&	0.4 $\pm$ 0.1  	&	0.6 $\pm$ 0.1  &	0.5 $\pm$ 0.1  	&	0.5 $\pm$ 0.1\\[1pt]
        \hline
      \end{tabular}
    \end{center}
    {\bf Notes.} LM0: low-mass Class~0 protostars; LMI: low-mass Class~I sources; IM: intermediate-mass YSOs; HM: high-mass objects.\\
    $^{(a)}$ As in \citet{14Mottram}, the envelope fraction, $W_{\rm envelope}$, is calculated subtracting the cavity shock (entrained outflowing contribution in the case of \twco) and spot shock contribution from the total integrated intensity, $W_{\rm tot}$, for each water and CO line, i.e,  $W_{\rm envelope}$ = $W_{\rm tot}$ -- $W_{\rm broad}$ -- $W_{\rm spot\,shock}$, where $W_{\rm broad}$ = $W_{\rm cavity\,shock}$ for \water\ and $W_{\rm broad}$ = $W_{\rm entrainment}$ for \twco\ \ten9. 
  \end{table*}
}


\def\TableFWZIFWHM{
\begin{table*}[!h]
  \begin{center}
    \caption{\label{tbl4:Averaged_FWZIFWHM} Averaged rms value in 0.27~\kms\ bin and mean (dash) values of the \FWHMb\ and \FWZI\ for the three water lines and the \twco\ \ten9 spectra for each sub-type of YSO. The averaged \FWHMb\ and \FWZI\ ratios calculated from the \twco\ and \water\ observations are also indicated in the last two columns.}    
    \begin{tabular}{l c |cc |cc |cc ||ccc || cc}
      \hline\hline
      \rule{0pt}{2.4ex}&  	\water  &	 \multicolumn{2}{c|}{ \eight} 	&\multicolumn{2}{c|}{  \seven}	& \multicolumn{2}{c||}{ \tennine} &\multicolumn{3}{c||}{ \twco\ \ten9} &\multicolumn{2}{c}{ \twco\ / \water}\\
	\rule{0pt}{2.4ex}& $\overline{FWHM_{\rm b}}$ 	& $\overline{\sigma_{rms}}$		& $\overline{FWZI}$ & $\overline{\sigma_{rms}}$ 		& $\overline{FWZI}$ 	& $\overline{\sigma_{rms}}$ 			&  $\overline{FWZI}$ 		& $\overline{FWHM_{\rm b}}$ 	& $\overline{\sigma_{rms}}$		& $\overline{FWZI}$ 		& \FWHMb\ & \FWZI\\\
   \rule{0pt}{2.4ex}     & (\kms) &      (mK)  & (\kms) &   (mK)  & (\kms) &     (mK)  & (\kms) &    (\kms)  &  (mK)  & (\kms) &      &\\    
      \hline
      \rule{0pt}{2.8ex}LM0		&	30 $\pm$ 12	& 23	 	& 86 $\pm$ 45	& 20		& 87 $\pm$ 47	&	17	& 86 $\pm$ 31	&	20 $\pm$ 9	& 112	 	& 36 $\pm$ 14		& 	0.7 $\pm$ 0.3 & 0.5 $\pm$ 0.2 \\	
	LMI		&	18 $\pm$ 5	& 22	 	& 34 $\pm$ 13	&17		& 33 $\pm$ 15	&	9	& 42$^a$			&	11 $\pm$ 1	&  115	 	& 24 $\pm$ 8		&	0.6 $\pm$ 0.1 & 0.8 $\pm$ 0.4\\
	IM		&	29 $\pm$ 5	& 24	 	& 69 $\pm$ 19	& 22		& 63 $\pm$ 30	&19	& 79$^a$			&	20 $\pm$ 4	& 105	 	& 63 $\pm$ 21		&	0.7 $\pm$ 0.2 & 0.9 $\pm$ 0.1 \\
	HM		&	27 $\pm$ 10	& 78	 	& 47 $\pm$ 10	&68		& 70 $\pm$ 32	&	45	& 71 $\pm$ 35	&	22$^a$		& 180	 	& 81 $\pm$ 30$^b$	&  0.9$^a$ &   1.9 $\pm$ 0.2$^b$\\[1pt]
      \hline
    \end{tabular}
  \end{center}  
  {\bf Notes.} LM0: low-mass Class~0 protostars; LMI: low-mass Class~I sources; IM: intermediate-mass YSOs; HM: high-mass objects.\\
  See specification in the \FWZI\ calculation for the low-mass objects in \citealt{14Mottram}. \\
  $^{(a)}$ No standard deviation is given because the number of observed or detected sources is less than four. \\   
   $^{(b)}$ For the high-mass YSOs, the \twco\ \three2 spectra were included in the calculation of the mean value of the \FWZI\ parameter. \\  
\end{table*}
}

\def\TableCorrelationFits{
\begin{table*}[t]
  \begin{center}
    \caption{\label{tbl4:Correlation_fits} {\it Top half}: Slope ($m$), intercept ($n$), and Pearson correlation coefficient ($r$) of the power-law fit to the correlation between the logarithm of the \water\ line luminosity and the logarithm of the bolometric luminosity ($L_\mathrm{bol}$, left-columns) and the logarithm of the envelope mass ($M_\mathrm{env}$, right-columns). 
{\it Bottom half:} Same as above, but for the correlation between the luminosity corresponding to the cavity shock component and \Lbol\ or $M_\mathrm{env}$.}
    \begin{tabular}{l ccc | ccc}
      \hline\hline
        \rule{0pt}{2.8ex}Line    & \multicolumn{3}{c|}{$\log$\,(\Lwater) $=n+m\,\cdot$\,$\log$\,($L_{\rm{bol}}$)} &   \multicolumn{3}{c}{$\log$\,(\Lwater) $=n+m\,\cdot$\,$\log$\,($M_{\rm{env}}$)} \\   
       \cline{2-4}\cline{5-7}
        \rule{0pt}{2.8ex}(GHz) &        $m$       &        $n$       & $r$ &        $m$       &        $n$       & $r$ \\ 
      \hline
      \rule{0pt}{2.8ex}988        	&  0.85 $\pm$ 0.05  &  -3.48 $\pm$ 0.14  &  0.94 &  1.19 $\pm$ 0.05  &  -2.91 $\pm$ 0.10  & 0.96\\
      752         	&  0.94 $\pm$ 0.05  &  -3.59 $\pm$ 0.18  & 0.94 &  1.30 $\pm$ 0.06  &  -2.91 $\pm$ 0.12  & 0.96\\
      1097 		&  0.76 $\pm$ 0.05  &   -3.23 $\pm$ 0.18 &  0.95 &  1.11 $\pm$ 0.08  &  -2.94 $\pm$ 0.18  & 0.93\\[1pt]
      \hline
       \rule{0pt}{2.8ex}& \multicolumn{3}{c|}{$\log$\,($L_{\mathrm{broad\,H}_{2}\mathrm{O}}$) $=n+m\,\cdot$\,$\log$\,($L_{\rm{bol}}$)} & 
        \multicolumn{3}{c}{$\log$\,($L_{\mathrm{broad\,H}_{2}\mathrm{O}}$) $=n+m\,\cdot$\,$\log$\,($M_{\rm{env}}$)} \\
         \cline{2-4}\cline{5-7}
           \rule{0pt}{2.8ex}988       	&  0.79 $\pm$ 0.04  &  -3.43 $\pm$ 0.14  &  0.94 &  1.11 $\pm$ 0.05  &  -2.90 $\pm$ 0.10  & 0.96\\
        752      	&  0.83 $\pm$ 0.05  &  -3.37 $\pm$ 0.18  & 0.94 &  1.15 $\pm$ 0.06  &  -2.80 $\pm$ 0.13  & 0.95\\
        1097     	&  0.73 $\pm$ 0.05  &   -3.34 $\pm$ 0.19 &  0.95 &  1.08 $\pm$ 0.08  &  -3.06 $\pm$ 0.18  & 0.93\\[1pt]
      \hline
    \end{tabular}
  \end{center}  
\end{table*}
}

\def\TableRatiolines{
\begin{table*}[!t]
  \begin{center}
    \caption{\label{tbl4:Averaged_ratios_lines} Averaged values of \water\ line intensity ratios for the shocked gas along the outflow cavity (broad component), beam size ratios of those transitions and the ratios of the optically thin and thick solutions at $T_{\rm ex}$=300~K. }
         \begin{tabular}{l cccc c c}
      \hline\hline
      	\rule{0pt}{2.4ex}Transitions 	&  		 \multicolumn{3}{c}{Observed ratio}& $\theta_{\rm 1}$/$\theta_{\rm 2}$ & Thin$^a$ & Thick$^a$\\	
	\rule{0pt}{2.2ex}			&		LM	&	IM		&	HM		& & (LTE) & (LTE)\\    
      \hline
	\rule{0pt}{2.2ex}\water\ \seven/\eight	 	&	0.57 $\pm$ 0.14	&	0.60 $\pm$ 0.18	& 1.2 $\pm$ 0.2 	& 1.31 	& 1.85	& 1.02	\\
	\rule{0pt}{2.0ex}\water\ \tennine/\eight		&	0.52 $\pm$ 0.12	&	0.71 $\pm$ 0.16 	& 0.8 $\pm$ 0.2 	&  0.90 	& 17.58 	& 2.97	\\
	\rule{0pt}{2.0ex}\water\ \tennine/\seven		&	1.0 $\pm$ 0.3	&	0.94 $\pm$ 0.08 	& 0.7 $\pm$ 0.2 	&  0.68 	&  3.16	& 0.97	\\
      \hline
    \end{tabular}
  \end{center}  
  {\bf Notes.} LM: low-mass protostars; IM: intermediate-mass YSOs; HM: high-mass objects. The ratios are not corrected for different beam sizes.\\
 $^{(a)}$ Line intensity ratio calculated for an excitation condition of 300~K and assuming LTE .
\end{table*}
}


\def\TableObdIDs{
\begin{table}[!h]
  \begin{center}
 \caption{\label{tbl4:ObsIDs} Observation identification numbers for the \water\ \eight, \seven\ and \tennine\ lines of the intermediate- and high-mass YSOs. }
    \begin{tabular}{l ccc}
    	\noalign{\smallskip}
      \hline\hline
      \noalign{\smallskip}
	Source  &   \eight\ &  \seven\ &  \tennine \\
      	\noalign{\smallskip}
	\hline
	\noalign{\smallskip} 
	{\bf Inter.-mass} & &  & \\
   L1641\,S3MMS1   & 1342203147   & 1342203220   &        $-$   \\
     Vela\,IRS19   & 1342197952   & 1342201540   &        $-$   \\
     Vela\,IRS17   & 1342197951   & 1342201541   &        $-$   \\
  NGC7129\,FIRS2   & 1342191613   & 1342191747   & 1342227393   \\
        NGC2071   & 1342204503   & 1342194682   & 1342227395   \\
        AFGL490   & 1342204511   & 1342217717   &        $-$   \\
        OMC-2-FIR\,4 	& 1342218629 &  1342194681 & 1342217719\\
      	\noalign{\smallskip}
	\hline
	\noalign{\smallskip} 
         {\bf High-mass} & &  & \\
 IRAS05358+3543   & 1342204510   & 1342194684   & 1342206123   \\
 IRAS16272-4837   & 1342203168   & 1342205845   & 1342214418   \\
    NGC6334-I-1   & 1342204520   & 1342205846   & 1342206386   \\
        W43-MM1   & 1342191616   & 1342194565   & 1342194806   \\
        DR21-OH   & 1342195026   & 1342194574   & 1342196427   \\
        W3-IRS5   & 1342191612   & 1342201548   & 1342201592   \\
 IRAS18089-1732   & 1342215911   & 1342217712   & 1342218914   \\
           W33A   & 1342191636   & 1342191746   & 1342208090   \\
 IRAS18151-1208   & 1342218211   & 1342194679   & 1342218911   \\
       AFGL2591   & 1342195019   & 1342192335   & 1342194796   \\
       G327-0.6   & 1342203170   & 1342205844   & 1342214424   \\
   NGC6334I-N-1   & 1342204519   & 1342205847   & 1342206384   \\
    G29.96-0.02   & 1342191617   & 1342194563   & 1342194807   \\
    G31.41+0.31   & 1342191615   & 1342194566   & 1342219241   \\
     G5.89-0.39   & 1342218120   & 1342217707   & 1342218917   \\
    G10.47+0.03   & 1342215914   & 1342217711   & 1342218915   \\
    G34.26+0.15   & 1342194995   & 1342215950   & 1342219245   \\
        W51N-e1   & 1342195014   & 1342194568   & 1342196433   \\
   NGC7538-IRS1   & 1342201599   & 1342201546   & 1342200760   \\[1pt]
        \hline
  \end{tabular}
  \end{center}  
\end{table}
}


\def\TableFWHMcteWaterFst{
\begin{landscape}
	\begin{table}[t]
	\begin{center}	
	 \caption{\label{tbl4:FWHM_cte1} Gaussian decomposition results for the intermediate-mass objects.}
	\begin{tabular}{l c ccc cc cc cc} 
	\noalign{\smallskip}
	 \hline\hline
	\rule{0pt}{2.4ex}Source &  Comp.$^a$  & \FWHM\ & $\varv_{\rm peak}$ & $\varv_{\rm LSR}$ & \multicolumn{2}{c}{\water\ \eight}  & \multicolumn{2}{c}{\water\ \seven}  &  \multicolumn{2}{c}{\water\ \tennine} \\
	\rule{0pt}{2.4ex}	& &    &   && $T_{\rm{MB}}^{\rm{peak}}$  & ${\int{T_{\rm{MB}}{\rm{d}}\varv^{b}}}$ &  $T_{\rm{MB}}^{\rm{peak}}$  & ${\int{T_{\rm{MB}}{\rm{d}}\varv^{b}}}$ &  $T_{\rm{MB}}^{\rm{peak}}$  & ${\int{T_{\rm{MB}}{\rm{d}}\varv^{b}}}$ \\
	\rule{0pt}{2.4ex}	 & & (\kms)  & (\kms)  & (\kms) &(K)  & (K~\kms) &  (K)  & (K~\kms) &  (K)  & (K~\kms) \\		
	\hline
	\noalign{\smallskip}
	NGC7129 FIRS2  	&  C & 29.3 $\pm$ 0.3 	& -13.99 $\pm$ 0.17 	&  -9.8  	&   0.192 $\pm$ 0.004 	&    5.99 $\pm$ 0.14 	&    0.160 $\pm$ 0.004 &    4.99 $\pm$ 0.13 &    0.166 $\pm$ 0.003 &    5.18 $\pm$ 0.11  \\
                    			&  E &  8.46 $\pm$ 0.11 &  -7.20 $\pm$ 0.04 	&  -9.8  	&   0.368 $\pm$ 0.005 	&    3.31 $\pm$ 0.06 	&    0.282 $\pm$ 0.006 &    2.54 $\pm$ 0.06  &   0.285 $\pm$ 0.004 &    2.57 $\pm$ 0.05  \\
	L1641 S3MMS1	&  C & 34 $\pm$ 3 		&   6.4 $\pm$ 0.9 		&   5.3  	&   0.073 $\pm$ 0.013 	&    2.6 $\pm$ 0.5 		&    0.084 	&    3.04   & $-$ & $-$ \\
                    			&  E & 15.3 $\pm$ 0.5 	&   8.57 $\pm$ 0.16 	&   5.3  	&   0.197 $\pm$ 0.014 	&    3.2 $\pm$ 0.3 		&    0.155 $\pm$ 0.005 	&    2.53 $\pm$ 0.11 	& $-$ & $-$ \\
         NGC2071    	&  C & 29.60 $\pm$ 0.12 &  9.14 $\pm$ 0.05 	&   9.6  	&   1.722 $\pm$ 0.008 	&   54.3 $\pm$ 0.3 		&    1.42 $\pm$ 0.007 	&   44.7 $\pm$ 0.3 	&    1.311 $\pm$ 0.006 &  41.3 $\pm$ 0.3  \\
                    			&  E &  6.122 $\pm$ 0.04	&  10.17 $\pm$ 0.02 	&   9.6  	&   2.039 $\pm$ 0.017 	&   13.3 $\pm$ 0.2 	&    1.924 $\pm$ 0.012 &      12.5 $\pm$ 0.1  &   1.283 $\pm$ 0.009 &   8.36 $\pm$ 0.08  \\
					&  S	&	31.2 $\pm$ 0.7 &  46.9 $\pm$ 0.3 	&   9.6  	&   0.277 $\pm$ 0.005 	&   9.2 $\pm$ 0.3 		&    0.199 $\pm$ 0.004 	&   6.6 $\pm$ 0.2 	&    0.229 $\pm$ 0.003 &   7.6 $\pm$ 0.2  \\			
					&  A &  3.65 $\pm$ 0.09	&   6.54 $\pm$ 0.04 	&   9.6  	&  -0.814 $\pm$ 0.018	&   -3.16 $\pm$ 0.11    & $-$ & $-$ & $-$ & $-$ \\	
	Vela IRS17		&  C & 23.8 $\pm$ 0.4 	&   7.53 $\pm$ 0.14 	&   3.9  	&   0.277 $\pm$ 0.005 	&    7.01 $\pm$ 0.16 	&    0.158 $\pm$ 0.005 	&    4.00 $\pm$ 0.14   & $-$ & $-$ \\
                    			&  E &  4.19 $\pm$ 0.07 &   4.47 $\pm$ 0.03 		&   3.9  	&   0.462 $\pm$ 0.009 	&    2.06 $\pm$ 0.05 	&    0.608 $\pm$ 0.011 	&    2.71 $\pm$ 0.07   & $-$ & $-$ \\
	Vela IRS19    	&  C & 33.3 $\pm$ 0.8 	&  16.7 $\pm$ 0.3 		&  12.2  	&   0.141 $\pm$ 0.003 	&    5.00 $\pm$ 0.16 	&    0.068 $\pm$ 0.004 	&    2.41 $\pm$ 0.15   & $-$ & $-$ \\
                    			&  E &  4.14 $\pm$ 0.18 &  11.92 $\pm$ 0.07 	&  12.2  	&   0.183 $\pm$ 0.008 	&    0.81 $\pm$ 0.05 	&    0.215 $\pm$ 0.012 	&    0.95 $\pm$ 0.07   & $-$ & $-$ \\                			
   	AFGL490		&  C &  31.9 $\pm$ 1.0 	& -11.7 $\pm$ 0.4 		& -13.5 	&   0.107 $\pm$ 0.004 	&    3.63 $\pm$ 0.18 	&    0.033 $\pm$ 0.004 	&    1.12 $\pm$ 0.14   & $-$ & $-$ \\
                    			&  E &  4.56 $\pm$ 0.11 & -13.40 $\pm$ 0.04 	& -13.5	&   0.299 $\pm$ 0.007 	&    1.45 $\pm$ 0.05 	&    0.364 $\pm$ 0.011 	&    1.77 $\pm$ 0.07   & $-$ & $-$ \\
        OMC-2-FIR\,4    	&  C & 18.62 $\pm$ 0.07 &  11.33 $\pm$ 0.03 	&   11.3  	&   2.085 $\pm$ 0.018 	&   41.3 $\pm$ 0.4 	&    1.210 $\pm$ 0.009 &   24.0 $\pm$ 0.2 &    1.031 $\pm$ 0.010 &   20.4 $\pm$ 0.2  \\
                    			&  E &  6.26 $\pm$ 0.08 &   13.40 $\pm$ 0.02 	&   11.3  	&   1.03 $\pm$ 0.03 	&    6.87 $\pm$ 0.19 	&    0.677 $\pm$ 0.009 &    4.51 $\pm$ 0.08  &  0.566 $\pm$ 0.014 &    3.77 $\pm$ 0.11  \\[1pt]
	\hline
      \end{tabular}
      \end{center}
      {\bf Notes.} {$(^a)$} The types of components are: C = cavity shock; E = envelope emission; S = spot shock;  A = envelope absorption.  {$(^b)$} Integrated intensity of each velocity component approximated to the gaussian fit. \\      
  \end{table}
  \end{landscape}
}

\def \TableFWHMcteWaterSnd{
\begin{landscape}
	\begin{table}[]
 	\begin{center}
	 \caption{\label{tbl4:FWHM_cte2} Gaussian decomposition results for the high-mass YSOs.}
	\begin{tabular}{l c ccc cc cc cc} 
	\noalign{\smallskip}
	 \hline\hline
	\rule{0pt}{2.4ex}Source &  Comp.$^a$  & \FWHM\ & $\varv_{\rm peak}$ & $\varv_{\rm LSR}$ & \multicolumn{2}{c}{\water\ \eight}  & \multicolumn{2}{c}{\water\ \seven}  &  \multicolumn{2}{c}{\water\ \tennine} \\
	\rule{0pt}{2.4ex}& &    &   && $T_{\rm{MB}}^{\rm{peak}}$  & ${\int{T_{\rm{MB}}{\rm{d}}\varv^{b}}}$ &  $T_{\rm{MB}}^{\rm{peak}}$  & ${\int{T_{\rm{MB}}{\rm{d}}\varv^{b}}}$ &  $T_{\rm{MB}}^{\rm{peak}}$  & ${\int{T_{\rm{MB}}{\rm{d}}\varv^{b}}}$ \\
	\rule{0pt}{2.4ex}& & (\kms)  & (\kms)  & (\kms) &(K)  & (K~\kms) &  (K)  & (K~\kms) &  (K)  & (K~\kms) \\
	\hline
	\noalign{\smallskip}              
  IRAS05358+3543	&  C & 23.0 $\pm$ 0.2 & -13.56 $\pm$ 0.07 & -17.6  &   0.786 $\pm$ 0.015 &   19.2 $\pm$ 0.4 &    0.812 $\pm$ 0.014 &   19.9 $\pm$ 0.4 &    0.550 $\pm$ 0.009 &   13.5 $\pm$ 0.3  \\
                    		&  E &  6.49 $\pm$ 0.09 & -15.05 $\pm$ 0.03 & -17.6  &   1.14 $\pm$ 0.03 &    7.9 $\pm$ 0.2 &    1.079 $\pm$ 0.017 &    7.46 $\pm$ 0.16  &   0.617 $\pm$ 0.012 &    4.26 $\pm$ 0.10  \\
                    		&  A &  3.69 $\pm$ 0.14 & -18.01 $\pm$ 0.06 & -17.6  &  -0.85 $\pm$ 0.03 &   -3.32 $\pm$ 0.16    & $-$ & $-$ & $-$ & $-$ \\
IRAS16272-4837 	&  C & 23.0 $\pm$ 0.4 & -46.15 $\pm$ 0.13 & -46.2  &   0.531 $\pm$ 0.018 &   13.0 $\pm$ 0.5 &    0.72 $\pm$ 0.02 &   17.5 $\pm$ 0.6 &    0.300 $\pm$ 0.008 &    7.4 $\pm$ 0.2  \\
                    		&  E &  5.61 $\pm$ 0.11 & -46.78 $\pm$ 0.04 & -46.2  &   0.72 $\pm$ 0.03 &    4.3 $\pm$ 0.2 &    1.59 $\pm$ 0.03 &    9.6 $\pm$ 0.3  &   0.420 $\pm$ 0.012 &    2.51 $\pm$ 0.09  \\
    NGC6334I-N-1	&  C & 36.4 $\pm$ 0.3 &  -5.24 $\pm$ 0.08 &  -3.3  &   0.803 $\pm$ 0.012 &   31.1 $\pm$ 0.5 &    1.191 $\pm$ 0.015 &   46.1 $\pm$ 0.7 &    0.508 $\pm$ 0.006 &   19.7 $\pm$ 0.3  \\
                    		&  E & 10.41 $\pm$ 0.07 &  -2.99 $\pm$ 0.02 &  -3.3  &   2.39 $\pm$ 0.04 &   26.5 $\pm$ 0.5 &    2.703 $\pm$ 0.018 &   30.0 $\pm$ 0.3  &   0.763 $\pm$ 0.009 &    8.46 $\pm$ 0.12  \\
                    		&  A &  3.95 $\pm$ 0.05 &  -3.07 $\pm$ 0.01 &  -3.3  &  -3.60 $\pm$ 0.04 &  -15.2 $\pm$ 0.2    & $-$ & $-$ & $-$ & $-$ \\
         W43-MM1    	&  C & 48.8 $\pm$ 0.5 & 101.65 $\pm$ 0.19 &  98.8  &   0.429 $\pm$ 0.008 &   22.3 $\pm$ 0.5 &    0.436 $\pm$ 0.007 &   22.7 $\pm$ 0.4 &    0.152 $\pm$ 0.003 &    7.90 $\pm$ 0.17  \\ 
                    		&  E &  9.10 $\pm$ 0.06 &  98.50 $\pm$ 0.02 &  98.8  &   1.59 $\pm$ 0.09 &   15.3 $\pm$ 0.9 &    2.150 $\pm$ 0.012 &   20.82 $\pm$ 0.18  &   0.563 $\pm$ 0.006 &    5.45 $\pm$ 0.07  \\
                    		&  A &  5.79 $\pm$ 0.12 &  99.30 $\pm$ 0.03 &  98.8  &  -2.28 $\pm$ 0.08 &  -14.0 $\pm$ 0.6    & $-$ & $-$ & $-$ & $-$ \\
         DR21-OH    	&  C & 37.6 $\pm$ 0.3 &  -1.99 $\pm$ 0.09 &  -3.1  &   0.640 $\pm$ 0.016 &   25.6 $\pm$ 0.7 &    0.827 $\pm$ 0.017 &   33.1 $\pm$ 0.7 &    0.629 $\pm$ 0.010 &   25.2 $\pm$ 0.5  \\
                    		&  E & 12.34 $\pm$ 0.04 &  -2.81 $\pm$ 0.01 &  -3.1  &   5.63 $\pm$ 0.03 &   74.0 $\pm$ 0.5 &    5.46 $\pm$ 0.02 &   71.8 $\pm$ 0.3  &  2.971 $\pm$   0.010 &   39.04 $\pm$ 0.17  \\
                    		&  A &  3.28 $\pm$  0.03 &  -1.49 $\pm$ 0.01 &  -3.1  &  -4.74 $\pm$ 0.04 &  -16.6 $\pm$ 0.2    & $-$ & $-$ & $-$ & $-$ \\
         W3-IRS5    	&  C & 31.57 $\pm$ 0.13 & -37.62 $\pm$ 0.05 & -38.4  &   1.948 $\pm$ 0.014 &   65.5 $\pm$ 0.5 &    2.140 $\pm$ 0.013 &   71.9 $\pm$ 0.5 &    1.976 $\pm$ 0.014 &   66.4 $\pm$ 0.5  \\
                    		&  E &  5.86 $\pm$ 0.02 & -37.49 $\pm$ 0.01 & -38.4  &   4.21 $\pm$ 0.03 &   26.21 $\pm$ 0.19 &    6.55 $\pm$ 0.02 &   40.8 $\pm$ 0.2  &    4.63 $\pm$ 0.03 &   28.84 $\pm$ 0.19  \\
  IRAS18089-1732	&  C & 20.2 $\pm$ 0.3 &  35.41 $\pm$ 0.10 &  33.8  &   0.455 $\pm$ 0.018 &    9.8 $\pm$ 0.4 &    0.534 $\pm$ 0.015 &   11.5 $\pm$ 0.4 &    0.459 $\pm$ 0.008 &    9.8 $\pm$ 0.2  \\
                    		&  E &  4.27 $\pm$ 0.09 &  33.30 $\pm$ 0.03 &  33.8  &   0.58 $\pm$ 0.04 &    2.65 $\pm$ 0.18 &    1.20 $\pm$ 0.03 &    5.47 $\pm$ 0.16  &   0.537 $\pm$ 0.012 &    2.44 $\pm$ 0.07  \\
            W33A    	&  C & 26.8 $\pm$ 0.2 &  36.02 $\pm$ 0.08 &  37.5  &   0.512 $\pm$ 0.008 &   14.6 $\pm$ 0.3 &    0.728 $\pm$ 0.013 &   20.7 $\pm$ 0.4 &    0.410 $\pm$ 0.005 &   11.68 $\pm$ 0.17  \\
                    		&  E &  5.81 $\pm$ 0.03 &  38.01 $\pm$ 0.01 &  37.5  &   1.568 $\pm$ 0.016 &    9.70 $\pm$ 0.11 &    1.83 $\pm$ 0.03 &   11.35 $\pm$ 0.17  &   1.347 $\pm$ 0.007 &    8.34 $\pm$ 0.07  \\
IRAS18151-1208  	&  C & 18.6 $\pm$ 0.2 &  33.6 $\pm$ 0.2 &  32.8  &  0.220 $\pm$ 0.007 &    4.36 $\pm$ 0.15 &   0.269 $\pm$ 0.02 &  5.33 $\pm$ 0.14  &   0.1 $\pm$ 0.3 &   2.64 $\pm$ 0.18  \\
                    		&  E & 3.3  $\pm$ 0.3 &   32.9 $\pm$ 0.3 &  32.8  &   0.426 $\pm$ 0.015 &    1.52 $\pm$ 0.11 &   0.57  $\pm$ 0.04 &  2.02 $\pm$ 0.17  &   0.3 $\pm$ 0.3 &    1.1 $\pm$ 0.2  \\
        AFGL2591    	&  C & 15.23 $\pm$ 0.17 &  -5.32 $\pm$ 0.05 &  -5.5  &   0.48 $\pm$ 0.02 &    7.8 $\pm$ 0.4 &    0.607 $\pm$ 0.011 &    9.8 $\pm$ 0.2 &    0.456 $\pm$ 0.007 &    7.40 $\pm$ 0.14  \\  
                    		&  E &  3.50 $\pm$ 0.02 &  -5.48 $\pm$ 0.08 &  -5.5  &   2.87 $\pm$ 0.04 &   10.7 $\pm$ 0.17 &    3.071 $\pm$ 0.016 &   11.45 $\pm$ 0.08  &   2.006 $\pm$ 0.011 &    7.48 $\pm$ 0.06  \\[1pt]
	\hline
	\end{tabular}
	\end{center}
	{\bf Notes.} {$(^a)$} The types of components are: C = cavity shock; E = envelope emission; S = spot shock; A = envelope absorption.  {$(^b)$} Integrated intensity of each velocity component approximated to the gaussian fit.\\\  
	\end{table}
\end{landscape}
}

\def \TableFWHMcteWaterTrd{
\begin{landscape}
	\begin{table}[t]
 	\begin{center}
	 \caption{\label{tbl4:FWHM_cte3} Gaussian decomposition results for the high-mass YSOs (continuation).}
	\begin{tabular}{l c ccc cc cc cc} 
	\noalign{\smallskip}
	 \hline\hline
     \rule{0pt}{2.4ex}Source &  Comp.$^a$  & \FWHM\ & $\varv_{\rm peak}$ & $\varv_{\rm LSR}$ & \multicolumn{2}{c}{\water\ \eight}  & \multicolumn{2}{c}{\water\ \seven}  &  \multicolumn{2}{c}{\water\ \tennine} \\
	\rule{0pt}{2.4ex}& &    &   && $T_{\rm{MB}}^{\rm{peak}}$  & ${\int{T_{\rm{MB}}{\rm{d}}\varv^{b}}}$ &  $T_{\rm{MB}}^{\rm{peak}}$  & ${\int{T_{\rm{MB}}{\rm{d}}\varv^{b}}}$ &  $T_{\rm{MB}}^{\rm{peak}}$  & ${\int{T_{\rm{MB}}{\rm{d}}\varv^{b}}}$ \\
  	\rule{0pt}{2.4ex}& & (\kms)  & (\kms)  & (\kms) &(K)  & (K~\kms) &  (K)  & (K~\kms) &  (K)  & (K~\kms) \\
	\hline
	\noalign{\smallskip}   
G327-0.6 	&  C & 24.3 $\pm$ 0.2 	& -42.28 $\pm$  0.08 & -45.0  &   0.89 $\pm$ 0.02 &   23.0 $\pm$ 0.6	&    0.616 $\pm$   0.011 &   15.9 $\pm$   0.3 &    0.649 $\pm$   0.010 &   16.8 $\pm$ 0.3  \\
                    	&  E &  6.45 $\pm$ 0.06 	& -44.32 $\pm$  0.02 & -45.0  &   7.4 $\pm$ 0.4 &   51 $\pm$   3 		&    1.618 $\pm$   0.014 &   11.11 $\pm$   0.14  &   1.042 $\pm$   0.015 &    7.16 $\pm$   0.12  \\
                    	&  A &  4.83 $\pm$ 0.07 	& -43.79 $\pm$  0.03 & -45.0  &  -7.8 $\pm$ 0.4 &  -40 $\pm$   2    	& $-$ & $-$ & $-$ & $-$ \\
NGC6334-I-1 	&  C & 17.83 $\pm$ 0.14 	&   1.43 $\pm$  0.09 &  -7.4  &   1.88 $\pm$ 0.02 &   35.7 $\pm$ 0.5 	&    1.478 $\pm$   0.014 &   28.0 $\pm$   0.3 &    1.76 $\pm$   0.02 &   33.3 $\pm$   0.5  \\
                    	&  E &  6.27 $\pm$  0.03	&  -6.40 $\pm$  0.01 &  -7.4  &   4.82 $\pm$ 0.07 &   32.2 $\pm$ 0.5 	&    7.31 $\pm$   0.02 &   48.8 $\pm$   0.3  &  2.09 $\pm$   0.04 &   14.0 $\pm$   0.3  \\
                    	&  A &  2.86 $\pm$  0.10	&  -7.77 $\pm$  0.04 &  -7.4  &  -1.99 $\pm$ 0.07 &   -6.1 $\pm$ 0.3    & $-$ & $-$ & $-$ & $-$ \\
G29.96-0.02 	&  C & 29.1 $\pm$  0.2 	&  97.74 $\pm$  0.07 &  97.6  &   1.059 $\pm$ 0.014 & 32.8 $\pm$ 0.5	&    0.810 $\pm$  0.016 &   25.1 $\pm$  0.5 &    0.705 $\pm$   0.008 &   21.9 $\pm$ 0.3  \\
                    	&  E &  6.20 $\pm$0.03 	&  98.56 $\pm$  0.01 &  97.6  &   4.36 $\pm$ 0.02 &   28.7 $\pm$ 0.2 	&    3.13 $\pm$   0.03 &   20.7 $\pm$   0.2  &   2.058 $\pm$   0.012 &   13.58 $\pm$   0.10  \\
G31.41+0.31	&  C & 27.7 $\pm$  0.4 	&  99.62 $\pm$  0.16 &  97.4  &   0.413 $\pm$ 0.013 &   12.2 $\pm$ 0.4 &    0.627 $\pm$   0.016 &   18.5 $\pm$ 0.5 &    0.317 $\pm$   0.008 &    9.3 $\pm$   0.3  \\
                    	&  E &  6.73 $\pm$ 0.09 	&  96.46 $\pm$  0.03 &  97.4  &   1.21 $\pm$ 0.04 	&    8.6 $\pm$ 0.3 	&    1.65 $\pm$   0.02 &   11.8 $\pm$   0.2  &   0.931 $\pm$   0.012 &    6.67 $\pm$   0.12  \\
                    	&  A &  4.57 $\pm$ 0.16 	&  99.09 $\pm$  0.08 &  97.4  &  -1.02 $\pm$ 0.03	&   -4.9 $\pm$ 0.2  	& $-$ & $-$ & $-$ & $-$ \\
G5.89-0.39    	&  C & 45.06 $\pm$ 0.05 	&  17.81 $\pm$  0.02 &  10.0  &   4.154 $\pm$ 0.011 & 199.2 $\pm$ 0.6 &    4.057 $\pm$   0.009 &  194.6 $\pm$   0.5 &    3.474 $\pm$   0.007 &  166.6 $\pm$   0.4  \\
                    	&  E & 10.93 $\pm$ 0.02	&  11.62 $\pm$  0.01 &  10.0  &   5.564 $\pm$ 0.019 &   64.7 $\pm$ 0.3 &    6.452 $\pm$   0.015 &   75.1 $\pm$ 0.2  &   4.918 $\pm$   0.010 &   57.22 $\pm$   0.16  \\
                    	&  A &  8.92 $\pm$ 0.11	&   1.42 $\pm$  0.05 &  10.0  &  -1.741 $\pm$ 0.018 &  -16.5 $\pm$ 0.3    & $-$ & $-$ & $-$ & $-$ \\
G10.47+0.03 	&  C & 13.79 $\pm$ 0.13 	&  70.16 $\pm$  0.11 &  67.3  &   1.53 $\pm$ 0.03 &   22.4 $\pm$ 0.5 	&    2.162 $\pm$   0.02 &   31.7 $\pm$   0.5 &    0.903 $\pm$   0.011 &   13.3 $\pm$   0.2  \\
                    	&  E &  5.64 $\pm$ 0.06	&  64.48 $\pm$  0.02 &  67.3  &   3.22 $\pm$ 0.09 &   19.4 $\pm$ 0.6 	&    3.17 $\pm$   0.05 &   19.0 $\pm$   0.3  &   1.67 $\pm$   0.02 &   10.01 $\pm$   0.16  \\
                    	&  A &  4.5 $\pm$  0.2 	&  66.95 $\pm$  0.12 &  67.3  &  -1.32 $\pm$ 0.06 &   -6.4 $\pm$ 0.5 	& $-$ & $-$ & $-$ & $-$ \\
G34.26+0.15 	&  C & 18.39 $\pm$ 0.12	&  63.52 $\pm$  0.08 &  58.0  & 1.186 $\pm$ 0.015 & 23.2 $\pm$ 0.3 	&    1.492 $\pm$   0.017 &   29.2 $\pm$   0.4 &    1.091 $\pm$   0.010 &   21.4 $\pm$ 0.2  \\
                    	&  E &  5.50 $\pm$ 0.02	&  57.53 $\pm$  0.01 &  58.0  &  52 $\pm$ 18 	&  305 $\pm$ 109 	&    8.69 $\pm$   0.03 &   50.9 $\pm$   0.2  &  3.99 $\pm$   0.02 &   23.38 $\pm$   0.13  \\
                    	&  A &  5.60 $\pm$  0.03	&  57.71 $\pm$  0.07 &  58.0  & -48 $\pm$ 18 	& -286 $\pm$ 111  	& $-$ & $-$ & $-$ & $-$ \\
W51N-e1    	&  C & 36.1 $\pm$ 0.2 	&   59.58 $\pm$  0.07 &  59.5  &   1.47 $\pm$ 0.05 	&  56.27 $\pm$  0.11 	&    1.72 $\pm$   0.02 &  66.1  $\pm$   0.6 &    1.28 $\pm$ 0.02 &   49.0 $\pm$   0.4  \\
                    	&  E &  8.5 $\pm$ 0.2	&  54.30 $\pm$  0.01 &  59.5  &   5.31 $\pm$ 0.07 &   47.9 $\pm$ 0.7 	&    7.59 $\pm$   0.03 &   68.5 $\pm$   0.4  &   3.12 $\pm$   0.03 &   28.1 $\pm$   0.2  \\
                    	&  A & 19.4 $\pm$ 0.4 	&  62.08 $\pm$  0.14 &  59.5  &  -5.2 $\pm$ 0.5 	& -107 $\pm$ 9    	& $-$ & $-$ & $-$ & $-$ \\
NGC7538-IRS1&  C &15.37 $\pm$ 0.16	& -58.02 $\pm$  0.05 & -56.2  &   1.02 $\pm$ 0.03	&   16.7 $\pm$ 0.4 	&    1.37 $\pm$   0.03 &   22.4 $\pm$   0.5 &    0.781$\pm$   0.014 &   12.8 $\pm$   0.3  \\
                 	&  E &  4.42 $\pm$  0.02	& -57.38 $\pm$  0.01 & -56.2  &   4.51 $\pm$ 0.04	&   21.2 $\pm$ 0.2 	&    5.48 $\pm$   0.03 &   25.77 $\pm$   0.19  &   3.030 $\pm$   0.018 &   14.2 $\pm$ 0.11  \\[1pt]
	\hline
	\end{tabular}
	\end{center}
	{\bf Notes.} {$(^a)$} The types of components are: C = cavity shock; E = envelope emission; S = spot shock; A = envelope absorption. 
	{$(^b)$} Integrated intensity of each velocity component approximated to the gaussian fit.\\  
	\end{table}
\end{landscape}
}

\def\TableMainValuesA{
  \begin{table}[t]
    \begin{center}
	 \caption{\label{tbl4:Mainparameters_202-111} Observed and fitted properties of the H$_2$O 2$_{02}$-1$_{11}$ line for the 
	 detected intermediate- and high-mass sources.}
      \begin{tabular}{l ccrc}
      	\noalign{\smallskip}
        \hline\hline
	\rule{0pt}{2.4ex}Source     &   rms$^a$  & $T_{\rm{MB}}^{\rm{peak}}$  &	${\int{T_{\rm{MB}}{\rm{d}}\varv^{b}}}$ & \FWZI\ \\
         \rule{0pt}{2.4ex}  &      (mK)  & (K)   &(K km\,s$^{-1}$) &  (km\,s$^{-1}$) \\
	\hline
	\noalign{\smallskip} 
	{\bf Inter.-mass} 	&		&		&& \\	
 	NGC7129 FIRS2 	&     18 	&     0.58 	&     9.50 $\pm$  0.02 	&    48 \\
   	L1641\,S3MMS1 	&     23 	&     0.30 	&     6.28 $\pm$  0.03	&    48 \\
       NGC2071 			&     38 	&     3.83 	&    76.98 $\pm$  0.04	&    108 \\
      	Vela IRS17 		&     22 	&     0.74 	&     9.97 $\pm$  0.02 	&    63 \\
     	Vela IRS19 		&     21 	&     0.34 	&     6.23 $\pm$  0.02 &    72 \\         
       	AFGL490 			&     21 	&     0.44 	&     6.94 $\pm$  0.02	&    75 \\
	OMC-2-FIR\,4		&	28	&	3.15		&	52.31$\pm$  0.04 &	66\\
      	\noalign{\smallskip}
	\hline
	\noalign{\smallskip} 
	{\bf High-mass}  			&		&		&& \\
 	IRAS05358+3543	&     57 	&     1.93		&    24.69 $\pm$  0.09 	&    54 \\
 	IRAS16272-4837 	&     81 	&     1.34 	&    17.34 $\pm$  0.11 	&    39 \\
   	NGC6334I-N-1 		&     60 	&     2.25 	&    43.55 $\pm$  0.09 	&    57 \\
      	W43-MM1 		&     54 	&     1.02 	&    25.99 $\pm$  0.06 	&    84 \\
       	DR21-OH 			&     73 	&     5.86 	&    29.39 $\pm$  0.11 	&    51 \\
     	W3-IRS5 			&     95 	&     7.07 	&    88.32 $\pm$  0.12 	&    63 \\             
 	IRAS18089-1732 	&    107 	&     1.27 	&    12.01 $\pm$  0.14 	&    27 \\
    	W33A 			&     47 	&     2.23	 	&    24.13 $\pm$  0.05 	&    63 \\
 	IRAS18151-1208 	&     96 	&     0.82 	&     4.88 $\pm$  0.12 		&    30 \\
      	AFGL2591 		&    100 	&     3.41 	&    17.21 $\pm$  0.13 	&    24 \\
      	G327-0.6 			&     91 	&     3.23 	&    33.50 $\pm$  0.14 	&    51 \\
     	NGC6334-I-1 		&    106 	&     5.32 	&    69.44 $\pm$  0.14 	&    60 \\  
     	G29.96-0.02 		&     67 	&     5.36 	&    62.34 $\pm$  0.07 	&    66 \\
    	G31.41+0.31 		&     61 	&     1.41 	&    18.74 $\pm$  0.07 	&    60 \\
     	G5.89-0.39 		&     67 	&     9.60 	&   265.34 $\pm$  0.07 	&   108 \\
    	G10.47+0.03 		&    104 	&     3.58 	&    34.75 $\pm$  0.12 	&    39 \\
    	G34.26+0.15 		&     64 	&     6.08 	&    48.49 $\pm$  0.07 	&    60 \\                   	
      	W51N-e1 			&     64 	&     6.75 	&   102.12 $\pm$  0.07 	&    81 \\
  	NGC7538-IRS1 	&     91 	&     5.56 	&    37.20 $\pm$  0.12 	&    24 \\[1pt]	 
	\hline
      \end{tabular}
      \end{center}
       {\bf Notes.} \FWZI\ has been calculated by binning the spectra to 3 km\,s$^{-1}$. \\
       	$^{(a)}$ In 0.27 km\,s$^{-1}$ bin. $^{(b)}$ Integrated over the interval of velocities defined by the \FWZI.\\ 
  \end{table}
}

\def\TableMainValuesB{
  \begin{table}[t]
    \begin{center}
	 \caption{\label{tbl4:Mainparameters_211-202} Observed and fitted properties of the H$_2$O 2$_{11}$-2$_{02}$ line for the 
	 detected intermediate- and high-mass YSOs.}
      \begin{tabular}{l ccrc}
      	\noalign{\smallskip}
        \hline\hline
	\rule{0pt}{2.4ex}Source     &   rms$^a$  & $T_{\rm{MB}}^{\rm{peak}}$  &	${\int{T_{\rm{MB}}{\rm{d}}\varv^{b}}}$ & \FWZI\ \\
         \rule{0pt}{2.4ex}  &      (mK)  & (K)   &(K km\,s$^{-1}$) &  (km\,s$^{-1}$) \\
	\hline
	\noalign{\smallskip} 
	{\bf Inter.-mass} 	&		&		&& \\	
	NGC7129 FIRS2 	&     22 	&     0.52 	&     7.76 $\pm$  0.02 	&    48 \\
	L1641\,S3MMS1 	&     28	&     0.24 	&     2.69 $\pm$  0.05	&    22\\
	NGC2071 		&     37 	&     3.78 	&    62.30 $\pm$  0.04 	&    123 \\
	Vela IRS17	 	&     30 	&     0.82 	&     8.13 $\pm$  0.03 	&    39 \\
     	Vela IRS19	 	&     36 	&     0.41 	&     3.74 $\pm$  0.04 	&    63 \\
  	AFGL490		 	&     100 	&     0.51 	&    22.46 $\pm$  0.03 	&    33 \\
  	OMC-2-FIR\,4		&	15	&    1.87	&    29.92 $\pm$  0.05	&    69\\
      	\noalign{\smallskip}
	\hline
	\noalign{\smallskip} 
	{\bf High-mass}  		&		&		&& \\
	IRAS05358+3543 		&     44 	&     2.01 &    27.57 $\pm$  0.05 	&    100 \\
 	IRAS16272-4837 		&     81 	&     2.37 &    21.70 $\pm$  0.08 	&    60 \\
 	NGC6334I-N-1 			&     92 	&     4.38 &   117.17 $\pm$  0.17 	&    90 \\
	W43-MM1 			&     48 	&     2.77 &    45.14 $\pm$  0.05 	&   108 \\
	DR21-OH 				&     68 	&     7.48 &   104.18 $\pm$  0.07 	&    90 \\
	W3-IRS5 				&     68 	&     9.09 &   112.14 $\pm$  0.07 	&    105 \\
 	IRAS18089-1732 		&     68 	&     1.72 &    15.62 $\pm$  0.07 	&    36 \\
         W33A 				&     79 	&     2.76 &    30.17 $\pm$  0.08 	&    60 \\
 	IRAS18151-1208 		&     36 	&     0.86 &     6.41 $\pm$  0.04 	&    40 \\
   	AFGL2591 			&     37 	&     3.76 &    21.37 $\pm$  0.04 	&    42 \\
      	G327-0.6 				&    147 	&     5.19 &    60.08 $\pm$  0.15 	&    66 \\
   	NGC6334-I-1 			&    152 	&     9.69 &    78.44 $\pm$  0.06 	&   117 \\
      	G29.96-0.02 			&     96 	&     4.05 &    42.21 $\pm$  0.10 	&    69 \\
    	G31.41+0.31 			&     74 	&     2.28 &    23.93 $\pm$  0.08 	&    57 \\
    	G5.89-0.39 			&     59 	&   10.67	&   285.22 $\pm$  0.06 	&   144 \\
       	G10.47+0.03 			&     73 	&     4.47 &    45.81 $\pm$  0.07 	&    27 \\
      	G34.26+0.15 			&    225 	&   10.24 &    93.03 $\pm$  0.10 	&   42 \\
      	W51N-e1 				&    242 	&     9.46 &   150.81 $\pm$  0.24 	&   42 \\
     	NGC7538-IRS1 		&     61 	&     6.97 &    49.10 $\pm$  0.06 	&    42 \\[1pt]             
	\hline
      \end{tabular}
      \end{center}
       {\bf Notes.} \FWZI\ has been calculated by binning the spectra to 3 km\,s$^{-1}$. \\
       	$^{(a)}$ In 0.27 km\,s$^{-1}$ bin. $^{(b)}$ Integrated over the interval of velocities defined by the \FWZI.\\ 
  \end{table}
}

\def\TableMainValuesC{
  \begin{table}[t]
    \begin{center}
	 \caption{\label{tbl4:Mainparameters_312-303} Observed and fitted properties of the H$_2$O 3$_{12}$-3$_{03}$ line for the 
	 detected intermediate- and high-mass  objects.}
      \begin{tabular}{l ccrc}
      	\noalign{\smallskip}
        \hline\hline
	\rule{0pt}{2.4ex}Source     &   rms$^a$  & $T_{\rm{MB}}^{\rm{peak}}$  &	${\int{T_{\rm{MB}}{\rm{d}}\varv^{b}}}$ & \FWZI\ \\
         \rule{0pt}{2.4ex}  &      (mK)  & (K)   &(K km\,s$^{-1}$) &  (km\,s$^{-1}$) \\
	\hline
	\noalign{\smallskip} 
	{\bf Inter.-mass} 	&		&			&					& \\		
  	NGC7129-FIRS2 	&     14 	&     0.46 		&     7.48 $\pm$  0.02 	&    66 \\
        NGC2071 		&     28 	&     2.68 		&    58.82 $\pm$  0.03 	&   117 \\
        OMC-2-FIR\,4		&	30	&	1.589	&	25.10 $\pm$  0.11	&    54\\
      	\noalign{\smallskip}
	\hline
	\noalign{\smallskip} 
	{\bf High-mass}  	&		&		&					& \\
 	IRAS05358+3543	&     30 &     1.20 	&    19.29 $\pm$  0.03 	&    60 \\
 	IRAS16272-4837 	&     32 &     0.77 	&    11.11 $\pm$  0.04 	&    42 \\
    	NGC6334I-N-1 	&     28 &     1.34 	&    28.34 $\pm$  0.03 	&    90 \\
      	W43-MM1 		&     24 &     0.83 	&    13.07 $\pm$  0.03 	&    85 \\
      	DR21-OH 		&     25 &     3.66 	&    63.78 $\pm$  0.03 	&    102 \\
       	W3-IRS5 		&     81 &     6.78 	&    99.98 $\pm$  0.08 	&    78 \\
 	IRAS18089-1732 	&     30 &     1.04 	&    15.99 $\pm$  0.03 	&    66 \\
     	W33A 			&     19 &     1.84 	&    20.24 $\pm$  0.02 	&    66 \\
 	IRAS18151-1208 	&     30 &     0.47 	&     5.85 $\pm$  0.03 	&    40 \\
      	AFGL2591 		&     22 &     2.54 	&    16.02 $\pm$  0.03 	&    57 \\
       	G327-0.6 		&     39 &     1.88 	&    24.97 $\pm$  0.04 	&    57 \\
    	NGC6334-I-1 		&    128 &     3.73	&    48.12 $\pm$  0.12 	&    30 \\
    	G29.96-0.02 		&     34  &     2.88	&    36.15 $\pm$  0.04 	&    81 \\
    	G31.41+0.31 		&     36 &     1.28 	&    13.79 $\pm$  0.04 	&    45 \\
     	G5.89-0.39 		&     36 &     8.38 	&   237.69 $\pm$  0.04 	&   144 \\
      	G10.47+0.03 		&     32 &     2.25 	&    24.35 $\pm$  0.04 	&    27 \\  
    	G34.26+0.15 		&     42 &     5.01 	&    45.95 $\pm$  0.05 	&    48 \\	    	
       	W51N-e1 		&     63 &     4.40 	&    82.48 $\pm$  0.07 	&    110 \\
   	NGC7538-IRS1 	&     37 &     3.96 	&    27.53 $\pm$  0.04 	&    81 \\[1pt]	
	\hline
      \end{tabular}
      \end{center}
       {\bf Notes.} \FWZI\ has been calculated by binning the spectra to 3 km\,s$^{-1}$. \\
       	$^{(a)}$ In 0.27 km\,s$^{-1}$ bin. $^{(b)}$ Integrated over the interval of velocities defined by the \FWZI.\\ 
  \end{table}
}



\begin{abstract}{
	\textit{Context.} Water probes the dynamics in young stellar objects (YSOs) effectively, especially shocks in molecular outflows. It is therefore a key molecule for exploring whether the physical properties of low-mass protostars can be extrapolated to massive YSOs, an important step in understanding the fundamental mechanisms regulating star formation.
	\\
    	\textit{Aim.} As part of the WISH key programme, we investigate excited water line properties as a function of source luminosity, in particular the dynamics and the excitation conditions of shocks along the outflow cavity wall.
	\\	
  	\textit{Methods.} Velocity-resolved {\it Herschel}-HIFI spectra of the \water\ \eight\ (988 GHz), \seven\ (752 GHz) and \tennine\ (1097 GHz) lines were analysed, together with \twco\ \ten9 and 16--15, for 52 YSOs with bolometric luminosities ranging from $<$ 1 to $>$ 10$^5 L_\odot$. 
	 The \water\ and \twco\ line profiles were decomposed into multiple Gaussian components which are related to the different physical structures of the protostellar system. 
	 The non-LTE radiative transfer code {\sc radex} was used to constrain the excitation conditions of the shocks along the outflow cavity. 
	 \\	
	\textit{Results.} The profiles of the three excited water lines are similar, indicating that they probe the same gas. 
	Two main emission components are seen in all YSOs: a broad component associated with non-dissociative shocks in the outflow cavity wall (``cavity shocks'') and a narrow component associated with the quiescent envelope material. 
	More than 60\% of the total integrated intensity in the excited water lines comes from the broad cavity shock component, while the remaining emission comes mostly from the envelope for low-mass Class I, intermediate- and high-mass objects, and dissociative ``spot shocks'' for low-mass Class 0 protostars. 
	The widths of the water lines are surprisingly similar from low- to high-mass YSOs, whereas \twco\ \ten9 line widths increase slightly with \Lbol.  
	The excitation analysis of the cavity shock component shows stronger 752 GHz emission for high-mass YSOs, most likely due to pumping by an infrared radiation field. 
	Finally, a strong correlation with slope unity is measured between the logarithms of the total \water\ line luminosity, \Lwater, and \Lbol, which can be extrapolated to extragalactic sources. 
	This linear correlation, also found for CO, implies that both species primarily trace dense gas directly related to star formation activity. 
	 \\	 
	\textit{Conclusions.} The water emission probed by spectrally unresolved data is largely due to shocks. 
	Broad water and high-$J$ CO lines originate in shocks in the outflow cavity walls for both low- and high-mass YSOs, whereas lower-$J$ CO transitions mostly trace entrained outflow gas. 
	The higher UV field and turbulent motions in high-mass objects compared to their low-mass counterparts may explain the slightly different kinematical properties of \twco\ \ten9 and \water\ lines from low- to high-mass YSOs.
	}
\end{abstract}

   \maketitle



\section{Introduction}\label{ch4_Introduction}

The physical and chemical conditions present in low- and high-mass star-forming regions differ significantly. 
Massive star-forming regions are found to have higher UV radiation fields and levels of turbulence than their low-mass counterparts \citep[see][]{07Stauber, 12Herpin}. 
The temperatures, feedback mechanisms, magnetic fields, accretion rates, and outflow forces are different between low- and high-mass young stellar objects \citep[for more details see][]{96Bontemps, 01Behrend, 02Beuther, 07Beuther,  93PallaStahler, 05Cesaroni, 07ZinneckerYorke}.

However, many studies have shown that high-mass YSOs behave in certain aspects as scaled-up versions of their low-mass counterparts (\citealt{99vanderTak, 00vanderTak, 03Shepherd,12Johnston, 13SanJoseGarcia,14Karska}; \citealt{14SanJoseGarciaSub}, subm.). 
In addition, the lifetime of the embedded phase of high-mass YSOs (0.07-0.4 Myrs, \citealt{11Mottram}) is comparable to that of low-mass YSOs (0.15 Myr for Class 0, 0.5 Myr for Class 0+I, \citealt{14Dunham}), even if massive objects evolve more in this period.
The line luminosity of molecules like CO, HCO$^+$, and OH scales with bolometric luminosity and envelope mass, as well as the degree of turbulence in the warmer inner regions of protostellar envelopes (\citealt{13SanJoseGarcia, 13vanderTak,13Wampfler}; \citealt{15BenzSub}, subm.). 
Moreover, the kinematics of outflows and envelopes seem to be linked independently of the mass of the forming star \citep{13SanJoseGarcia}.
Therefore, a further characterisation of the physical conditions and dynamics of these components will help to identify the differences and similarities between low- and high-mass YSOs and better understand the fundamental processes in the formation of stars. 

Water is unquestionably a key molecule for studying the energetics and dynamical properties of protostellar environments \citep{11vanDishoeck}. 
In particular, the analysis of the velocity-resolved water data provided by the Heterodyne Instrument for the Far-Infrared (HIFI; \citealt{10deGraauw}) on board of {\it Herschel} Space Observatory \citep{10Pilbratt} allows us to characterise the emission from molecular outflows, which play a crucial role in the formation of young stars and in the feedback on their surroundings \citep{12Kristensen, 13vanderTak, 14Mottram}. 
Given that the bulk of the water data on extragalactic sources out to the highest redshifts and the data provided by the other {\it Herschel} instruments (the Photodetector Array Camera and Spectrometer, PACS, \citealt{10Poglitsch}; and the the Spectral and Photometric Imaging Receiver, SPIRE, \citealt{10Griffin}) on galactic sources are spectrally unresolved, it is important to quantify the different components that make up the observed lines. 

Outflows remove angular momentum effectively, which is necessary for the formation of a disk and mass accretion onto the forming star \citep[see review by][]{99Lada}.
The power agent of these structures (either jets or winds from the star/disk system) triggers not only the formation of the outflows, but also extreme and complex physical and chemical processes across the protostellar environment. 
In particular, different types of shocks take place in the outflow cavity wall at the interface between the cavity and the envelope. 
Non-dissociative outflow-cavity shocks are localised along the outflow cavity wall \citep{12Visser, 12Kristensen, 14Mottram}. 
On the other hand, dissociative shocks take place either along the jet, revealed through perturbations known as extremely high velocity components (EHV) \citep{90Bachiller, 10Tafalla}, or at the base of the outflow cavity wall where the jet or wind impacts directly \citep{12Kristensen, 13Kristensen, 14Mottram}.
These shocks are also called spot shocks. 
Therefore, shocks and turbulent motions injected into the cavity wall propagating within this physical structure are products linked to the activity of the molecular outflow \citep{07Arce}. 
The dynamical nature of these two phenomena (turbulence and shocks) is different, and they also differ from that characterising the entraining gas in classical outflows. 
To comprehensibly interpret the molecular emission of \water\ and \twco\ within the outflow-cavity system, it is important to investigate whether the dynamical properties of low-mass objects can be extrapolated to more massive YSOs.

This was one of the goals of the ``Water In Star-forming regions with {\it Herschel}'' key programme (WISH; \citealt{11vanDishoeck}), which observed several water lines, as well as high-$J$ CO and isotopologue transitions, for a large sample of YSOs covering early evolutionary stages over a wide range of luminosities. 
An extensive study has been performed on all HIFI water observations for low-mass protostars \citep{14Mottram} and for low-lying water transitions within the high-mass sub-sample \citep{13vanderTak}. 
The line profiles of the water transitions were analysed and decomposed into multiple velocity components, which are associated to different physical structures of the protostellar system.
These studies investigated trends with luminosity, mass, and evolution and explored the dynamical and excitation conditions probed by these lines. 
In addition, observations with PACS reveal that \twco\ $J$$>$20 transitions originate mostly in shocks for both low- and high-mass YSOs \citep{12Herczeg, 13Manoj, 13Green, 13Karska, 14Karska, 14Karska_b}. 
The excitation of warm CO is also similar across the luminosity range, but rotational temperatures in high-mass objects are higher than in their low-mass counterparts in the case of \water, due to their higher densities \citep[][]{14Karska}. 

In order to link these two studies, this paper focuses on the analysis of the excited water lines across the entire WISH sample of YSOs. 
The ground-state water transitions of high-mass sources show absorption features from foreground clouds which complicates the extraction of velocity information from these lines \citep{13vanderTak}, a reason why these lines are not considered in this study. 
Results from the line profile, line luminosity and excitation condition analysis are connected from low- to high-mass YSOs and interpreted together with those obtained from high-$J$ \twco\ observations ($J$$\geq$10).
In addition, the obtained results may help to interpret and understand those of extragalactic sources. 
The aim is to better constrain the dynamical properties of molecular outflows across a wide range of luminosities and complement the study presented in \citet{13SanJoseGarcia} based on the analysis of high-$J$ CO and isotopologue transitions for the same sample of YSOs. 

We start by introducing the selected sample, the studied \water\ and \twco\ observations and the reduction and decomposition methods in Sect.~\ref{ch4_Observations}.
The results from the water line profile and line luminosity analysis are presented in Sect.~\ref{ch4_Results} and compared to those obtained for CO. 
In this section, the excitation conditions are also derived from the line intensity ratios.
The interpretation of these results are discussed in Sect.~\ref{ch4_Discussion}, and extrapolated to extragalactic sources.
Finally, in Sect.~\ref{ch4_Conclusions} we summarise the main conclusions of this work.



\section{Observations}\label{ch4_Observations}

\subsection{Sample}\label{ch4_Sample}

The sample of 51 YSOs is drawn from the WISH survey and is composed of 26 low-mass, six intermediate-mass and 19 high-mass YSOs. 
In addition, the intermediate-mass object OMC-2 FIR\,4 \citep{13Kama} taken from the ``Chemical {\it HErschel} Surveys of Star forming regions'' key programme (CHESS; \citealt{10Ceccarelli}) is added to enlarge the number of sources of this sub-group. 
The intrinsic properties of each source such as coordinates, source velocity ($\varv_{\rm{LSR}}$), bolometric luminosity, distance ($d$), and envelope mass ($M_{\rm{env}}$) can be found in \citet{14Mottram}, \citet{13Wampfler} and \citet{13vanderTak} for the low-, intermediate- and high-mass YSOs respectively.

The sample covers a wide interval of luminosity and each sub-group of YSOs contains a mix of different evolutionary stages: both low- and intermediate-mass objects range from Class~0 to Class~I; and the high-mass YSOs from mid-IR-quiet/mid-IR-bright massive young stellar objects (MYSOs) to ultra-compact \ion{H}{ii} regions (UC \ion{H}{ii}).
The focus of this paper is to analyse different physical properties across the luminosity range; trends within the low-mass sample are discussed in \citet{14Mottram}; the intermediate- and high-mass samples are too small to search for trends within their several evolutionary stages.

\subsection{Water observations}\label{ch4_Water_observations}

The targeted para-H$_2$O \eight\ (988 GHz) and \seven\ (752 GHz) lines and the ortho-H$_2$O \tennine\ (1097 GHz) line were observed with the HIFI instrument. 
The upper energy level ($E_{\rm u}$), frequency, HIFI-band, beam efficiency ($\eta_{\mathrm{MB}}$), beam size ($\theta$) and spectral resolution of each water transition are given in Table~\ref{tbl4:H2O_lines}.
The beam efficiencies of the different HIFI-bands have been recently updated\footnote{\label{beameff}Further information regarding the updated beam efficiencies values is presented in the technical note ``Measured beam efficiencies on Mars (revision v1.1, 1 October 2014)'' of the HIFI wikipage: 
 http://herschel.esac.esa.int/twiki/bin/view/Public/HifiCalibrationWeb? template=viewprint} and in general the values decrease by 15-20\% for the band considered here. 
 The presented observations have not been corrected with the new $\eta_{\mathrm{MB}}$ parameters because the analysis in this paper was completed before the new numbers were available and for consistency with our previous CO study. 
For completeness, the new beam correction factor of each HIFI-band are listed in Table~\ref{tbl4:H2O_lines}.

The H$_2$O \eight\ line was observed for the entire WISH sample; the \seven\ line for 24 out of the 26 studied low-mass protostars and for all intermediate- and high-mass YSOs; and the \tennine\ transition was observed for only 10 low-mass protostars, two out of six intermediate-mass sources and all high-mass YSOs. 
These water lines are detected for all observed intermediate- and high-mass YSOs and for 75\% of the observed low-mass protostars \citep[see][]{14Mottram}. 

\TableOverviewWATER

The data were observed simultaneously by the Wide Band Spectrometer (WBS) and the High Resolution Spectrometer (HRS), in both vertical (V) and horizontal (H) polarisation (more details in \citealt{12Roelfsema}).   
We present the WBS data because the baseline subtraction for the HRS data becomes less reliable due to the width of the water lines, which is comparable to the bandwidth of the HRS setting. 
Single pointing observations were performed for all targets in dual-beam-switch (DBS) mode with a chop throw of 3$\arcmin$. 
Contamination from emission by the off-position has only been identified in the \water\ \eight\ spectrum of the low-mass protostar BHR71 \citep[further information in][]{14Mottram}.
The allocated observation numbers for each source and line, designated with the initial obsIDs, are indicated in Table A.2 of \citet{14Mottram} for the low-mass protostars, and in Table~\ref{tbl4:ObsIDs} of this manuscript for the intermediate- and high-mass YSOs.

\subsection{Additional $^{12}$CO observations}\label{ch4_Additional_data}

Complementing the water observations, \twco\ \ten9 and 16--15 spectra are included in this study to extend the comparison to other components of the protostellar system traced by this molecule and set a reference for abundance studies.  
The \ten9 transition was observed as part of the WISH key programme for the entire low- and intermediate-mass sample and for the high-mass object W3-IRS5 \citep[see][]{13SanJoseGarcia}.
The \ten9 spectrum was obtained for AFGL\,2591 from the work of \citet{14Kamierczak}.
For the other high-mass sources, \twco\ \three2 spectra are used as a proxy \citep{13SanJoseGarcia}. 

\twco\ \sixteen15 observations of 13 low-mass Class~0 protostars were observed within the OT2\_lkrist01\_2 {\it Herschel} programme (Kristensen et al. in prep.).
Finally, this transition was obtained for three high-mass YSOs: W3-IRS5 (OT2\_swampfle\_2 {\it Herschel} programme; Wampfler et. al 2014), and for AFGL\,2591 \citep{14Kamierczak} and NGC6334-I \citep{12Zernickel} as part of the CHESS key programme \citep{10Ceccarelli}.

\subsection{Reduction of the H$_2$O data}\label{ch4_Reduction}

The calibration process of the water observations was performed in the \textit{Herschel} Interactive Processing Environment (HIPE\footnote{HIPE is a joint development by the Herschel Science Ground Segment Consortium, consisting of ESA, the NASA Herschel Science Centre, and the HIFI, PACS and SPIRE consortia.}; \citealt{10Ott}) using version 8.2 or higher. 
The intensity was first converted to the antenna temperature $T_{A}^{*}$ scale and velocity calibrated with a $\varv_{\mathrm{LSR}}$ precision of a few m\,s$^{-1}$.  
Further reduction was performed with the {\small{GILDAS-\verb1CLASS1}}\footnote{{http://www.iram.fr/IRAMFR/GILDAS/}} package. 
The spectra observed in the H and V polarisations were averaged together to improve the signal-to-noise ratio ({\it S/N}) and the intensity scale converted to main-beam brightness temperature scale, $T_{\mathrm{MB}}$, using the specific beam efficiencies for each band \citep{12Roelfsema}.
Finally, a constant or linear baseline was subtracted.

All data were then resampled to 0.27 km\,s$^{-1}$ in order to compare the results among the water lines and to those from the high-$J$ CO lines \citep{13SanJoseGarcia} in a systematic manner. 
The rms noise of the spectra at that resolution, the maximum peak brightness temperature, $T_{\rm{MB}}^{\rm{peak}}$, the integrated intensity, \textit{W}=${\int{T_{\rm{MB}}{\rm{d}}\varv}}$, and the full width at zero intensity, \FWZI, are presented in Tables~\ref{tbl4:Mainparameters_202-111} to \ref{tbl4:Mainparameters_312-303}. 
The latter parameter was measured as in \citet{14Mottram}: first by resampling all spectra to 3~\kms\ to improve the {\it S/N}, then re-calculating the rms and finally considering the ``zero intensity" where the intensity of the spectrum drops below 1$\sigma$ of that rms.  
The velocity range constrained by the \FWZI\ is used to define the limits over which the integrated intensity of the line is calculated. 

Since the spectra have not been corrected with the recently released beam efficiencies of the different HIFI-bands$^1$ (Sect.~\ref{ch4_Water_observations}), the results presented in Tables~\ref{tbl4:Mainparameters_202-111} to \ref{tbl4:Mainparameters_312-303}, as well as those shown in Figs.~\ref{fig4:LH2O_vs_Lbol} and \ref{fig4:LH2O_vs_Lbol_extra}, should be divided by the correction factor indicated in Table~\ref{tbl4:H2O_lines}.
The line profiles do not change if the new $\eta_{\mathrm{MB}}$ values are applied and the variation of the line ratios is of the order of 1\%.

Finally, the \cei\ \ten9 emission line is detected in the line wing of the H$_2$O \tennine\ spectrum for five low-mass protostars (NGC\,1333\,IRAS2A and IRAS4B, Ser\,SMM1, GSS30 and Elias\,29) and four high-mass YSOs (G5.89-0.39, W3-IRS3, NGC6334-I and W51N-e1). 
Therefore, a Gaussian profile with the same {\it FWHM}, position and amplitude as those constrained in \citet{13SanJoseGarcia} was used to remove the contribution of \cei\ \ten9 line from the reduced H$_2$O \tennine\ spectrum for each of these sources.  
The data of these specific YSOs are then analysed and plotted after subtracting the \cei\ line.

\subsection{Decomposition method}\label{ch4_Decomposition_method}

As shown by \citet[][]{10Kristensen,12Kristensen}, \citet{13vanderTak} and \citet{14Mottram}, the water line profiles are complex and can be decomposed into multiple velocity components. 
The purpose of decomposing the line profile is to disentangle the different regions probed within the protostar, which are characterised by specific physical conditions and kinematics. 

Generally, these velocity components can be well reproduced by Gaussian-like profiles; other types of profiles do not give improved fits \citep[][]{14Mottram}.
Depending on the water transition and luminosity of the source, the number of components needed to fit the profile varies. 
For most of the low-mass protostars, the spectra can be decomposed into a maximum of four different Gaussian components, as shown in \citet{10Kristensen,12Kristensen,13Kristensen} and \citet{14Mottram}. 
In order to determine the number of velocity components of the water lines for the intermediate- and high-mass YSOs, these spectra were initially fit with one Gaussian profile using the {\sc idl} function \textit{mpfitfun}. 
Then, the residual from this fit was analysed and since it was larger than 3 sigma rms for all lines, an extra Gaussian component was added to the decomposition method to improve the fitting. 
The procedure is the same but now considering two independent Gaussian profiles. 
A self-absorption feature at the source velocity was detected in the \water\ \eight\ line for 9 out of 19 high-mass objects, so for those objects an extra Gaussian component in absorption was added in the decomposition method. 
In some high-mass sources this component is weaker or non-detected in the other studied transitions and it can be negligible (for an example, see the DR21(OH) observations).
Therefore, the number of components is determined by the spectrum itself and its {\it S/N} and not by the assumed method. 

As in \citet{14Mottram}, we force the {\it FWHM} and central position of each component to be the same for all \water\ transitions of a given sources.  
This procedure is adopted because, as for the water observations of the low-mass protostars, the width of the line profiles does not change significantly between the different observed transitions of a source \citep[see also Fig.~2 of][for several low-mass protostars]{13Kristensen}, suggesting that in each case the emission from the excited water lines comes from the same parcels of gas.
While these two parameters are constrained simultaneously by all available spectra of a given YSO, the intensities of each Gaussian component are free parameters that can vary for each transition. 
The resulting fits were examined visually as a sanity check. 
The values of the \FWHM, $T_{\rm peak}$, $\varv_{\rm peak}$, and integrated intensity of each Gaussian component are summarised in Appendix~\ref{ch4_Spectra_Excited_H2O_appendix}, Tables~\ref{tbl4:FWHM_cte1} to \ref{tbl4:FWHM_cte3}.

\subsection{Association with physical components}

The multiple velocity components needed to reproduce the \water\ line profiles can be related to physical components in protostellar systems \citep{11Kristensen, 12Kristensen, 13vanderTak, 14Mottram}. 

Quiescent inner envelope gas produces a Gaussian profile in emission with the smallest \FWHM\ centred at the source velocity (see Sect. 3.2.2 of \citealt{14Mottram}). 
In previous studies these velocity components were called narrow components.
The cold outer protostellar envelope can cause a self-absorption, which is more common in ground-state \water\ lines and in objects with massive and cold envelopes (e.g. \citealt{13vanderTak}, \citealt{13Mottram}). 

The chemical and physical conditions present in shocks increase the abundance of water molecules by sputtering from the grain mantles \citep{10Codella, 13VanLoo, 14Neufeld} and/or by the action of the high-temperature water formation route in the warm post-shock gas \citep{13vanDishoeck,14Suutarinen}.  
The line profile resulting from shocked water emission depends on the nature and kinematical properties of the shocks generating it, which translate into velocity components with different features (see Table 3 and Sect.~3.2 of \citealt{14Mottram}). 

The emission from non-dissociative shocks in layers along the outflow cavity wall produces velocity components with the largest \FWHM\ ($>$20~\kms) and are generally centred near the source velocity \citep{10Kristensen, 13Kristensen, 10vanKempen, 10Nisini, 14Suutarinen, 14Santangelo}. 
However, these broad Gaussian-like profiles, named cavity shock components \citep{14Mottram} or simply broad components, should be differentiated from the broad velocity component identified in low- and mid-$J$ ($J$$<$10) CO spectra, even if shape and width are similar.
The reason is that the water emission originates in shocks in the cavity while the CO emission comes from cooler material deeper in the wall and closer to the quiescent envelope \citep[][]{95Raga,13Yildiz}. 

In contrast, spot shocks occur in small localised regions and are associated to hotter and more energetic dissociative shocks \citep{14Mottram}. This emission may originate in extremely-high velocity (EHV) gas along the jets \citep{90Bachiller, 10Tafalla, 11Kristensen} or at the base of the outflow cavity (previously referred to as either the medium or the offset component; \citealt{13Kristensen}). 
These Gaussian profiles show smaller {\it FWHM}s than those measured for the cavity shock component and are in general more offset from the source velocity. 
A more detailed characterisation and discussion can be found in \citet{14Mottram}, \citet{13vanderTak} and Kristensen et al. in prep. 

The contribution of the cavity shock and envelope components with respect to the total integrated intensity of the water lines for the low-, intermediate- and high-mass YSOs are presented in Table~\ref{tbl4:ContributionCavity} together with the analogous contribution from the entrained outflowing material (broad) and envelope gas (narrow) components for the \twco\ \ten9 line (ratios derived from \citealt{13SanJoseGarcia}).
The values for the low-mass Class~0 and Class~I protostars were calculated by \citet{14Mottram} for different water transitions as well as the fraction corresponding to the spot shock component (see Table~4 of that manuscript). 

\TableContributionCavity

In the following, the different velocity components of the water and CO lines are distinguished according to the terminology based on the probable physical origin of the emission and the width of the profile.



\FigAveragedWater

\section{Results}\label{ch4_Results}  

The basic properties of the spectra and their decomposition are introduced in Sect.~\ref{ch4_Line_profile}. 
In Sect.~\ref{ch4_Comparison_H2OvsCO} the results of the line profile decomposition are compared to those obtained for the high-$J$ CO observations described in \citet{13SanJoseGarcia}. 
The water line luminosity properties are also compared to those of CO in Sect.~\ref{ch4_Line_luminosity_study}. 
In Sect.~\ref{ch4_Integrated_intensity_ratios} integrated intensity ratios calculated for different pairs of water transitions are presented and these line ratios are further studied as across the line profiles in Sect.~\ref{ch4_Intensity_ratios_vs_velocity}.
Finally, the excitation conditions of the water lines are analysed with the non-LTE radiative transfer code {\sc radex} in Sect.~\ref{ch4_Excitation_conditions}.

\subsection{Water line profile characterisation}\label{ch4_Line_profile}

The observed \water\ \eight, \seven, and \tennine\ spectra for the intermediate- and high-mass YSOs are presented in Appendix~\ref{ch4_Spectra_Excited_H2O_appendix}, Figs.~\ref{fig4:988GHzspectra}, \ref{fig4:752GHzspectra} and \ref{fig4:1097GHzspectra}, respectively. 
The Gaussian profile fitting the broad (cavity shock) component of each water transition and source is indicated with a pink line. 
The spectra of the low-mass protostars are shown in Appendix A of \citet{14Mottram}. 

In order to easily compare all the data, normalised averaged spectra of the \water\ \eight, \seven\ and \tennine\ transitions are computed for the low-mass Class~0 and Class~I protostars, the intermediate-mass sources and high-mass YSOs, as shown in the three first panels of Fig.~\ref{fig4:AveragedWater}. 
These spectra are calculated for each transition by shifting each spectrum to zero velocity, normalising it to its peak intensity and averaging it together with the observations of the corresponding sub-group of objects. 
The presence of self-absorption features for some of these sources, which are stronger for certain water transitions, prevents the normalised averaged spectra to reach unity for several of these lines. 
Independently of this issue, the normalised averaged spectra obtained for the three \water\ transitions are similar for each sub-type of YSOs.
Only the \water\ \eight\ high-mass averaged spectrum shows a slightly different profile with respect to the other two water lines because a larger number of YSOs show strong and deep self-absorption features. 
Except for the Class~I protostars, the averaged spectrum for a given transition seems to be broader for the low-mass objects, but at the base of the spectra the widths are similar, independent of the luminosity. 

In the right-hand panel of Fig.~\ref{fig4:AveragedWater} the three \water\ transitions (\eight\ in red, \seven\ in blue and \tennine\ in purple) are over-plotted for four different sources: a low-mass Class~0 (NGC1333\,IRAS4B), a low-mass Class~I (GSS\,30 \,IRS), an intermediate-mass objects (NGC2071) and a high-mass YSO (W33A). 
For each source, the shapes of the three water line profiles are similar but scaled-up in intensity, a result that is confirmed from the visual inspection of the water line profiles of all YSOs. 
In particular, the line wings are very similar.  
This indicates that the three water transitions are probing the same dynamical properties in each source.

Moving to the outcomes from the line decomposition explained in Sect.~\ref{ch4_Decomposition_method}, the analysis suggests that the quiescent envelope and cavity shock components are the only two physical components consistently present in the \water\ \eight, \seven\ and \tennine\ spectra of all low-, intermediate- and high-mass YSOs. 
The spot shock components are not detected for the excited water lines presented here towards high-mass YSOs and six out of seven intermediate-mass objects, though they have been seen in absorption against the outflow in some ground-state \water\ lines for some high-mass sources \citep[for more information see][]{13vanderTak}. 

As shown in Table~\ref{tbl4:ContributionCavity}, for a given sub-sample of YSOs the averaged contribution of the cavity shock component with respect to the total integrated intensity of the line is the same for the three water transitions. 
This fraction seems to decrease from low- to high-mass, but no statistically significant trend with \Lbol\ can be claimed because the specific contribution of the cavity shock emission varies from source to source. 
The remaining emission comes from the envelope in the case of the low-mass Class I, intermediate- and high-mass YSOs and from spot shock components for low-mass Class~0 protostars \citep[][]{14Mottram}. 
This picture is consistent with the average spectra presented in Fig.~\ref{fig4:AveragedWater}, where the envelope component of the water lines is more prominent for the high-mass sources than for their low-mass Class 0 counterparts. 
In addition, this narrow component associated with the envelope is also less prominent in the \water\ \tennine\ transition regardless of the YSO mass, as expected since the envelope is presumably composed of cool quiescent gas. 

Independently of these numbers, in this paper we focus on characterising the physical conditions causing the line-wing emission in the water line profiles, i. e., the broader velocity component associated to the shock emission along the outflow cavity.

\subsection{Comparison of the H$_2$O and $^{12}$CO line profiles}\label{ch4_Comparison_H2OvsCO}
\FigFWZICOwatervsLbol

The fourth panel in Fig.~\ref{fig4:AveragedWater} includes the normalised averaged \twco\ \ten9 spectrum of each sub-sample of YSOs. 
The procedure followed to obtain these spectra is the same as that used for the water data. 
These averaged profiles are clearly narrower than those of water, especially for the low-mass Class~0 protostars, and the width of the spectra seems to increase from low- to high-mass.
Therefore, just from a basic visual inspection of the water and the high-$J$ CO normalised averaged spectra we can point to differences in the shape of the line profiles of these two molecules and a different trend in the width from low- to high-mass.

To consistently compare the dynamical conditions of the entrained outflowing material traced by CO and the shocked gas along the outflow cavity, the line-wing emission of these two species is analysed using two parameters: the \FWHM\ and the \FWZI\ (see Sections~\ref{ch4_Reduction} and \ref{ch4_Decomposition_method}). 
Both variables are used because \FWHM\ characterises the average extent of emission from the source velocity while \FWZI\ characterises the fastest material. 
For simplification, the \FWHM\ of the Gaussian profile reproducing the cavity shock component is differentiated from the other velocity components by using the sub-script $b$ to indicate that this is the broader velocity component obtained from the line decomposition. 

The top panel of Fig.~\ref{fig4:FWZI_12CO_H2O_vsLbol} shows the \FWZI\ for the \water\ \eight\ transition as a function of bolometric luminosity (in Fig.~\ref{fig4:FWZIvsLbolMenv} the \FWZI\ of the other water lines are also plotted versus \Lbol\ and envelope mass). 
Similarly, the constrained \FWHMb\ from the cavity shock component (same for all three lines) versus \Lbol\ and \Menv\ are presented in Fig.~\ref{fig4:FWHM_12CO_H2O_vsLbol&Menv}. 
The \FWZI\ values vary from 15 to 189~\kms, while the \FWHMb\ range from 13 to 52~\kms. 
Due to the large scatter and dispersion of the data points, no trend or correlation with luminosity can be claimed in either case. 
The smaller \FWZI\ and \FWHMb\ values are those of the low-mass Class~I protostars, consistently lying at the bottom of these figures. 
In addition, the low- and intermediate-mass YSOs which show EHV components are marked with squares in Fig.~\ref{fig4:FWZI_12CO_H2O_vsLbol} to indicate that their \FWZI\ was calculated including the spot shock emission and to investigate if there is any particular trend for these objects.
The spectra of the marked high-mass object do not have EHV components but their line profiles are characterised with broad and triangular shapes. 
More information about these specific sources can be found in Appendix~\ref{ch4_Specific_sources_appendix}. 

As indicated in Sect.~\ref{ch4_Additional_data}, the $^{12}$CO \ten9 transition was not observed for most of the high-mass YSOs.
However, those sources for which both \twco\ \ten9 and \three2 transitions were available, the values of the constrained \FWHMb\ and also \FWZI\ are similar within the uncertainty \citep[see][]{13SanJoseGarcia}.  
Therefore, the \three2 transition is used as a proxy of the \ten9 spectra for the study of the kinematical structure of the outflowing gas. 
The \FWZI\ and \FWHMb\ for the $^{12}$CO observations (middle panels of Figs.~\ref{fig4:FWZI_12CO_H2O_vsLbol} and \ref{fig4:FWHM_12CO_H2O_vsLbol&Menv} respectively) are spread across a smaller velocity range than that for water. 
In addition, the \FWZI\ shows less scatter than the \FWHMb. 
There is a statistically significant trend of larger \FWZI\ for more luminous sources (4.7$\sigma$\footnote{The significance of a correlation for a given number of data points and Pearson correlation coefficient is calculated as described in \citet{10Marseille}.}) with a Pearson correlation coefficient $r=$ 0.72, which is also seen to a lesser extent for the \FWHMb. 

Table~\ref{tbl4:Averaged_FWZIFWHM} presents the mean \FWHMb\ and \FWZI\ values for \water\ and \twco\ and the averaged rms in a 0.27~\kms\ bin, $\sigma_{\rm rms}$. 
In the case of the high-mass YSOs, the derived values are not affected by the higher $\sigma_{\rm rms}$ in those data (as left panels of Fig.~\ref{fig4:AveragedWater} already show) since the actual signal to noise, {\it S/N}, on the water spectra themselves given by the peak intensity relative to the rms are higher (averaged {\it S/N} value of $\sim$60) than those of their low-mass counterparts (averaged {\it S/N} of $\sim$20).  

Without considering the low-mass Class~I protostars, which are more evolved and therefore have weaker, less powerful outflows \citep{14Mottram}, the average \FWHMb\ and \FWZI\ values derived for \water\ are similar from low- to high-mass (Table~\ref{tbl4:Averaged_FWZIFWHM}). 
A decrease of the mean \FWZI\ values with increasing luminosity is only hinted at for the \water\ \eight\ transition. 
Combining the results from both \FWHMb\ and \FWZI\ we conclude that the extent of the water line emission is similar for the entire sample. 
In contrast, and as suggested by the middle panel of Fig.~\ref{fig4:FWZI_12CO_H2O_vsLbol}, the averaged values of both \FWZI\ and \FWHMb\ for the \twco\ observations seem to increase with luminosity.

\TableFWZIFWHM

The dispersion observed for the \FWZI\ and \FWHMb\ in both \water\ and CO could be related to the intrinsic properties of the source, such as its inclination, evolutionary stage, clustering, etc.
In order to minimise possible effects caused by these inherent characteristics, the ratio of the \FWZI\ derived for the \twco\ observations and the \FWZI\ of the water lines is plotted versus the bolometric luminosity in the bottom panel of Fig.~\ref{fig4:FWZI_12CO_H2O_vsLbol}. 
The same procedure is followed for \FWHMb\ of the \twco\ and water spectra (see bottom panels of Fig.~\ref{fig4:FWHM_12CO_H2O_vsLbol&Menv} in Appendix~\ref{ch4_Additional_figures_appendix}).

Independently of the use of \FWZI\ or \FWHMb, a correlation between these ratios and \Lbol\ is measured for each of the three water lines with statistical significance between 3.3$\sigma$ and 5.0$\sigma$ (Pearson correlation coefficients between 0.50 and 0.75). 
While \twco\ \three2 and 10--9 may trace different layers in the outflow \citep{13Yildiz, 13Santangelo}, the \FWZI, i. e., the maximum offset velocity ($\varv_{\rm max}$), of the CO lines does not change with the $J$ transition (Kristensen et al. in prep.). 
Therefore, the use of \twco\ \three2 as a proxy for \twco\ \ten9 for the high-mass sources will not affect this trend.

The average ratios of \FWZI\ and \FWHMb\ are also given in Table~\ref{tbl4:Averaged_FWZIFWHM} and generally increase with increasing \Lbol. 
The \FWZI\ ratio indicates that the velocity of the material traced by the wings of \twco\ is larger than that of water for the high-mass YSOs. 
The \FWHMb\ values are basically the same for \twco\ and \water, independently of whether these molecules are probing different regions and physical conditions within the outflow. 
However, this is not the case for the low-mass protostars, which show larger line-wings for the water observations than for the \twco\ observations, consistent with \citet{12Kristensen}. 
This trend is further analysed and interpreted in Sect.~\ref{ch4_Disentangling_dynam_prop_H2OvsCO}.

\subsection{Line luminosity study}\label{ch4_Line_luminosity_study}

The integrated intensity of the water emission lines, \textit{W}=${\int{T_{\rm{MB}}{\rm{d}}\varv}}$, is calculated as described in Sect.~\ref{ch4_Reduction} and presented, together with its uncertainty, in Tables~\ref{tbl4:Mainparameters_202-111} to \ref{tbl4:Mainparameters_312-303}. 
For those \water\ \eight\ spectra with an absorption at the source velocity, the integrated intensity was determined by masking the absorption feature and using the area traced by the two Gaussian emission profiles fitting the spectrum. 

Since the studied sample of YSOs covers a wide range of luminosities and distances (from 0.13 to 7.9 kpc), the integrated intensity is converted to line luminosity, \Lwater.
This parameter takes into account the distance and can be compared to values obtained in extragalactic studies.
These quantities have been calculated using Eq. (2) of \citet{05Wu}, assuming a Gaussian beam with size according to Table~\ref{tbl4:H2O_lines} and point source objects.
The uncertainties in the line luminosity are calculated from the rms of the spectrum and assuming a distance uncertainty of $\sim$20\%. 

Figure~\ref{fig4:LH2O_vs_Lbol} presents the logarithm of the line luminosity for each water transition, log(\Lwater), versus the logarithm of the bolometric luminosity, log($L_{\rm{bol}}$). 
A strong correlation between the logarithm of these quantities is measured (solid black line) for each of the three lines.
The Pearson correlation coefficient, $r$, is larger than 0.9 and the trends extend over more than six orders of magnitude in both axes. 
Similarly, a strong correlation is also measured between the logarithm of \Lwater\ and the logarithm of the envelope mass, $M_{\rm env}$ (see Fig.~\ref{fig4:LH2O_vs_Menv} in Appendix~\ref{ch4_Additional_figures_appendix}). 
The parameters of the fit and correlation coefficient for all water lines and for both \Lbol\ and \Menv\ are presented in Table~\ref{tbl4:Correlation_fits}. 

\FigLwatervsLbol
\TableCorrelationFits

The slope of the correlations are close to unity within the uncertainty. 
As there is no saturation of the line luminosity, the emission must be optically thin or at least effectively thin (i.e. optically thick but sub-thermally excited, see \citealt{77Linke}). 
In either case, the intensity scales as the product of the beam-averaged column density and average volume density in the emitting gas. 
Assuming that the volume and column density in the regions probed by water in low- and high-mass protostars and the source geometry are similar, a correlation between log(\Lwater) and log(\Lbol) with slope close to unity suggests that the emitting region size increases proportionally with \Lbol\ (see Sections~\ref{ch4_Excitation_conditions} and \ref{ch4_Correlations_with_Lbol} for further discussion). 
The \tennine\ transition shows the smallest number of detections for the low-mass protostars, which could explain the difference in the value of the slope with respect the other two water lines. 
This linear correlation of log(\Lwater) and log(\Lbol) should not be extrapolated to the ground-state water transitions due to the presence of broad and deep absorption features which complicate the comparison of the water line luminosity, particularly in the case of massive YSOs \citep[see][]{13vanderTak}.

The same correlation between the $^{12}$CO 10--9 line luminosity and \Lbol\ was measured by \citet[][their Fig.~6]{13SanJoseGarcia}. 
The calculated slope of the linear fit is the same within the uncertainty to those derived for the water lines \citep[see Table 4 of][for comparison]{13SanJoseGarcia}. 
As also noted by \citet{14Mottram}, the low-mass Class~I sources have lower water line luminosities than the Class~0 sources, due to decreases in both \FWHM\ and \FWZI\ (see Table 2), as well as peak intensities, causing a clear separation between these two classes of low-mass protostars. 
However, in the case of CO, the separation between the low-mass Class~0 and Class~I was small. 
Both $^{12}$CO \ten9 and especially water are more sensitive to fast moving material in the outflows, which differentiate sources at different evolutionary stages, while the CO isotopologues are mostly probing the bulk of quiescent envelope material. 
This confirms the conclusions of \citet{14Mottram} that the difference in water emission between Class~0 and Class~I sources is related to the decreasing strength of the outflow and not to the removal of the envelope. 

The integrated intensity emission coming from the cavity shocks,
i. e., the area of the Gaussian profile fitting the line-wings, is
also converted to line luminosity,
$L_{\mathrm{broad\,H}_{2}\mathrm{O}}$.
Figure~\ref{fig4:LH2Onor_vsLbol} of Appendix~\ref{ch4_Additional_figures_appendix}
shows the logarithm of this quantity as a function of log(\Lbol).  As
for the total line luminosity, a strong correlation which extends
across the entire luminosity range is measured but with slightly
smaller values of the slope (averaged slope of 0.78$\pm$0.04).
Figure~\ref{fig4:LH2O_vs_Lbol} includes for each water transition the
relation between the logarithm of
$L_{\mathrm{broad\,H}_{2}\mathrm{O}}$ and \Lbol\ (dashed black line)
for comparison with the total line luminosity, \Lwater, (solid black
line).  The parameters from the log-log correlations of the broad
component are also presented in
Table~\ref{tbl4:Correlation_fits}. Consistent with the larger envelope
contribution for high-mass sources (see
Table~\ref{tbl4:ContributionCavity}) the differences between the fits
for \Lwater\ and $L_{\mathrm{broad\,H}_{2}\mathrm{O}}$ are larger for
all three lines for the high-mass sub-sample.

The fact that we found the same strong correlation between log(\Lbol) and log(\Lwater) or log($L_{\mathrm{broad\,H}_{2}\mathrm{O}}$), i.e., similar values for the slope close to unity, confirms that the total water line luminosity is dominated by the broad component (outflow cavity shocks). 

Finally, we found that the line width of the cavity shocked gas seems independent of the source luminosity (see previous section), i.e, the kinetic energy of the outflow increases little from low- to high-mass YSOs.
Thus, this suggests that the physical processes originating and powering the outflows (the dynamical conditions) are similar from low- and high-mass and it is the mass of the outflow, i.e, the total cavity shocked water gas, what increases at higher \Lbol\ through a larger emitting area. 

\subsection{Integrated intensity ratios}\label{ch4_Integrated_intensity_ratios}

The integrated intensity ratios of two different water transitions across the luminosity range can be used to analyse the excitation and physical conditions of protostellar systems. 
\citet{14Mottram} show that there is little variation in line ratio between outflow-related components for the low-mass sample.
They showed that the lines lie in the optically thick but effectively thin regime.
This means that the water lines presented in that study are sub-thermally excited due to their high critical densities compared to the actual gas density of the emitting region.
This condition lowers the chance of collisional de-excitation, so photons will scatter and eventually escape as if the lines were optically thin. 
We will extend this study to intermediate- and high-mass YSOs, first considering the total integrated intensity ratios and then focusing on just the cavity shock component. 

The observations of the different water transitions have different beam sizes which must be corrected by the corresponding beam-size ratio, $\theta_{\rm 1}$/$\theta_{\rm 2}$. 
However, the correction depends on the size and shape of the emitting region within the protostellar system covered by each beam. 
If the emission comes from a point source, the correction factor is ($\theta_{\rm 1}$/$\theta_{\rm 2}$)$^2$; if the emitting area is conical, meaning that it fills the beam in one axis and is point-like in the other, then the correction applied should be ($\theta_{\rm 1}$/$\theta_{\rm 2}$); and if the emitting region covers both axes the factor used is 1 \citep[][]{10Tafalla}.
For the considered water lines, the maximum scaling correction would be 2.1, corresponding to a point-source scaling of the \seven\ (752 GHz) and  \tennine\ (1097 GHz) beams. 

The emission from the excited water transitions for the low-mass protostars mostly comes from the outflow cavity, which extend beyond a single {\it Herschel} beam \citep[see][]{10Nisini, 12Santangelo, 14Santangelo,14Mottram}.
In this case, the emitting area could be taken as conical or even assumed to fill both beams. 
So, either the beam correction factor is ($\theta_{\rm 1}$/$\theta_{\rm 2}$) or 1 (or something in between for other, more complex geometries). 
For the intermediate- and high-mass YSOs it is more complicated to define the area covered by the beams as the larger distances mean that the outflows are often unresolved in the {\it Herschel} beam. 
For this reason, we consider the three scenarios introduced in the previous paragraph and the entire range of intensity ratios (applying the three possible beam correction factors) are evaluated together with the results obtained from their low-mass counterparts. 
The $\theta_{\rm 1}$/$\theta_{\rm 2}$ value for each pair of water lines are indicated in Table~\ref{tbl4:Averaged_ratios_lines}.

The ratios of the total integrated intensity of \water\ \seven\ / \eight; \tennine\ / \eight; and \tennine\ / \seven\ are shown as a function of bolometric luminosity in the top-, middle- and bottom-left panels of Fig.~\ref{fig4:ratioLH2OvsLbol}, respectively. 
The value of these ratios without applying the beam corrections are indicated by filled dots. 
The dashed lighter line connected to each dot represents the value of the ratio after applying the larger beam size correction assumed for each sub-sample of objects: $\theta_{\rm 1}$/$\theta_{\rm 2}$ linear correction for the low-mass protostars and ($\theta_{\rm 1}$/$\theta_{\rm 2}$)$^2$ for the intermediate- and high-mass YSOs.
Therefore, the lighter vertical lines illustrate the range of values that the integrated intensity ratio could take for each source depending on the shape of the emission region. 

Figure~\ref{fig4:ratioLH2OvsLbol} also shows the analytical ratios of
the integrated intensity derived for two extreme scenarios: when both
water lines are optically thin (purple dotted-dashed line) and when
they are optically thick (orange dashed line).  In both cases local
thermodynamical equilibrium (LTE), an excitation temperature, $T_{\rm
  exc}$, of 300~K and the same beam size for the compared transitions
are assumed.  The specific numbers corresponding to each regime for
each pair of water lines are presented in
Table~\ref{tbl4:Averaged_ratios_lines}.  We take $T_{\rm exc}$$\approx$300~K because this is the temperature found for the cavity
  shock component by \citep{14Mottram}. Also, \citet[][and references
  cited]{13Karska} observe with {\it Herschel} PACS a warm component
  for CO with an excitation temperature around 300 K and water is
  found to be spatially associated with this component in the
  outflow cavity \citep{13Santangelo}. Note that for high densities,
  the excitation temperature approaches the kinetic temperature. 

\FigRatiosLuminoLbol

The ratio of the \seven\ (752 GHz) and \eight\ (988 GHz) transitions (top-left panel of Fig.~\ref{fig4:ratioLH2OvsLbol}) shows a significant increase with \Lbol. 
The trend of the non-beam-corrected ratios (dots) is indicated by the dotted black line and confirmed by a Pearson correlation coefficient, $r$, equal to 0.83.
This increase with luminosity is also seen when the maximum beam correction factor for each sub-group of YSOs (delineated by the dash) is applied ($r=$ 0.85). 
Therefore, the result is consistent across the luminosity range and regardless of emitting region shape. 
Furthermore, the intensity ratio of these \water\ transitions is above 1 only for those sources with \Lbol\ $>$ 10$^3$ $L_{\odot}$.  

The \tennine\ (1097 GHz) to \eight\ (988 GHz) intensity ratio also seems to increase slightly with luminosity (middle-left panel Fig.~\ref{fig4:ratioLH2OvsLbol}).
However, no trend can be claimed on a statistical basis because the significance of the Pearson correlation coefficient is less than 2.5$\sigma$ ($r$$<$0.5 and $N=$ 32). 
These ratios lie below 1 for all sources, independent of whether the beam correction factor has been applied (which is small for this combination of lines). 
The intensity ratio of the \tennine\ (1097 GHz) to \seven\ (752 GHz) (bottom-left panel) is generally smaller than 1 for the three sub-sample of YSOs and it seems to decrease with \Lbol.
Once again, a trend cannot be claimed because the significance of this correlation is less than 2.3$\sigma$ ($r$$<$-0.5 and $N=$ 31). 

Comparison of the left and  right panels of Fig.~\ref{fig4:ratioLH2OvsLbol} show that similar results are obtained if the integrated intensity ratio corresponding to only the broad component is used 
for the same pair of water transitions as a function of \Lbol.
A trend with luminosity is also only observed for the \seven\ / \eight\ intensity ratio (with a smaller correlation coefficient $r=$ 0.66).
The increase of the \seven\ / \eight\ ratio with \Lbol\ suggests that the emission from the shocked gas along the outflow cavity becomes more excited, i.e., warmer and/or denser, for more massive objects. 
The \tennine\ / \eight\ ratio seems to increase with \Lbol\ while the \tennine\ / \seven\ ratio decreases with luminosity, both of them tendencies not statistically significant. 

\TableRatiolines

The mean and standard deviation of the intensity ratios for the broad component are presented in Table~\ref{tbl4:Averaged_ratios_lines}, together with the beam size ratios and the estimated intensity ratios for the optically thin and thick solutions calculated assuming LTE and $T_{\rm exc}=$ 300~K. 
These calculated values ratify the results and tends from low- to high-mass described in the previous paragraph. 

We can rule out the optically thin LTE solution for all sources and lines, except when the largest beam size correction is applied to the massive YSOs in the case of the \seven\ / \eight\ integrated intensity ratio.
The optically thick LTE solution for the \tennine\ / \eight\ ratios can be also excluded for the entire sample of objects. 
In order to further constrain the excitation conditions, the non-LTE radiative transfer code {\sc radex} will be used in Sect.~\ref{ch4_Excitation_conditions}.

\subsection{Intensity ratios versus velocity for H$_2$O and $^{12}$CO}\label{ch4_Intensity_ratios_vs_velocity}

Here we investigate whether the excitation conditions change with velocity.
Figure~\ref{fig4:Intensity_ratios} shows the calculated line ratios of the \water\ \seven\ (752 GHz; top-panel) and \tennine\ (1097 GHz; bottom-panel) transitions over the \eight\ (988 GHz) line as a function of velocity with respect to the source velocity. 
The grey shaded areas correspond to the averaged line ratio and standard deviation of the low-mass Class~0 protostars presented in \citet{14Mottram}.
The green and red lines are associated with the intensity ratios of the intermediate- and high-mass YSOs respectively.

To obtain these ratios, the spectra of each set of water transitions for a given source are resampled to 1~\kms\ bins.
Then, the studied line (either the 752 or 1097 GHz transition) is divided by the spectrum of the 988 GHz line within the range of velocities over which the intensities of both lines are above 3$\sigma$ of the re-binned data.
Finally, the red and blue wings are averaged together and the obtained intensity ratio as a function of velocity is averaged over all sources that compose each sub-sample of YSOs.
Only velocity ranges well offset from the quiescent envelope emission are plotted, typically $|\varv-\varv_{\rm lsr}|$ $>$ 3-5 \kms.

\FigIntensityRatios
\FigIntensityRatiosTwo

As found for the low-mass Class~0 protostars, the line intensity ratios seem to be constant as a function of velocity for both intermediate- and high-mass YSOs. 
This result confirms the fact that the shape of the three water line profiles is similar for a given source and just scales in intensity (Fig.~\ref{fig4:AveragedWater}, right).   
For the high-mass objects, the ratio increases slightly for offset velocities $>$25~\kms.
However, the significance of the increase in the ratio is of the order of the uncertainty of those channels. 
The \seven\ / \eight\ ratio is above 1 for this sub-sample of objects at all velocities, which is consistent with the result obtained in the previous sub-section for this set of water lines.
A constant line ratio as a function of velocity rules out the optically thick LTE solution for all lines as this would require high column densities even at high velocities, and thus a flatter line profile than observed, or high opacity across the entire profile. 

A different outcome is obtained if the CO line intensity is compared to that of water as a function of velocity.
Figure~\ref{fig4:Intensity_ratios2} shows the \twco\ \ten9 and $J$=16--15 over the \water\ \eight\ (988 GHz) line ratio as a function of velocity offset (top and bottom panel respectively).
In this case, we observe two different behaviours for the two sub-samples of YSOs.
For the low-mass Class~0 protostars the \twco\ \ten9 / \water\ \eight\ line ratio decreases with velocity, while the \twco\ $J$=16--15 / \water\ \eight\ ratio is invariant with velocity. 

On the other hand, for the high-mass YSOs the line ratio of both CO transitions with \water\ \eight\ is constant with velocity across the studied velocity interval.
This agrees with the ratio of the \FWZI\ and \FWHMb\ for these two molecules being slightly greater and around unity respectively (see red crosses in Figs.~\ref{fig4:FWZI_12CO_H2O_vsLbol} and \ref{fig4:FWHM_12CO_H2O_vsLbol&Menv}).
Therefore, in massive objects the emission of these CO and water transitions may originate in regions of the outflow cavity wall not too distant from each other and characterised by similar excitation and dynamical conditions. 
For the low-mass Class~0 sources, the \twco\ 16--15 line seems to trace the same material as \water\ but the 10--9 transition does not (Kristensen et al. in prep.).

\subsection{Excitation conditions}\label{ch4_Excitation_conditions}

In order to characterise the excitation conditions responsible for the broad cavity shock water emission, we compared the line fluxes presented above, together with the fluxes for the \water\ 1$_{11}-$0$_{00}$ transition from M$^c$Coey et al. in prep. and \citet{13vanderTak}, for the intermediate and high-mass YSOs to a grid of non-LTE \textsc{radex} calculations \citep{07vanderTak}.
We use the same modifications and parameters as adopted by \citet{14Mottram} for the low-mass cavity shock components  
and we assume 1-D slab geometry, an ortho-to-para ratio of H$_2$ of 3 and a circularly emitting region at the distance of the source. 
The area of this emitting region in the plane of the sky is derived by comparing the modelled and observed \water\ \eight\ integrated intensity and by determining the fraction of the beam for which the modelled and integrated intensity values match. 

We find a good fit with post-shock volume densities $n_{\mathrm{H}_{2}}$=10$^{5}-$4$\times$10$^{6}$\,cm$^{-3}$ and water column densities $N_{\mathrm{H}_{2}\mathrm{O}}$=10$^{17}-$10$^{18}$\,cm$^{-2}$ for 5/7 of the intermediate-mass sources. 
The results for NGC7129 are shown as an example in the top panels of Fig.~\ref{fig4:Excitation}. 
These results are similar to lower-density solution obtained for the low-mass sources by \citet{14Mottram} but with slightly larger emitting regions sizes, equivalent to radii of 300$-$500\,AU if circular. 
All studied water transitions are optically thick but effectively thin in these conditions. 
In one of the remaining two intermediate-mass sources (L1641\,S3MMS1) we cannot constrain the fit very well, while for the other source (NGC2071) a good fit cannot be found within the grid ($\chi^{2}_{\mathrm{best}} >$ 100), with the emission in the \water\ 2$_{11}-$2$_{02}$ line underproduced and emission in the 1$_{11}-$0$_{00}$ line overproduced by the models. 
For the high-mass YSOs, good fits can only be found for 6 of the 19 sources, with the best fits again giving similar densities and column densities and emitting regions ranging from 800 to 6000\,AU. 
However, even in these cases the best-fit models consistently underproduce the \water\ 2$_{11}-$2$_{02}$ transition. 
To conclude, if we assume a circular geometry for all sources, the radius of this emitting region is $\sim$10--200 AU for low-mass protostars, $\sim$200--600 AU for intermediate-mass YSOs and $\sim$600--6000 AU for non-pumped high-mass sources. 

In order to find a good solution for these high-mass sources, we first ran additional grids of models for the poorly fit source G10.47+0.03 where the kinetic temperature was increased from the default value of 300\,K up to 1500\,K. 
However, the {\sc radex} models continued underproducing the \water\ \seven\ transition, independently of the adopted temperature, and no better solutions were found. 
This suggests that the excitation of these observed water lines (with $E_{\rm u}$ in range up to $\sim$250~K) is not sensitive to temperature, as also found for their low-mass counterparts (c.f. Fig. 11 of \citealt{14Mottram}). 
Although a realistic shock environment has a range of temperatures this would provide similar line ratios within the uncertainty.
This scenario is confirmed by the fact that the intensity ratios do not change with velocity (see Fig.~\ref{fig4:Intensity_ratios}), even if the line profile is composed of cavity and spot shock emission, i.e, gas which is at different temperatures (see Fig.~9 of \citealt{14Mottram}). 
In addition, this conclusion is consistent with the results of \citet{14Karska}, who find similar rotational temperatures for warm CO emission observed with PACS for both low- and high-mass sources. 

\FigExcitation

Since the observed \water\ \seven\ / \eight\ line ratio for most high-mass YSOs is larger than unity, an alternative scenario is explored for one of these massive YSOs (G10.47+0.03). 
In this case, we study the effect of pumping by an infrared radiation field with the same shape as the source SED from \citet{13vanderTak}, and scaled assuming that the source of infrared radiation is at varying distances from the water emitting region.  
Such an approach was found by \citet{14Mottram} to be inconsistent with the observations of low-mass protostars, but \citet{14Gonzalez-Alfonso} show that this is needed to reproduce \water\ observations towards some galaxies. 
For the high-mass YSO G10.47+0.03, we find that pumping by infrared photons can indeed provide a better fit to the data (compare middle and lower panels of Fig.~\ref{fig4:Excitation}). 
This result is consistent with previous studies based on observations of vibrational excited molecules such as HC$_{\rm 3}$N and HCN \citep{99Wyrowski, 11Rolffs}.
We explored a range of distances between the source of the infrared radiation and the H$_{2}$O emitting region, from 10 to 1000\,AU, and find the best fit for a distance of $\sim$100\,AU. 
The best fit for this strong radiation field has both high densities ($\sim$10$^{9}$--10$^{11}$\,cm$^{-3}$) and column densities ($\sim$10$^{18}$\,cm$^{-2}$) for emitting regions with sizes of order 500\,AU. 
All other high-mass sources were then tested over a more limited grid of distances. 
For 10 out of 19 sources, inclusion of an infrared radiation field improved the fit to the line ratios, while for the remaining 9 YSOs the fit was worse. 
This latter group contains 3/5 of the mid-IR quiet HMPOs and 3/4 of the hot molecular cores within the WISH sample, but also 3/5 of the UCH{\sc ii} regions. 
More detailed investigation is needed to ascertain why certain sources seem to shield water from infrared radiation better than others. 
The implications and reasonableness of these results will be discussed further in Sect.~\ref{ch4_Excitation_conditions_across_Lbol}.



\section{Discussion}\label{ch4_Discussion}

\subsection{Disentangling the dynamical properties of H$_2$O and CO emission}\label{ch4_Disentangling_dynam_prop_H2OvsCO}

The data presented in Sect.~\ref{ch4_Results} suggest that \water\
  appears in a different physical component from some of the CO
  lines. It is now well established that the broad component
  seen in low- ($J$$<$5) and mid-$J$ (5$\leq$$J$$<$10) \twco\ lines
  probes the colder entrained outflow gas, whereas only the high-$J$
  ($J$$\geq$10) \twco\ transitions trace the currently shocked gas also
  seen in water 
  \citep{09vanKempen_a, 10vanKempen, 10Nisini,
    12Santangelo, 13Tafalla, 12Goicoechea, 13Yildiz,13SanJoseGarcia}. 
The cartoon presented in Fig.~\ref{fig4:cartoonOutflowCavityWall}
illustrates a simplified version of a possible framework for the
origin of \twco\ (yellow, orange and red areas) and water (blue lines
and opaque region) within the outflow cavity wall for a low- (LM) and
high-mass (HM) YSO.  Turbulent motions, likely driven by the outflow
(\citealt{13SanJoseGarcia}; \citealt{14SanJoseGarciaSub}, subm.) are 
represented with spiral symbols and the entrainment with swirls.

The yellow areas are related to colder layers of the outflow cavity wall directly attached to the envelope where the broad velocity component of low-$J$ \twco\ transitions originates. 
The orange regions are associated with warmer entrained outflowing material within the outflow cavity wall and the red areas indicate material that is currently mixing and undergoing shocks along the outflow cavity. 
Finally, the velocity of the layers forming the outflow cavity wall and the temperature of the gas decrease with distance from the cavity and the central source \citep[][]{97Raga&Canto}. 
The shocked gas along the layers closest to the outflow cavity should have higher average velocities than that originating near the envelope.  

\FigcartoonOutflowCavityWall

In this context, Fig.~\ref{fig4:FWZI_12CO_H2O_vsLbol} and Fig.~\ref{fig4:FWHM_12CO_H2O_vsLbol&Menv} show that both the values of \FWZI\ and \FWHMb\ behave differently with \Lbol\ for \water\ and high-$J$ \twco. 
In the case of the low-mass protostars, the \FWHMb\ of the water lines (which probes the shocked gas 
along the outflow cavity) is larger than that of \twco\ \ten9 (which traces the warm entrained outflowing material). 
In addition, the decrease of the \twco\ \ten9 / \water\ \eight\ ratio as a function of velocity (Fig.~\ref{fig4:Intensity_ratios2}) suggests that these two transitions are kinematically distinct, and originate in different layers of the outflow cavity wall. 
On the other hand, the ratio of the CO 16--15 / \eight\ line intensity is constant with velocity, so water and \twco\ 16--15 emission go together along the outflow cavity wall. 

For the high-mass YSOs, the \FWHMb\ for high-$J$ \twco\ transitions is comparable to that of \water\ (see bottom panel of
Fig.~\ref{fig4:FWHM_12CO_H2O_vsLbol&Menv}).
Furthermore, the line intensity ratio of these two species is constant as a function of velocity for both the \twco\ \ten9 and 16--15 transitions. 
This implies that the emission of the \twco\ \ten9 transition already comes from a region with dynamical properties that are similar to those of water. 
This is in contrast to what is seen for their low-mass counterparts where \twco\ \ten9 and water do not go together \citep{13Yildiz} suggesting that the water emission arises in warmer gas as probed by higher-$J$ CO lines, e.g. \twco\ \sixteen15 ($E_{\rm up}/k_{\rm B} = 750$~K), consistent with what is seen at off-source shock positions \citep[e.g.][]{13Santangelo}. 
In the following we will discuss two scenarios to explain these results.

The first scenario is based on the fact that high-mass objects have stronger UV fields than those measured for their low-mass counterparts \citep{07Stauber}. 
The strong UV radiation would principally destroy water molecules in the warmer layers of the outflow cavity wall closer to the cavity because colder entrained material will be shielded. 
Therefore, water may be effectively photo-dissociated (and photo-desorbed) deeper into the outflow cavity wall for high-mass objects than for low-mass protostars. 
CO molecules are not as easily photo-dissociated as \water, but still the destruction of CO will be more efficient in the layers nearest to the outflow cavity. 
This means that for massive YSOs, the water molecules present in layers of the outflow cavity wall closer to the wall, which move at higher velocities, are destroyed while the water molecules located in regions near to the entrained layer, which move at slower velocities, may survive. 

Due to the weaker UV fields, the photo-dissociation of water molecules along layers closer to the outflow cavity is less efficient for low-mass protostars. 
Therefore, a higher percentage of fast shocked gas emission along the regions closest to the cavity may survive, emission which contributes to the broadening of the line-wings \citep[quantitative models in][]{12Visser}. 

An alternative scenario is based on the fact that the protostellar environments of high-mass YSOs are more turbulent than those of their low-mass counterparts (see \citealt{12Herpin}, \citealt{13SanJoseGarcia}, \citealt{14SanJoseGarciaSub}, subm. and bottom panel of Fig.~\ref{fig4:cartoonOutflowCavityWall}).
Outflow systems inject turbulent motions into the structure forming the outflow cavity wall and these motions could propagate to other physical components of the protostellar system such as the entrained outflowing material.
In this case, the material from both components may have a high level of mixing and the separation between the entrained and shocked material within the outflow cavity wall would not be as clearly delineated as for the low-mass protostars. 
Since the turbulent motions in low-mass protostars are weaker, the different layers constituting the cavity wall are better defined.
Thus, the kinematical properties of each $J$ transition are better differentiated and the emission of CO \ten9 material is separated from that originating in shocked water gas along the cavity. 
Therefore, the dynamical properties of these two structures would be more closely linked for the high-mass YSOs than for their less massive counterparts, being harder to disentangle. 
This could explain why the calculated ratio of the \FWZI\ or \FWHMb\ for the \twco\ \ten9 and \water\ profiles is around 1 for the high-mass sub-sample but is lower than 1 for their less massive counterparts.

Finally, we can rule out that high-mass YSOs are analogous to low-mass Class~I protostars, for which the line profiles are narrower due to decreased wind velocities \citep[c.f.][]{12Kristensen, 14Mottram}. 
The high-mass sources considered here are closer to the low-mass Class~0 objects in terms of evolutionary stage because the ratios of the envelope mass to gravitational stellar mass, $M_{\rm env}$/$M_{\rm g}$, are $>$20 (\citealt{14SanJoseGarciaSub}, subm.). 
Massive YSOs are therefore still deeply embedded and follow the relation with luminosity of the Class~0 rather than to the Class~I low-mass sources.

\subsection{Excitation condition across the luminosity range}\label{ch4_Excitation_conditions_across_Lbol}

As discussed in Sect.~\ref{ch4_Excitation_conditions}, the excitation conditions of water in low- and intermediate-mass sources seem to be similar.
However, for high-mass YSOs it is difficult to reproduce the observed line ratios, particularly those involving the \water\ 2$_{11}-$2$_{02}$ line (752 GHz), with a similar model.
Since water excitation is not sensitive to temperature \citep{14Mottram}, the other plausible scenario is that for the majority of the high-mass sources the lines may be pumped by infrared radiation, a process probably not present in low- and intermediate-mass objects due to lower luminosities. 
However, the best-fit excitation conditions obtained for the high-mass source G10.47+0.03 with the inclusion of a radiation field require very high densities, enough that water emission should be in LTE, over considerable regions. 
Note that from the grid without radiative pumping we can rule out that high densities alone can reproduce the observed water line ratios. 
The regions around these high-mass YSOs undoubtably have a complex geometry and in several cases, including G10.47+0.03, are known to harbour multiple sources within the \textit{Herschel} beam. 
Therefore, the assumption of a slab geometry, while instructive, may not accurately constrain the range of excitation conditions present.   
In addition, we have only performed a limited test using a scaled version of the SED as the radiation field, while the ``true'' radiation field irradiating the water emitting gas in these regions may include shorter-wavelength photons. 
Furthermore, we assume all the gas is irradiated by the same radiation field, when the geometry of the water emitting region could lead to variation in the field experienced by different parts of the gas. 

With all these caveats, we therefore can conclude two things. 
Firstly, the excitation of water in most high-mass sources is different to low- and intermediate-mass objects. 
Secondly, while radiative pumping can be ruled out in low-mass protostars, it is a plausible explanation for high-mass sources given that higher kinetic temperatures do not lead to strong enough emission in the \seven\ transition.
In addition, a similar gas temperature for low- and high-mass YSOs is consistent with the results obtained from PACS observations, which show a comparable CO rotation temperature for low- and high-mass objects \citep{14Karska}.

\subsection{Correlations with bolometric luminosity}\label{ch4_Correlations_with_Lbol}

Figure~\ref{fig4:LH2O_vs_Lbol} illustrates the relation between the luminosities of the
water lines and the bolometric luminosity of the sources from low- to
high-mass.  The inferred slope on the log-log scale ranges from 0.76
to 0.94. \citet{13SanJoseGarcia} found a similar relation for the various
high-$J$ CO isotopolog lines, with slopes around unity.

What is the origin of these strong correlations? 
As shown in \citealt{15BenzSub} (submitted, Appendix D), the luminosity of any optically thick line originating from the protostellar envelope scales with the radius, $r_{\rm thick}$, where the line becomes optically thick (i.e. the $\tau$=1 photosphere) and the temperature at that radius:

\begin{equation}
L_{\rm line} \propto r_{\rm thick}^2 B_{\nu}(T(r_{\rm thick})).
\end{equation}

It can be shown that $r_{\rm thick}$ scales roughly as $r_{\rm thick}\propto L_{\rm bol}^{0.25-0.4}$.  
For a power-law density structure, $T \propto L_{\rm bol}^{0.2}\,r^{-0.4}$ in the outer envelope. 
Inserting these relations into the above equation one can show that $L_{\rm line} \propto L_{\rm bol}^{\alpha}$ with the exponent $\alpha$ in the 0.60--0.82 range. 
This is entirely consistent with the slope of 0.76 observed for species such as HCO$^+$ through its $J$=6--5 line (\citealt{15BenzSub}, subm.). 

This analysis should hold for any molecule or line that is optically thick and whose emission arises primarily from the envelope. 
The slope depends on where the line becomes optically thick, so it will vary with different molecules and transitions. 
These arguments could thus explain the strong relation found, for example, for the \thco\ \ten9 lines by \citet{13SanJoseGarcia}, which are dominated by narrow envelope emission.  
The somewhat steeper relation of $\sim$1.0 found in that case could be due to additional heating for high-mass sources due to UV radiation, enhancing the temperature. 

For \twco\ \ten9, a significant fraction of the emission comes from the outflow rather than the envelope. 
Table~\ref{tbl4:ContributionCavity} shows that the fraction of envelope emission is rather constant at around 50\% across the luminosity range, so that a similar relation could still hold, albeit with extra scatter due to the variation in outflow contribution for each source.

The situation is in principle different for the \water\ lines, which are dominated by outflow material and which are effectively optically thin \citep[see][and Table~\ref{tbl4:ContributionCavity}]{14Mottram}. 
In this case, the line luminosity should scale with the total mass of water. 
The (swept-up) outflow mass as measured from low-$J$ CO lines has been found observationally to scale as $M_{\rm outflow} \propto L_{\rm bol}^{0.56}$ \citep{05Wu}. 
If we assume that the same relation holds for the shocked gas where \water\ emission originates and if the gas temperature still has a weak dependence on \Lbol, one could imagine that various factors conspire to give a slope around 0.75. 
As argued in Sections~\ref{ch4_Line_luminosity_study} and \ref{ch4_Excitation_conditions}, the emitting area increases from low- to high-mass YSOs. 
Table~\ref{tbl4:ContributionCavity} also shows that the fraction of water emission from the envelope increases strongly from low- to high-mass sources, from 0\% to close to 40\%. 
This envelope water emission is optically thick and thus the above analysis holds, which would introduce an extra slope on top of the outflow relation.

\subsection{From Galactic to extragalactic sources}\label{ch4_Galactic_to_extragalactic_sources}
 
\citet{13SanJoseGarcia} found that the correlation between the logarithms of the \twco\ \ten9 line luminosity and \Lbol\ for the WISH sample held when they extended their analysis to include a sample of extragalactic sources. 
Similarly, we investigate if the significant correlation measured between the logarithm of the water luminosities, \Lwater, and \Lbol\ continues in the regime of extragalactic sources.

\FigLwatervsLbolExtrag

Figure~\ref{fig4:LH2O_vs_Lbol_extra} shows the values of \Lwater\ for a series of AGN and other extragalactic objects presented in \citet{13Yang}. 
A correlation between the logarithm of these parameters, which extends over more than 12 orders of magnitude in both axis, is measured for the three transitions.
The value of the slope is the same for each \water\ line and is equal to 1, which in the case of the \tennine\ transition is notably larger than the slope obtained considering only the WISH sample.
Therefore, if the extragalactic sources are added, the proportionality between the logarithms of the line luminosity and \Lbol\ is emphasised for all water lines and the slope becomes unity within the uncertainties, as for the \twco\ \ten9 analysis of \citet{13SanJoseGarcia}.
As previously discussed in Sect.~4.3 of that paper, this result suggests that both high-$J$ CO and excited water transitions can be used as a tracer of dense gas, similar to HCN \citep{04Gao&Solomon, 04Gao&Solomon_a} and that the power-law index does not become sub-linear for higher excitation transitions.
Indeed, \citet{15Liu} also show that for \twco\ \ten9 this relation continues to be linear even up to \Lbol=10$^{13}$ \Lsun. 
Regardless of the underlying physical cause, the fact that the line and bolometric luminosities are strongly correlated across such a wide range of \Lbol\ means that water and mid-$J$ CO are good and relatively clean tracers of star formation. 

The observed line intensity ratios of the cavity shock component are determined for the three water lines with respect to the \eight\ transition (see Fig.~\ref{fig4:ratioLH2O_vs_Eu_extrag}). 
The results are presented in the context of the extragalactic framework by comparing the excitation conditions of low-, intermediate- and high-mass YSOs to those investigated in the extragalactic sources discussed in \citet{13Yang}. 

\FigLwaterNorvsEuExtrag

Surprisingly, the 752 GHz / 988 GHz line intensity ratio only lies above 1 for the high-mass YSOs. 
This peculiar result, already highlighted within our studied sample of YSOs, also stands out when the line ratios are compared with those measured for extragalactic sources. 
However, as discussed in Sect.~\ref{ch4_Excitation_conditions_across_Lbol}, the modelling used in this paper is quite simple and is only for lines with $E_{\rm u}\lesssim$250\,K. 
Better characterisation of the excitation conditions of more \water\ lines than modelled here is therefore required to conclusively interpret this result.

To summarise, the fact that the slope of the correlation between \Lwater\ and \Lbol\ (Fig.~\ref{fig4:LH2O_vs_Lbol_extra}) gets closer to unity, that the significance of the relation improves, and that the dashed line drops below the low-mass Class~0 protostars might suggest a plausible scenario in which the \water\ emission from extragalactic sources comes from more evolved and likely massive sources, and therefore indirectly measures the star formation potential. 
However, a more fundamental question is whether the extragalactic emission comes from star formation, PDRs, AGN or some combination thereof. 
Once there is better understanding of where extragalactic water emission originates, further progress can be made on comparing to Galactic observations such as those presented in this paper and on explaining the measured difference between the dashed and solid lines.



\section{Conclusions}\label{ch4_Conclusions}

From  the analysis of excited \water\ and high-$J$ \twco\ observations across the entire WISH sample of YSOs we derive the following conclusions:

\begin{itemize}
  \renewcommand{\labelitemi}{$\bullet$}

	\item The shape of the \water\ \eight, \seven\ and \tennine\ line profiles for a given source are similar but scaled in intensity.
	In addition, their averaged line intensity ratios are constant with increasing offset from the source velocity. 
	These results suggest that the dynamical properties probed by the three water transitions are similar for a given source.
	
	\item On average, more than 60\% of the total integrated \water\ line intensity comes from the cavity shock component for all YSOs. 
	The calculated contribution fractions can be used to interpret and compare velocity resolved and unresolved observations of water. 
	
	\item \FWZI\ and \FWHMb\ change little between low- and high-mass YSOs for the \water\ spectra, while the \FWZI s of the \twco\ \ten9 and 3--2 transitions increase slightly with \Lbol. 
	For both molecules, there is significant scatter in the values of \FWZI\ and \FWHMb\ over the studied luminosity interval, particularly in the case of water (\FWZI\ ranges from 15 to 189~\kms).	
	
	\item This scatter, related to the intrinsic properties of the sources and the varying line-of-sight orientation of the source, is reduced when the ratio of the \twco\ and \water\ \FWZI\ or \FWHMb\ were considered. 
	The averaged ratio of their \FWZI\ increases with the source bolometric luminosity ranging from 0.5 to 1.9 and from 0.6 to 0.9 in the case of the \FWHMb. 
	This suggests that for low-mass protostars the entrained material traced by the wings of the \twco\ \ten9 (and lower-$J$ lines) is kinematically different to that traced by water in the cavity shocked gas. 
	In the case of the high-mass YSOs, the CO \ten9 emitting region seems to coexist with that of water, so both may come from the same or from closer layers of the outflow cavity wall (Fig.~\ref{fig4:cartoonOutflowCavityWall}). 
	
	\item The line intensity ratio of \twco\ \ten9 to water decreases with increasing velocity for low-mass YSOs but is approximately constant for high-mass YSOs (Fig.~\ref{fig4:Intensity_ratios2}, {\it top}), confirming the previous conclusion. 
	However, the line intensity ratio of \twco\ \sixteen15 and \water\ remains constant for both low- and high-mass YSOs as a function of velocity (Fig.~\ref{fig4:Intensity_ratios2}, {\it bottom}). 
	This suggests that higher-$J$ \twco\ transitions ($J$$>$10) trace deeper warmer layers of the outflow cavity characterised by dynamical properties likely linked to those of the shocked water emission, independently of the mass of the object. 
		
	\item Two scenarios were proposed to explain the previous result. 
	The first scenario assumes that the increased UV radiation in high-mass YSOs leads to water being more effectively photo-dissociated in those layers of the cavity wall closer to the outflow cavity than for the low-mass protostars. 
	An alternative scenario was also discussed, where the stronger turbulence in high-mass star-forming regions, injected by the outflows, propagate into layers of the outflow cavity wall further from the outflow cavity and cause a larger mixing of the material forming those layers than in low-mass protostars. 
	Therefore, the high level of turbulence helps the interaction of warm entrained and cavity shocked material.

	\item The water excitation for the high-mass YSOs is somewhat different from that of their low- and intermediate-mass counterparts. 
	Since the excitation of the observed \water\ lines is not sensitive to temperature for any of the three sub-types of objects, pumping from 
	an infrared radiation field is the most likely explanation for how the results are reproduced for the majority of the massive YSOs. 

	\item As for CO, the logarithm of the \water\ line luminosity (as well as the emission from the cavity shock component) strongly correlates with the logarithm of the bolometric luminosity and envelope mass of the object with the slope of the correlations being unity within the uncertainty. 
	Although not yet fully understood, the relation seems to imply that as \Lbol\ increases, whatever drives the outflow (wind/jet from the source) can heat material further from the source and/or deeper into the cavity wall, and that the shock cooling structure (and thus $n$ and source-averaged $N$) is universal. 
This then means that the emission is always dominated by material at similar conditions but more luminous sources keep more gas at those conditions. 
	
	\item The correlation between the total \water\ line luminosity can be extended to extragalactic sources. 
	In addition, the values of the line intensity ratios of the extragalactic objects are similar to those calculated for the low-mass protostars, but these ratios differ from those measured for the intermediate- and high-mass YSOs. 
	Thus, both high-$J$ CO and \water\ trace star formation. 
 
\end{itemize}

Overall, despite some differences in excitation, the properties of excited water line emission are similar between low- and high-mass YSOs, suggesting that a common physical mechanism is at work in the outflows of sources across a wide range of luminosity. 
High spatial resolution and fidelity imaging of tracers probing both the entrained and currently shocked outflow material will help us to draw a more comprehensive picture of the mechanisms driving the outflows of all YSOs as well and the physical conditions of the material constituting their outflow cavity wall, something ALMA and JWST will enable us to do in the coming years for the first time.


\begin{acknowledgements} 
The authors are grateful to the anonymous referee for the constructive and thorough reviews, to
Sylvie Cabrit for the useful feedbacks that helped to improve the manuscript and to Mihkel Kama for providing data and suggestions to the paper. 
We would like to thank Umut Y{\i}ld{\i}z for helping on the data reduction as well as the WISH team for many inspiring discussions. 
This work is supported by the European Community's Seventh Framework
Programme FP7/2007-2013 under grant agreement 238258 (LASSIE)
and by the Space Research Organisation Netherlands
(SRON). 
Astrochemistry in Leiden is supported by the Netherlands Research
School for Astronomy (NOVA), by a Spinoza grant, A-ERC grant 291141 CHEMPLAN and grant 614.001.008
from the Netherlands Organisation for Scientific Research (NWO).
HIFI has been designed and built by a consortium of 
institutes and university departments from across Europe, Canada and the 
United States under the leadership of SRON Netherlands Institute for Space
Research, Groningen, The Netherlands and with major contributions from 
Germany, France and the US. Consortium members are: Canada: CSA, 
U.Waterloo; France: CESR, LAB, LERMA, IRAM; Germany: KOSMA, 
MPIfR, MPS; Ireland, NUI Maynooth; Italy: ASI, IFSI-INAF, Osservatorio 
Astrofisico di Arcetri- INAF; Netherlands: SRON, TUD; Poland: CAMK, CBK; 
Spain: Observatorio Astron{\'o}mico Nacional (IGN), Centro de Astrobiolog{\'i}a 
(CSIC-INTA). Sweden: Chalmers University of Technology - MC2, RSS $\&$ 
GARD; Onsala Space Observatory; Swedish National Space Board, Stockholm 
University - Stockholm Observatory; Switzerland: ETH Zurich, FHNW; USA: 
Caltech, JPL, NHSC.
\end{acknowledgements}

\bibliographystyle{aa} 
\bibliography{bibdata}

\Online
\appendix

\section{ Spectra of the excited water lines}\label{ch4_Spectra_Excited_H2O_appendix}

The \water\ \eight, \seven\ and \tennine\ spectra for the intermediate- and high-mass YSOs and the observation number identification are presented in this section (see Figs.~\ref{fig4:988GHzspectra} to \ref{fig4:1097GHzspectra} and Table~\ref{tbl4:ObsIDs}).
The basic properties derived from these data, such as the rms of the spectra, $T_{\rm{MB}}^{\rm{peak}}$, integrated 
intensity and \FWZI\ are summarised from Table~\ref{tbl4:Mainparameters_202-111} to \ref{tbl4:Mainparameters_312-303}. 
In addition, the results from the Gaussian decomposition explained in Sect.~\ref{ch4_Decomposition_method} are shown in Tables~\ref{tbl4:FWHM_cte1} to \ref{tbl4:FWHM_cte3}. 

\TableObdIDs

\TableMainValuesA
\TableMainValuesB
\TableMainValuesC

\TableFWHMcteWaterFst
\TableFWHMcteWaterSnd
\TableFWHMcteWaterTrd

\FigNineGHz
\FigSevenGHz
\FigTenGHz

\section{Specific sources}\label{ch4_Specific_sources_appendix}

The low-mass Class~0 protostars indicated in Fig.~\ref{fig4:FWZI_12CO_H2O_vsLbol} are characterised for showing bullet emission in their water profiles. 
These already studied sources are: L1448-MM, NGC1333\,IRAS2A, BHR\,71, Ser\,SMM1 and L1157.
In the case of the low-mass Class~I object IRAS12496, a spot shock component significantly offset from the source velocity is identified in the 988 GHz water line in emission but in absorption in the other ground-based transitions. 
More information about these objects and their observations in \citet{14Mottram}.

The excited water spectra of the intermediate-mass YSO NGC2071 show an offset component at around 35~\kms\ from the source velocity, which could be considered as a spot shock component, in particular an EHV component.
This source is known for being formed by several YSOs, which line profile could be then composed by the emission of several sources and molecular outflows \citep{12vanKempen}.
Therefore, interferometric data is needed disentangle and better understand the emission. 

Finally, the line profiles of the high-mass YSO G5.89-0.39 are complex and composed by velocity components with non-Gaussian profiles. 
As for NGC2071, this region is crowded with several protostars and energetic outflows \citep{12Su} with some of the highest velocities measured for these structures \citep{93Choi}
Therefore, the interpretation of the emission should go together with extra single-dish and interferometric observations.

\section{Additional figures}\label{ch4_Additional_figures_appendix}

\FigFWHMvsMenvforCOwater
\FigFWZIvsLbolMenv

\FigFWZIcovsLbolMenv

\FigLwatervsMenv
\FigLbroadnorvsLbol

\end{document}